\documentclass[11pt,a4paper]{amsart}

\usepackage{hyperref}
\hypersetup{
    colorlinks=true,
    linkcolor=blue,
    citecolor=red,
    pdfpagemode=FullScreen,
    pdftitle={Aletti,Baldi,Crimaldi,Frieri,Ghiglietti}
    }

\numberwithin{equation}{section}

\usepackage{url}
\usepackage{bbm}

\usepackage{amssymb}
\usepackage{graphicx}
\graphicspath{{Fig/}}

\usepackage{amsmath}
\usepackage{rotating}
\usepackage{lscape}
\usepackage{rotfloat}
\usepackage{subfig}
\usepackage{times,dsfont,srcltx}
\usepackage{amsfonts}
\usepackage[bottom]{footmisc}
\usepackage{dsfont,srcltx}
\usepackage[english]{babel}
\usepackage{multirow}

\theoremstyle{plain}
\newtheorem{thm}{Theorem}[section]
\newtheorem{lem}[thm]{Lemma}
\newtheorem{prop}[thm]{Proposition}

\newtheorem*{rem}{Remark}
\theoremstyle{definition}

\allowdisplaybreaks[4]

\usepackage{tikz}
\foreach \a in {q,w,e,r,t,y,u,i,o,p,a,s,d,f,g,h,j,k,l,z,x,c,v,b,n,m,Q,W,E,R,T,Y,U,I,O,P,A,S,D,F,G,H,J,K,L,Z,X,C,V,B,N,M}{
  \expandafter\xdef\csname v\a\endcsname {
		{\noexpand\mathbf{\a}}
	}
}

\newcommand{\vdelta}{{\boldsymbol{\delta}}}
\newcommand{\vDelta}{{\boldsymbol{\Delta}}}
\newcommand{\diag}{\text{{diag}}}

\newcommand{\vtheta}{{\boldsymbol{\theta}}}
\newcommand{\valpha}{{\boldsymbol{\alpha}}}
\newcommand{\vbeta}{{\boldsymbol{\beta}}}

\begin{document}

\title[Multi-arm clinical trials with informational borrowing]{Design and inference for multi-arm clinical trials with informational borrowing: the interacting urns design}

\author[G. Aletti]{Giacomo Aletti}
\address{G. Aletti, Department of Environmental Science and Policy,
  Universit\`a degli Studi di Milano, Italy}
\email{giacomo.aletti@unimi.it}

\author[A. Baldi Antognini]{Alessandro Baldi Antognini}
\address{A. Baldi Antognini, Department of Statistical Sciences, University of Bologna, Bologna, Italy}
\email{a.baldi@unibo.it}

\author[I. Crimaldi]{Irene Crimaldi}
\address{I. Crimaldi, IMT School for Advanced Studies Lucca, Lucca, Italy}
\email{irene.crimaldi@imtlucca.it  (Corresponding author)}

\author[R. Frieri]{Rosamarie Frieri}
\address{R. Frieri, Department of Statistical Sciences, University of Bologna, Bologna, Italy}
\email{a.baldi@unibo.it}

\author[A. Ghiglietti]{Andrea Ghiglietti}
\address{A. Ghiglietti, Universit\`a degli Studi di Milano-Bicocca, Milan, Italy}
\email{andrea.ghiglietti@unimib.it}

\begin{abstract}
This paper deals with a new design methodology for stratified comparative experiments based on a system of interacting urns. The key idea is to model the interaction between urns for borrowing information across strata and to use it in the design phase in order to i) enhance the information exchange at the beginning of the study, when only few subjects have been enrolled and the stratum-specific information on treatments' efficacy could be scarce, ii) let the information sharing adaptively evolve via an update mechanism based on the observed outcomes, for skewing at each step the allocations towards the stratum-specific most promising treatment and iii) make the contribution of the strata with different treatment efficacy vanishing as the stratum information grows. In particular, we introduce the Interacting Urns Design, namely a new Covariate-Adjusted Response-Adaptive procedure, that randomizes the treatment allocations according to the evolution of the urn system. The theoretical properties of this proposal are described and the corresponding asymptotic inference is provided. Moreover, by a functional central limit theorem, we obtain the asymptotic joint distribution of the Wald-type sequential test statistics, which allows to sequentially monitor the suggested design in the clinical practice.

\end{abstract}
\maketitle

\section{Introduction}
The present paper introduces a new design methodology for stratified experiments aimed at comparing multiple treatments. In particular, consider a sequential study to compare the efficacy of $J\geq2$ experimental arms, where each enrolled subject is stratified based on a categorical covariate taking $H$ possible different values (e.g., tumor sub-type, genetic mutation or, more generally, a predictive risk factor) not under the investigator's control. Each statistical unit is randomized to one of the available treatments and a binary response is observed. In this setting, the general experimental problem is to assess the treatment efficacy in each of the $H$ sub-populations or strata identified by the levels of the covariate. 

In many circumstances, the stratum-specific effectiveness of a treatment is related to its performance in other strata, for example, when an experimental drug targets a common risk factor shared across multiple disease subtypes. Thus, leveraging this relationship by sharing information across similar strata becomes essential to strengthen statistical conclusions about treatment effects. Indeed, enriching the information on the treatment efficacy in a given stratum through the performance of the same treatment in other strata could strongly improve the efficiency of the study. This concept of ``borrowing of information" has been increasingly adopted to improve inference in sub-group analyses by combining data in suitably pooled evaluations (see, e.g., \cite{Berry13,Chu18,Cun17,Thall03, Neu16} ) and, as shown by Lu et al. \cite{Lu21}, it induces increased efficiencies, accelerates drug development, reduces cost, and potentially improves patient care (see also \cite{Saville16,Hobbes18}).   
To account for the relationships among stratum-specific treatment effects, some statistical models have been proposed in the literature. For instance, let $\theta_{j,h}$ denote the success probability of treatment $j=1,\ldots,J$ in the $h$th stratum ($h=1,\ldots,H$), we can assume in a Bayesian framework that $\theta_{j,1},\ldots,\theta_{j,H}$ (eventually, after suitable transformations)  
are $H$ independent realizations of a common prior distribution. Then, the variance of such prior shows the stratum-specific oscillations around the mean, representing the overall efficacy of treatment $j$ (in this setting, the logit transformation is usually matched with a normal prior). Otherwise, adopting hierarchical models, $(\theta_{j,1},\ldots,\theta_{j,H})$ could be regarded as a vector of $H$ independent realizations of a random variable $\Theta_{j}$ following a Beta distribution $\mathcal{B}(\alpha_j,\beta_j)$. Alternatively, they could be modeled by an overall efficacy of treatment $j$ perturbed by additive stratum-specific effects, as in the fixed/random effects framework (see \cite{BLAST,Hobbes18,Thall03,Ventz17,Zheng22,Zhou24}).

While these approaches focus primarily on the inferential phase, the goal of this paper is to develop a theoretical framework for integrating information borrowing already at the design stage. 
Introducing borrowing mechanisms from the design stage can be
particularly relevant in clinical contexts, as they may improve inferential
efficiency and support early stopping decisions for either efficacy or futility.
In addition, although methods that explicitly adapt allocation ratios are still rarely used
in practice, combining borrowing with response-adaptive allocation rules may
also help pursue ethically desirable goals, such as assigning more patients to
better-performing treatments.

 In stratified experiments the best performing treatment possibly changes on the basis of the subjects' characteristics, thus the treatment decision depends on stratum-specific estimates. However, especially at the early stages of a study, when only a limited number of patients have been enrolled, stratum-specific estimates of treatment efficacy are highly unstable, making the exchange of information between strata particularly relevant.
Moreover, we do not assume a pre-specified model for the information sharing across strata, and instead provide a general template that can be adapted to any modeling framework, both frequentist or Bayesian.
\\

These challenges arise in a wide range of experimental scenarios involving multiple patients' subpopulations, among which master protocols stand out as an important example of application. 
Master protocols are a flexible class of experiments aimed at testing multiple treatments in different sub-populations within a single overarching protocol, guaranteeing a more efficient and rapid drug development.  They represent one of the most recent advancements in personalized medicine (\cite{Wood17, park19, fdaMP}) reflecting the current shift of biomedical research towards tailored therapies for specific subgroups of patients sharing a common characteristic (e.g. genetic/molecular aberrations or proteomic/metabolic factors).   Originally proposed after the Ebola virus epidemic of 2013-2014 and the COVID-19 pandemic outbreak,  these experiments currently represent
an important innovation in oncology and infectious diseases and are becoming increasingly widespread in other
clinical areas, particularly through basket and platform trials, allowing for the rapid evaluation of several treatments and with the ideal aim of ensuring the best possible care for the participants. Although the flexibility of master protocols is widely recognized in the literature (\cite{park19}), methodological developments for tackling this complex
experimental framework remain relatively limited, especially from the design of experiments perspective. In particular, while some
works have considered adaptive design features, such as sample size adaptation \cite{Zheng23, Zheng23biom}, the problem of adapting allocation ratios across
treatments and strata has received  less attention. 
 Excluding rare exceptions (such as \cite{Ventz17}), the great majority of these trials still adopt balanced allocation within strata, and a comprehensive theoretical framework integrating design and inference is missing; up to now, the proposed procedures are always evaluated just through simulation studies (see, e.g., \cite{BLAST,Zheng22}).\\

To address these issues, this paper proposes a novel sequential design that incorporates information borrowing through a system of randomly updated interacting urns.  

Urn models as adaptive designs in clinical trials have a long history in the statistical literature, dating back to classical works  in adaptive designs (see e.g. \cite{WeiLac88}, for a review) and reinforced processes (e.g., \cite{AlGhRo, BaiHu, May09}).  More recently, research has extended to urn models interacting on a finite graph, allowing for more complex dependency structures (see \cite{urne1,urne2,urne3}).  The design proposed in this work is extremely different from previously studied interacting urn models, as the updating mechanism is explicitly driven by the observed outcomes to meet the specific demands of clinical trials. Indeed, the proposed design does not present any of the asymptotic behaviors induced by the urn dynamics adopted in \cite{urne1,urne2,urne3} (e.g. the synchronization or the periodicity of the urn proportions).

In this paper, we introduce a new covariate-adjusted response-adaptive (CARA) design (see, e.g., \cite{baldi2015adaptive,Hu06,Zhang07}) for stratified multi-armed binary trials, called the interacting urns design (IUD), where the informational borrowing evolves as a system of interacting updated urns in order to:
\begin{itemize}
  \item enhance the information exchange especially at the early stage of the experiment, when stratum-specific information on the treatment performance is limited;
  \item let the information sharing evolve adaptively, for skewing at each step the allocations in each stratum towards the most promising treatment;
  \item ensure that the amount of informational borrowing from the strata with different treatment efficacy decreases as the number of enrolled patients grows.
\end{itemize}

To this aim, three alternative approaches are proposed, each corresponding to a specific update mechanism for the urns system. In the first approach, the informational sharing within each stratum vanishes as its sample size increases; the second approach allows for information borrowing across strata provided that the estimated treatment efficacies are sufficiently similar; in the last one, the borrowing of information naturally evolves with weights induced by a generative model for the treatment efficacy (avoiding subjective experimental choices).  For each of the update mechanisms, the theoretical properties of the IUD are described together with the corresponding asymptotic inference. To show the desirable properties of the proposed procedure, in particular for small samples, we include an extensive simulation study. 

In addition, the paper contributes with two further important methodological developments. First, by applying a functional central limit theorem for triangular arrays of multivariate martingale, we prove that, under the IUD, the Wald-type sequential test statistics converge in law to a Wiener process and asymptotically follow the canonical joint distribution of Jennison and Turnbull \cite{Jen00}. This result enables sequential monitoring of the proposed design, allowing type I error control  (and other possible experimental objectives) by choosing appropriate boundaries. This extends the current theory of sequential monitoring for adaptive designs, which, up to now, has been developed only for Response-Adaptive procedures \cite{Zhu10} and Covariate-Adaptive rules \cite{Zhu19} and results for CARA designs are not yet available in the literature.  Second, as a byproduct, new results about maximum likelihood inference for the beta-binomial distribution are also provided, for both iid observations and data accrued from adaptive designs. 

The major methodological advancements and novel aspects of this paper are the following. 
i) We provide a theoretical framework that addresses the design and inference. In contrast to the existing literature, which focuses on inference under standard balanced designs (e.g., completely randomized or permuted block designs), we propose a CARA design that yields non-iid observations, motivated by ethical considerations.
ii) At the finite sample level, the effect of information borrowing can be highly relevant, both from an ethical perspective (for all IUD procedures) and from an inferential one (in the case of IUD2, which aggregates information). This may allow the trial to be stopped earlier, either in favor of a treatment that has proven effective or to exclude ineffective ones.
iii) While the asymptotic results may appear standard, they are derived under non-standard CARA designs and rely on estimators that account for information borrowing, (different from classical MLEs), we establish a new asymptotic result that justifies the sequential monitoring of CARA designs, which has not yet been addressed in the literature.
Finally, iv) although we focus on the binary outcome case, our results are not limited to this setting and naturally extend to general types of responses through appropriate transformations.

\section{Notation and model}
Consider a clinical experiment with $n$ subjects enrolled sequentially. For each subject entering the trial we observe a qualitative covariate that is assumed to be a random variable not under the experimenters' control, but it can be measured before assigning a treatment. When the $i$th patient is ready to be enrolled, her/his covariate value $Z_i=h\in\{1,\ldots,H\}$ can be recorded; then she/he is assigned to treatment $j\in\{1,\dots, J\}$ according to a given randomization rule and a binary outcome $Y_i$ is observed. For each statistical unit, the treatment assignment is stored into a $J$-dimensional vector $\boldsymbol{\delta}_i=(\delta_{1,i},\ldots, \delta_{J,i})^\top$, with $\delta_{j,i}=1$ if the $i$th subject receives treatment $j$ and 0 otherwise (where, clearly, $\sum_{j=1}^J\delta_{j,i}=1$ for every $i\geq1$). We assume that $\{Z_i\}_{i\geq1}$ is a sequence of independent and identically distributed random variables with common probability distribution $\mathbf{p}=\left(p_1,\ldots,p_H\right)^\top$, where $p_h=\Pr(Z_i=h)>0$ for every $h=1,\ldots,H$ and $\sum_{h=1}^H p_h=1$. Conditionally on the covariates and the treatments, patients' responses are assumed to be independent binary variables with
\begin{equation*}
\Pr(Y_{i}=1\mid  \delta_{j,i}=1,Z_i=h)=\theta_{j, h} , \quad i\geq1, \; j=1,\dots, J \; \text{and} \; h=1,\ldots,H.
\end{equation*}
After $n$ assignments, let ${\vZ}_n=(Z_1,\ldots,Z_n)^\top$, $\boldsymbol{\Delta}_n=(\boldsymbol{\delta}_1^\top,\dots,\boldsymbol{\delta}_n^\top)$ and ${\vY}_n=(Y_1,\ldots,Y_n)^\top$ be the vectors of  covariates, allocations and responses, respectively, so ${\mathcal F}_n=\sigma(\vZ_n,\boldsymbol{\Delta}_n,\vY_n)$ denotes the sigma-algebra representing the information accrued up to that step. We assume that responses are observed immediately after treatment administration, so that when randomizing subject $n+1$, the information ${\mathcal F}_n$ is available.  Moreover, for each stratum $h=1,\ldots,H$, we let $S_{j,h,n}=\sum_{i=1}^n \mathbb{I}_{\{Z_{i}=h\}}\delta_{j,i}Y_{i}$, $F_{j,h,n}=\sum_{i=1}^n \mathbb{I}_{\{Z_{i}=h\}}\delta_{j,i}(1-Y_{i})$ and $N_{j,h,n}=S_{j,h,n}+F_{j,h,n}$ denote the number of successes, failures and assignments for  treatment $j$, respectively. Analogously, $S_{\centerdot, h,n}=\sum_{j=1}^J S_{j,h,n}=\sum_{i=1}^n \mathbb{I}_{\{Z_{i}=h\}}Y_i$, $F_{\centerdot, h,n}=\sum_{j=1}^JF_{j,h,n}=\sum_{i=1}^n \mathbb{I}_{\{Z_{i}=h\}}(1-Y_i)$ and $N_{\centerdot, h,n}=\sum_{j=1}^J N_{j,h,n}=\sum_{i=1}^n \mathbb{I}_{\{Z_{i}=h\}}$ represent the number of successes, failures and assignments in stratum $h$ after $n$ steps. Let ${\vS}_{j, n}=(S_{j, 1, n},\ldots,S_{j, H, n})^\top$, ${\vF}_{j, n}=(F_{j, 1, n},\ldots, F_{j, H, n})^\top$, ${\vN}_{j, n}=(N_{j, 1, n},\ldots, N_{j, H, n})^\top$ and $\mathbf{1}_{H}$ be the $H$-dim vectors of ones, then $S_{j,  \centerdot, n}={\vS}_{j, n}^\top \mathbf{1}_{H}$, $F_{j, \centerdot, n}={\vF}_{j, n}^\top \mathbf{1}_{H}$ and $N_{j,\centerdot, n}={\vN}_{j, n}^\top \mathbf{1}_{H}$ denote the total number of successes, failures and assignments for treatment $j$ after $n$ steps, while $S_{j,\bar{h},n}=S_{j,\centerdot, n}-S_{j,h,n}$, $F_{j,\bar{h},n}=F_{j,\centerdot, n}-F_{j,h,n}$ and $N_{j,\bar{h},n}=N_{j,\centerdot, n}-N_{j,h,n}$ represent the number of successes, failures and allocations for treatment $j$ outside the $h$th stratum (where, clearly, $N_{j,\bar{h},n}=S_{j,\bar{h},n}+F_{j,\bar{h},n}$ and $\sum_{j=1}^J N_{j,\bar{h},n}=N_{\centerdot, \bar{h},n}=n-N_{\centerdot, h,n}$).

Throughout the paper we denote by $\mathbb{I}_{E}$ the indicator function of the event $E$, while $\mathbf{0}_{k}$ is the $k$-dim vector of zeros and $\boldsymbol{I}_k$ is the $k$-dim identity matrix. Moreover, given a $k$-dim vector $\mathbf{a}$, $\diag(\mathbf{a})$ is the $k$-dim square matrix with diagonal elements $a_1,\ldots,a_k$ and $\{\ve_1,\dots,\ve_k\}$ denotes the canonical basis.
Sections reported in the supplementary document are referenced in this
paper with capital letters.

\section{Interacting Urns Design (IUD)}\label{sec3}

In this section, we introduce a general adaptive design that incorporates information borrowing. To better represent the mechanism through which information is shared, we adopt the metaphor of a system of interacting urns.

Let us consider a system with $H$ nodes, one for each stratum of patients enrolled in the trial, and in each node $h$ there is one urn for each treatment under comparison (i.e., in total $J\times H$ urns) containing white ($W$) and red ($R$) balls. The idea is to associate the proportion of white (resp. red) balls in the $j$th urn of node $h$ with the probability of success (resp. failure) of treatment $j$ in stratum $h$. For each $j$, let ${\vW}_{j, i}=(W_{j,1,i},\ldots, W_{j, H, i})^\top$ and ${\vR}_{j, i}=(R_{j,1,i},\ldots, R_{j, H, i})^\top$ denote the vectors containing the number of white and red balls of the urn of treatment $j$ in the $H$ strata at step $i$. We assume (wlog) the same and balanced initial composition for every stratum $h=1,\dots, H$, i.e. ${\vW}_{j,0}={\vR}_{j,0}=\varsigma\mathbf{1}_{H}$, where $\varsigma$ is an initializing positive constant. Obviously, the assumption of equal and balanced initial urn composition reflects the absence of a-priori knowledge on the treatments' performance and it could be suitably modified in the presence of available information before the experiment. At each step, the urns are updated on the basis of the patient's covariate (which selects the ``leading" node), the treatment assignment and the response: only the urns corresponding to the assigned treatment will be updated (while all the other urns will remain unchanged) and the update mechanism combines a (standard) component for the leading node depending on
the observed outcome and a part governing the borrowing of information across strata. More specifically, when $Z_i = h$ the $h$th stratum is the leading node and, if $\delta_{j,i}=1$, then the update of the type-$j$ urns consists of two different terms:
 
 \begin{itemize}
 \item[(i)] ({\em Standard Reinforcement}) for the leading node $h$, urn $j$ is updated by adding $Y_i$ white balls and $(1-Y_i)$ red balls;
\item[(ii)] ({\em Interaction/Informational borrowing})
  all type $j$ urns are updated with stratum-specific quantities that depend on the whole set of observations until patient $i$:
  urn $j$ of stratum $k$ (with $k=1,\dots, H$) is updated by $U^S_{j,k,i}$ white balls and $U^F_{j,k,i}$ red balls.
\end{itemize}

For each treatment $j$, let ${\vU}^S_{j,i}=(U^S_{j,1,i},\ldots, U^S_{j, H, i})^\top$
and ${\vU}^F_{j,i}=(U^F_{j,1,i},\ldots, U^F_{j, H, i})^\top$ (by convention, $U^S_{j,h,i}=U^F_{j,h,i}=0$ whenever $\delta_{j,i}=0$), this updating mechanism leads to the following urn composition at each step $n\geq 1$:
\begin{equation}\label{update0}
\left\{
\begin{aligned}
{\vW}_{j, n} &={\vW}_{j, n-1}+{\vU}^S_{j,n}+\delta_{j,n}Y_n{\ve}_{Z_n}\\
&= \varsigma \mathbf{1}_H +\sum_{i=1}^n {\vU}^S_{j,i}+ \sum_{i=1}^n \delta_{j,i}Y_i{\ve}_{Z_i}\\
{\vR}_{j, n} &= {\vR}_{j, n-1}+ {\vU}^F_{j,n}+ \delta_{j,n}(1-Y_n){\ve}_{Z_n}\\
&= \varsigma \mathbf{1}_H +\sum_{i=1}^n {\vU}^F_{j,i}+ \sum_{i=1}^n  \delta_{j,i}(1-Y_i){\ve}_{Z_i}.\\
\end{aligned}
\right.
\end{equation}
In order to treat successes/failures as well as the $J$ treatments in a symmetric way and to let the urn composition depend on ${\mathcal F}_n$ only through the number of successes and failures of the treatments in each stratum, 
we introduce a suitable vector of non-negative functions $\boldsymbol{\varphi}=(\varphi_1,\ldots, \varphi_H)^\top$ and we define the update as follows: for any $j$, we set 
${\vU}^S_{j,n}=\boldsymbol{\varphi}( {\vS}_{j, n}, {\vF}_{j, n}) - \boldsymbol{\varphi}( {\vS}_{j, n-1}, {\vF}_{j, n-1})$ and ${\vU}^F_{j,n}=\boldsymbol{\varphi}( {\vF}_{j, n}, {\vS}_{j, n}) - \boldsymbol{\varphi}( {\vF}_{j, n-1}, {\vS}_{j, n-1})$, with $\boldsymbol{\varphi}( \mathbf{0}_H, \mathbf{0}_H)=\varsigma\mathbf{1}_H$. From \eqref{update0}, the urn composition at step $n$ simply becomes
${\vW}_{j,n} =\boldsymbol{\varphi}({\vS}_{j, n},{\vF}_{j, n})+ {\vS}_{j, n}$ and ${\vR}_{j, n} =\boldsymbol{\varphi}({\vF}_{j, n},{\vS}_{j, n}) + {\vF}_{j, n}$, so that the proportion of white balls of the type-$j$ urn within the $h$th stratum is:
\begin{equation}\label{prop-urne}
\begin{split}
P_{j,h,n} & = \frac{W_{j,h,n}}{W_{j,h,n}+R_{j,h,n}}=\frac{\varphi_h({\vS}_{j,n},{\vF}_{j,n})+S_{j,h,n} }
{\varphi_h({\vS}_{j,n},{\vF}_{j,n})
+\varphi_h({\vF}_{j,n},{\vS}_{j,n})
+N_{j,h,n}}\\
& =
\rho_{j,h,n}\widehat{\theta}_{j,h,n}+(1-\rho_{j,h,n})\left\{
\frac{ \varphi_h({\vS}_{j,n},{\vF}_{j,n})}
{\varphi_h({\vS}_{j,n},{\vF}_{j,n})
+\varphi_h({\vF}_{j,n},{\vS}_{j,n})}\right\},
\end{split}
\end{equation}
where $\widehat{\theta}_{j,h,n}=S_{j,h,n}/N_{j,h,n}$ is the proportion of successes of treatment $j$ in the $h$th stratum after $n$ steps and
$$\rho_{j,h,n}= \frac{N_{j,h,n}}{\varphi_h({\vS}_{j,n},{\vF}_{j,n})
+\varphi_h({\vF}_{j,n},{\vS}_{j,n})+N_{j,h,n}}  \in [0,1]$$
represents the weight given to this information (by convention, $\widehat{\theta}_{j,h,n}=0$ whenever $N_{j,h,n}=0$), 
while $(1-\rho_{j,h,n})$ represents the weight given to the information coming from all the observations and shared across
strata. Thus, values of $\rho_{j,h,n}$ close to one correspond to a limited borrowing, whereas
smaller values correspond to a stronger borrowing. Within this setting, the updates are governed by the vector of non-negative functions $\boldsymbol{\varphi}=(\varphi_1,\ldots, \varphi_H)^\top$ where, at each stratum $h\in \{1,\ldots, H\}$, the function $\varphi_h$ will be chosen to let $P_{j,h,n}$ be a strongly consistent estimator of $\theta_{j, h}$ for every treatment $j\in \{1,\ldots, J\}$ (in Section \ref{update} we propose three different approaches to define the update mechanism of the urns).

When the $(n+1)$th subject belonging to stratum $h$ is ready to be randomized, he/she is assigned to treatment $j$ with probability
\begin{equation}\label{randfunction}
\Pr\left(\delta_{j,(n+1)}=1 | \mathcal{F}_n, Z_{n+1}=h\right)=\frac{ f( P_{j,h,n} ) }
{ \sum_{j=1}^J f( P_{j,h,n}) },
\end{equation}
where $f(\cdot)$ is a continuous increasing function with $f(0)>0$.
This allocation function has the general form typically adopted in the experimental design literature (see, e.g.,  \cite{baldi2015adaptive}), but is based on the urn proportions which model the informational borrowing.

\subsection{Updating mechanisms} \label{update}
In this section, we propose three different strategies for updating the urns according to the choice of $\boldsymbol{\varphi}$.
\subsubsection{Updating mechanism with asymptotically vanishing borrowing of information}
A first approach lets the update function $\varphi_h$ associated with stratum $h$ depend on the information (i.e., successes and failures) observed in the rest of the system. Formally, for any treatment $j$ and stratum $h$, at each step $n$ we set
\begin{equation}\label{updatefunc1}
\begin{split}
\varphi_h( {\vS}_{j, n}, {\vF}_{j, n})= &\widehat{\theta}_{j,\bar{h},n}\psi(N_{j,\bar{h},n})\\
\varphi_h({\vF}_{j, n}, {\vS}_{j, n})= &(1-\widehat{\theta}_{j,\bar{h},n})\psi(N_{j,\bar{h},n}),
\end{split}
\end{equation}
where $\widehat{\theta}_{j,\bar{h},n}=S_{j,\bar{h},n}/N_{j,\bar{h},n}$ is the proportion of successes of treatment $j$ observed outside the $h$th stratum and $\psi$ is a continuous and non-decreasing function such that $\psi(0)=0$ and $0<\psi(x)\leq \psi_{\max}$ for any $x>0$. Under this approach, at each step, the proportion of white balls in the type-$j$ urn of stratum $h$ is a convex combination of the information coming from this stratum and that accrued outside of $h$, namely
\begin{equation*}
P_{j,h,n}=\rho_{j,h,n}\widehat{\theta}_{j,h,n} +(1-\rho_{j,h,n})\widehat{\theta}_{j,\bar{h},n},
\end{equation*}
where $$\rho_{j,h,n}=\frac{N_{j,h,n}}{\psi( N_{j,\bar{h},n} ) + N_{j,h,n}} \in \left[\tfrac{N_{j,h,n}}{\psi_{\max}+ N_{j,h,n}},1\right].$$
The quantity  $\psi_{\max}$ 
controls the weight $(1-\rho_{j,h,n})$ of the influence coming from the other groups: 
the smaller $\psi_{\max}$ is, the higher the minimum value for $\rho_{j,h,n}$ (that is, the lower the maximum value for the weight 
$(1-\rho_{j,h,n})$) is. 
The boundedness of $\psi$ implies that the weight $(1-\rho_{j,h,n})$ of the information outside the $h$th stratum vanishes as $N_{j,h,n}$ grows; therefore, asymptotically, the information belonging to stratum $h$ tends to be dominant. Some useful examples are $\psi(x) = \min \{x , \psi_{\max}\}$, $\psi(x) =\psi_{\max}\left\{1-\exp(-x/\psi_{\max})\right\}$ and
\begin{equation}\label{psisim}
\psi(x) =\frac{x\psi_{\max} }{x +\psi_{\max}}\,.
\end{equation}
In particular, the function in \eqref{psisim} will be adopted in Section \ref{secsimulation}.

It could be worthwhile to underline that the update functions in (\ref{updatefunc1}) are derived by assuming $\varphi_h( {\vS}_{j, n}, {\vF}_{j, n})=\tilde{\varphi}{_h}(S_{j,\bar{h},n},N_{j,\bar{h},n})$, where $\tilde{\varphi}{_h}$ is chosen such that $\tilde{\varphi}{_h}(x,y) + \tilde{\varphi}{_h}(y,x) = \psi(x+y)$ and $y \tilde{\varphi}{_h}(x,y)=x \tilde{\varphi}{_h}(y,x)$.

\subsubsection{Updating mechanism based on treatment-similarity}\label{treatment-similarity}
An alternative approach is to share the information across strata only if the estimated success probabilities of a given treatment are similar. The rationale behind this choice is that strata sharing the same treatment efficacy could be aggregated. The similarity is assessed by using a decreasing sequence of positive real numbers converging to zero, representing the maximum distances between estimated success probabilities of a given treatment in the different strata; thus, asymptotically, the information borrowing occurs only across strata with the same efficacy. Note that the similarity assessment is treatment-specific, i.e., we could have two strata sharing information for one treatment and not sharing it for the others. In particular, by letting $\{c_n\}_{n\geq1}$ be a decreasing sequence of positive real numbers
with $\lim_{n\rightarrow\infty}c_n=0$, for any stratum $h\in\{1,\ldots,H\}$ we set
\begin{equation}\label{updatefunc2}
\begin{split}
  \varphi_h( {\vS}_{j, n}, {\vF}_{j, n})=&\sum_{k\neq h} S_{j,k,n}\mathbb{I}_{\{|\widehat{\theta}_{j,k,n} -\widehat{\theta}_{j,h,n}  |\leq c_n\}} \\
  \varphi_h({\vF}_{j, n}, {\vS}_{j, n})=&\sum_{k\neq h} F_{j,k,n}\mathbb{I}_{\{|\widehat{\theta}_{j,k,n} -\widehat{\theta}_{j,h,n}  | \leq c_n\}},
\end{split}
\end{equation}
so that
\begin{equation*}
P_{j,h,n}=\rho_{j,h,n}\widehat{\theta}_{j,h,n}+
(1-\rho_{j,h,n}) \frac{ \sum_{k\neq h} S_{j,k,n}\mathbb{I}_{\{|\widehat{\theta}_{j,k,n} -\widehat{\theta}_{j,h,n}|\leq c_n\}}}
{ \sum_{k\neq h} N_{j,k,n}\mathbb{I}_{\{|\widehat{\theta}_{j,k,n} -\widehat{\theta}_{j,h,n}|\leq c_n\}}},
\end{equation*}
with
$$\rho_{j,h,n}=\frac{N_{j,h,n}}{\sum_{k\neq h} N_{j,k,n}\mathbb{I}_{\{|\widehat{\theta}_{j,k,n} -\widehat{\theta}_{j,h,n}|\leq c_n\}}+ N_{j,h,n}}.$$
At each $h$, let $A_{j,h,n}$ be the set of strata which are similar to $h$ at step $n$ for treatment $j$,
i.e. $A_{j,h,n}=\{ k\ne h : \mid\widehat{\theta}_{j,k,n} -\widehat{\theta}_{j,h,n}\mid\leq c_n\}\subseteq\{1,\ldots,H\}$. Thus,
\begin{itemize}
\item if $A_{j,h,n}$ is empty, then $\rho_{j,h,n}=1$ and $P_{j,h,n}=\widehat{\theta}_{j,h,n}$;
\item if $A_{j,h,n}$ is not empty, then
$\rho_{j,h,n}=N_{j,h,n}/(\sum_{k \in A_{j,h,n}} N_{j,k,n}+ N_{j,h,n})$ and
$$P_{j,h,n}=\frac{\sum_{k \in A_{j,h,n}} S_{j,k,n}+ S_{j,h,n}}{\sum_{k \in A_{j,h,n}} N_{j,k,n}+ N_{j,h,n}},$$
namely $P_{j,h,n}$ combines the information coming from similar strata.
\end{itemize}
Since $c_n$ vanishes as $n$ grows, the sequence $\{c_n\}_{n\geq1}$ asymptotically does not aggregate
the information for a given treatment $j$ across strata $k$ and $h$ if $\theta_{j,h}\neq\theta_{j,k}$.
Indeed, $A_{j,h,n}$ is definitely included in $A^*_{j,h}$,
with $A^*_{j,h}=\{k\neq h:\, \theta_{j,k}=\theta_{j,h}\}$ and, if $A^*_{j,h}$ is not empty
(namely there exists at least one other stratum sharing with stratum $h$ the same efficacy of treatment $j$), then $\widehat{\theta}_{j,h,n}$ and every $\widehat{\theta}_{j,k,n}$ with $k\in A^*_{j,h}$
are consistent estimators of the common efficacy, as well as $P_{j,h,n}$ that combines their contributions (see Theorem~\ref{teoCons}).

\subsubsection{Model-based update mechanism}\label{modelbased}
To stress the relationship between the effects of the same treatment in the different strata, for each $j$ assume that $\boldsymbol{\theta}_j=(\theta_{j,1},\ldots,\theta_{j,H})^\top$ is a vector of $H$ independent realizations of the random variable $\Theta_j$ following a Beta distribution $\mathcal{B}(\alpha_j,\beta_j)$, where $\alpha_j,\beta_j\in \mathbb{R}^+$. In this case, the update functions can be defined recursively through a suitably chosen estimation process of the unknown model parameters $(\alpha_{j},\beta_{j})$ for $j=1, \dots, J$. Indeed, according to the computations done in Section \ref{app-model-based-likelihood} of the supplementary material, we can obtain the MLEs $(\widehat{\alpha}_{j,n},\widehat{\beta}_{j,n})$ of $(\alpha_{j},\beta_{j})$ based on the successes and failures observed for the treatments in all the strata until
the $n$th patient. Therefore, by letting for any $h\in\{1,\ldots,H\}$
\begin{equation}\label{updatefunc3}
\varphi_h( {\vS}_{j, n}, {\vF}_{j, n})=\widehat{\alpha}_{j,n} \quad \text{and} \quad \varphi_h({\vF}_{j, n}, {\vS}_{j, n})=\widehat{\beta}_{j,n},
\end{equation}
then
\begin{equation}\label{prop-urne-con-ipotesi}
P_{j,h,n}=\frac{\widehat{\alpha}_{j,n}+S_{j,h,n}}
{\widehat{\alpha}_{j,n} + \widehat{\beta}_{j,n} + N_{j,h,n}}=
\rho_{j,h,n}\widehat{\theta}_{j,h,n}+(1-\rho_{j,h,n}) \frac{\widehat{\alpha}_{j,n}}{ \widehat{\alpha}_{j,n} + \widehat{\beta}_{j,n}} ,
\end{equation}
where
\begin{equation*}
\rho_{j,h,n}=\frac{N_{j,h,n}}{ \widehat{\alpha}_{j,n} + \widehat{\beta}_{j,n} + N_{j,h,n}},
\end{equation*}
namely the borrowing of information across the strata naturally evolves with weights induced by the model 
(and so it does not require the preliminary choice of some items, such as, for instance, $\psi_{\max}$ or $(c_n)_n$ mentioned in the previous updating mechanisms). Indeed,  
 the quantity $\widehat{\alpha}_{j,n}/(\widehat{\alpha}_{j,n}+\widehat{\beta}_{j,n})$ estimates the central tendency of the treatment-$j$ success probabilities across strata and, hence, 
 especially in the early stages of the trial, the urn proportion $P_{j,h,n}$ shrinks the stratum-specific estimator $\widehat{\theta}_{j,h,n}$ toward this treatment-level mean. 
 Moreover, the magnitude $\widehat{\alpha}_{j,n}+\widehat{\beta}_{j,n}$ controls the strength of borrowing: larger values correspond to lower between-stratum variability and 
 hence stronger borrowing. This mechanism therefore adapts the amount of information sharing to the estimated similarity of treatment effects across strata.

\begin{rem}
Under the above model assumptions, for each treatment $j$, $\theta_{j,h}\neq \theta_{j,k}$ if $h\neq k$ with probability one. However, it is possible to extend the generating model in order to include the possibility that some strata share the same efficacy for one or more treatments (see Section \ref{sec:equal_treatments_model_based}
 of the supplementary material).
\end{rem}

\section{Theoretical results}\label{sectheory}
In this section, we describe the properties of the proposed Interacting Urns Design (IUD) by taking into account the three updating mechanisms previously discussed.
We start with some technical results guaranteeing that, besides $\widehat{\theta}_{j,h,n}$, also the proportion of white balls $P_{j,h,n}$ in each type-$j$ urn at each stratum $h$ is a consistent estimator of the corresponding treatment efficacy $\theta_{j,h}$, also deriving the limiting allocation proportions of each treatment induced by the IUD. From now on, at each stratum $h\in\{1,\ldots,H\}$, we set $\boldsymbol{\theta}_{(h)}=(\theta_{1,h},\ldots,\theta_{J,h})^\top$, ${\vN}_{(h)n}=(N_{1,h,n},\ldots,N_{J,h,n})^\top$,
$\boldsymbol{\widehat{\theta}}_{(h)n}=(\widehat{\theta}_{1,h,n},\ldots,\widehat{\theta}_{J,h,n})^\top$, $\boldsymbol{P}_{(h)n}=(P_{1,h,n},\ldots,P_{J,h,n})^\top$ and ${\boldsymbol{\pi}}_{(h) }=(\pi_{1,h},\ldots,\pi_{J,h})^\top$, where $$\pi_{j,h}=\frac{f(\theta_{j,h})}{\sum_{l=1}^J f(\theta_{l,h})}, \qquad \text{for every}\quad  j=1,\ldots,J\quad \text{and}\quad  h=1,\ldots,H.$$

\begin{rem}
With regard to the update mechanism~\eqref{updatefunc3}, the following a.s.-convergence and the convergence in distribution are meant with respect to any measure $\Pr_{\boldsymbol\theta}$, where $\{\Pr_{\boldsymbol\theta}\}_{\boldsymbol\theta}$ is a disintegration of $\Pr$ with respect to the possible realizations of $\boldsymbol{\theta}=(\theta_{j,h})$ see \cite{chang-pollard}.
\end{rem}

\begin{thm}\label{teoCons}
Under the IUD, assuming each of the update mechanisms (\ref{updatefunc1})-(\ref{updatefunc2})-(\ref{updatefunc3}), at each stratum $h\in\{1,\ldots,H\}$, $\lim_{n\rightarrow\infty}N_{j,h,n}\stackrel{a.s.}=\infty$ for every treatment $j\in\{1, \dots, J\}$ and
$$
\lim_{n\rightarrow\infty}\boldsymbol{\widehat{\theta}}_{(h)n}\stackrel{a.s.}= \boldsymbol{\theta}_{(h)}, \quad
\lim_{n\rightarrow\infty}\boldsymbol{P}_{(h)n}\stackrel{a.s.}= \boldsymbol{\theta}_{(h)} \quad \text{and }\; \lim_{n\rightarrow\infty}\frac{{\vN}_{(h)n}}{N_{\centerdot, h,n}}\stackrel{a.s.}=\boldsymbol{\pi}_{(h)}.
$$
In addition,
$$
\lim_{n\rightarrow\infty}\frac{{\vN}_{(h)n}}{n}\stackrel{a.s.}= p_h\boldsymbol{\pi}_{(h)} \quad \text{and} \quad \lim_{n\rightarrow\infty}\frac{{\vN}_{j,n}^\top \mathbf{1}_H}{n}\stackrel{a.s.}=\sum_{h=1}^H p_h {\pi}_{j,h}.$$
\end{thm}
\begin{proof}
See Section \ref{proofTeoCons}
 of the supplementary material.
\end{proof}
The next theorem provides the theoretical basis for the asymptotic inference under the newly suggested procedure.
\begin{thm}\label{teoAsiNorm}
Under the IUD, assuming each of the update mechanisms (\ref{updatefunc1})-(\ref{updatefunc2})-(\ref{updatefunc3}), at each fixed stratum $h\in\{1,\ldots,H\}$, as $n\to \infty$
\begin{equation}
 \diag\left( {\vN}_{(h)n} \right)^{1/2}\left(\boldsymbol{\widehat{\theta}}_{(h)n}-\boldsymbol{\theta}_{(h)}\right)
\stackrel{d}\hookrightarrow
\mathcal{N}\left(\boldsymbol{0}_J,\mathbf{\Sigma}_{(h)}\right),
\tag{CLT-$\widehat{\vtheta}$}\label{eq:CLT-theta}
\end{equation}
where $\mathbf{\Sigma}_{(h)}=\diag\left(\vtheta_{(h)}\right)\diag\left(\boldsymbol{1}_J-\boldsymbol{\theta}_{(h)}\right)$. Moreover, adopting the update mechanisms \eqref{updatefunc1} and \eqref{updatefunc3},
\begin{equation}
\diag\left( {\vN}_{(h)n} \right)^{1/2}\left(\boldsymbol{P}_{(h)n}-\boldsymbol{\theta}_{(h)}\right)\stackrel{d}\hookrightarrow
\mathcal{N}\left(\boldsymbol{0}_J,\mathbf{\Sigma}_{(h)}\right),
\tag{CLT-$\vP$}\label{eq:CLT-P}
\end{equation}
while under update \eqref{updatefunc2} with $\lim_{n\rightarrow\infty} c_n= 0$ and $\lim_{n\rightarrow\infty} c_n\sqrt{ n/\ln(\ln(n)) }=\infty$,
the limit \eqref{eq:CLT-P} holds by adding $(\sum_{k \in A_{1,h,n}} N_{1,k,n}, \ldots, \sum_{k \in A_{J,h,n}} N_{J,k,n})^\top$
to the vector ${\vN}_{(h)n}$.
\end{thm}
\begin{proof}
See Section \ref{proofTeoAsiNorm}  of the supplementary material.
\end{proof}

\begin{rem}
We highlight that the above  theorem shows that, under update~\eqref{updatefunc2} with suitable conditions on $(c_n)$, the convergence rate in the 
CLT-$\vP$ 
depends not only on the sample size 
of its own stratum, $N_{j,h,n}$, but also on the combined sample size of all strata with the same success probabilities, i.e.,
\[
\sum_{k \in A_{j,h,n}} N_{j,k,n}, \quad \text{where} \quad
A_{j,h,n} = \left\{ k \neq h : \left| \widehat{\theta}_{j,k,n} - \widehat{\theta}_{j,h,n} \right| \leq c_n \right\}.
\]
Consequently, in this case, borrowing of information also provides asymptotic benefit, that is a faster convergence rate, resulting in  
narrower confidence intervals and more powerful hypothesis tests.
\end{rem}

\begin{rem}
From a technical point of view, we wish to stress that, in general, our allocation function is not a special case of those considered in \cite{Zhang07} since function $f$ in \eqref{randfunction} is not necessarily differentiable under the expectation with finite derivatives (see \cite[condition A.2]{Zhang07}). Moreover, note that by taking as asymptotic limiting target $\boldsymbol{\pi}_{(h)}$ and estimating $\theta_{j,h}$ via $P_{j,h,n}$, \cite[condition 2.4]{Zhang07} could be verified only after proving all the technical results in Sections \ref{proofTeoCons}, \ref{proofTeoAsiNorm}, \ref{app-model-based-likelihood} and \ref{proof_beta}  of the supplementary material.
\end{rem}

In the context of multi-arm clinical trials, the inferential attention is usually devoted to the contrasts and one of the main goals is testing the hypothesis on the equality of the treatment effects within each specific stratum by the so-called null hypothesis of homogeneity $H_0: \theta_{1, h}=\ldots =\theta_{J, h}$. Let $\boldsymbol{A}^\top=[\mathbf{1}_{J-1}; -\boldsymbol{I}_{J-1}]$, $\boldsymbol{A}^\top\boldsymbol{\theta}_{(h)}$ defines the vector of contrasts with respect to the first treatment (considered as the reference, without loss of generality) and the problem is testing $H_0:\boldsymbol{A}^\top{\boldsymbol{\theta}}_{(h)}=\mathbf{0}_{J-1}$ against $H_1:\boldsymbol{A}^\top{\boldsymbol{\theta}}_{(h)}\ne\mathbf{0}_{J-1}$. The contrasts could be consistently estimated by $\boldsymbol{A}^\top\boldsymbol{\widehat{\theta}}_{(h)n}$ and, from Theorem \ref{teoAsiNorm}, as $n\rightarrow\infty$
$$
\left(\boldsymbol{A}^\top \diag\left( {\vN}_{(h)n} \right)^{-1} \mathbf{\Sigma}_{(h)} \boldsymbol{A} \right)^{-1/2}
\boldsymbol{A}^\top\left(\boldsymbol{\widehat{\theta}}_{(h)n}-\boldsymbol{\theta}_{(h)}\right)
\stackrel{d}\hookrightarrow
\mathcal{N} \left(\boldsymbol{0}_{J-1},\boldsymbol{I}_{J-1}\right).$$
Thus, by letting $\widehat{\boldsymbol{\Sigma}}_{(h)n}=\diag\left(\widehat{\boldsymbol{\theta}}_{(h)n}\right)\diag\left(\boldsymbol{1}_J-\widehat{\boldsymbol{\theta}}_{(h)n}\right)$,  the test statistic
$$\chi_n=\widehat{\boldsymbol{\theta}}_{(h)n}^\top\boldsymbol{A}\left[\boldsymbol{A}^\top  \diag\left( {\vN}_{(h)n} \right)^{-1} \widehat{\boldsymbol{\Sigma}}_{(h)n}
 \boldsymbol{A}\right]^{-1}\boldsymbol{A}^\top\widehat{\boldsymbol{\theta}}_{(h)n}$$
can be employed to test the null hypothesis of homogeneity. Indeed, under $H_0$, as $n$ grows $\chi_n \stackrel{d}\hookrightarrow \chi^2_{J-1}$, namely it converges to a (central) chi-squared distribution with $J-1$ degrees of freedom.

With regard to inference for comparing the efficacy of two treatments, asymptotic confidence intervals and statistical tests could be easily provided from Theorem \ref{teoAsiNorm}. Indeed, under the IUD, at each stratum $h\in\{1,\ldots,H\}$, as $n\to \infty$
\begin{equation*}
\begin{split}
\frac{ (\widehat{\theta}_{j,h,n}-\widehat{\theta}_{l,h,n}) - (\theta_{j,h}-\theta_{l,h}) }{ \sqrt{\frac{\theta_{j,h}(1-\theta_{j,h})}{N_{j,h,n}}+\frac{\theta_{l,h}(1-\theta_{l,h})}{N_{l,h,n}}}}
\stackrel{d}\hookrightarrow \mathcal{N}(0,1),
\end{split}
\end{equation*}
for every pair of treatments $j,l\in\{1, \dots, J\}$. Thus, an asymptotic confidence interval at level $1-\alpha$ for $(\theta_{j,h}-\theta_{l,h})$ is
$$\left(\widehat{\theta}_{j,h,n}-\widehat{\theta}_{l,h,n}\right) \pm z_{1-\alpha/2} \sqrt{\frac{\widehat{\theta}_{j,h,n}(1-\widehat{\theta}_{j,h,n})}{N_{j,h,n}}+\frac{\widehat{\theta}_{l,h,t}(1-\widehat{\theta}_{l,h,n})}{N_{l,h,n}}},$$
where $z_{\alpha}$ is the $\alpha$-level quantile of the standard normal distribution; moreover, the Wald-type test statistic
\begin{equation}\label{eq_ts}
\begin{split}
U_{h, n}=\frac{ \widehat{\theta}_{j,h,n}-\widehat{\theta}_{l,h,n} }{ \sqrt{\frac{\widehat{\theta}_{j,h,n}(1-\widehat{\theta}_{j,h,n})}{N_{j,h,n}}+\frac{\widehat{\theta}_{l,h,n}(1-\widehat{\theta}_{l,h,n})}{N_{l,h,n}}}}
\end{split}
\end{equation}
could be employed for testing
\begin{equation}\label{hypo}
H_0:\theta_{j,h}=\theta_{l,h} \quad  \text{versus} \quad H_1:\theta_{j,h}\neq\theta_{l,h}
\end{equation}
(or other specific alternatives); indeed, under $H_0$, $U_{h, n}\stackrel{d}\hookrightarrow \mathcal{N}(0,1)$ as $n\rightarrow \infty$ (i.e., $U^2_{h, n}$ follows asymptotically a chi-squared distribution with 1 degree of freedom).

As previously discussed, similar results hold by replacing $\widehat{\boldsymbol{\theta}}_{(h)n}$ by $\boldsymbol{P}_{(h)n}$, using \eqref{eq:CLT-P} instead of \eqref{eq:CLT-theta}. For instance, an analogous
Wald type statistic can be obtained from \eqref{eq_ts} by replacing
$\widehat{\theta}_{j,h,n}$ and $\widehat{\theta}_{l,h,n}$ with
$P_{j,h,n}$ and $P_{l,h,n}$, respectively. It is worth noticing that the effect of the borrowing of information across strata works by changing the allocation probability of patients to treatments through $P_{j,h,n}$. Thus, due to the nature of the design in \eqref{randfunction}, the effect of the borrowing is still present even when employing $\widehat{\theta}_{j,h,n}$.

\begin{rem}\label{rem_test}
With the above procedure, it is possible to test other hypotheses of interest according to the experimental context.
For instance, it is possible to employ any contrast matrix $\vA$ and/or specify superiority/non-inferiority alternatives.
The hypothesis of homogeneity of a given treatment across the strata may also be tested. Clearly, if the global null hypothesis of homogeneity of the $J$ treatments in the $h$-th stratum  $H_0: \theta_{1, h}=\ldots =\theta_{J, h}$ is rejected, then simultaneous pairwise comparisons could be tested by applying suitable multiplicity corrections, like Bonferroni or Tukey's procedure. 
\end{rem}

\subsection{Sequential monitoring}
Since clinical trials usually involve interim analyses to achieve specific design goals, such as controlling the type I error rate, we now provide a theoretical framework in order to sequentially monitor the suggested IUD. Thus, in addition to the asymptotic inference with fixed sample size previously discussed, we now focus on the sequential test statistics and, to formulate the problem in the Skorohod's topology, for any information time $t \in [0,1]$ let ${\lfloor nt \rfloor}$ be the largest integer not greater than $nt$.

To compare the efficacy of two treatments $j,l\in\{1,\ldots,J\}$ at a generic information time $t\in(0;1]$, at a fixed stratum $h$ the problem is testing the hypothesis in \eqref{hypo} and we consider the previously introduced Wald-type test
$${U}_{h,n,t}=\frac{\widehat{\theta}_{j,h,{\lfloor nt \rfloor}}-\widehat{\theta}_{l,h,{\lfloor nt \rfloor}}}{\sqrt{\frac{\widehat{\theta}_{j,h,{\lfloor nt \rfloor}}(1-\widehat{\theta}_{j,h,{\lfloor nt \rfloor}})}{N_{j,h,{\lfloor nt \rfloor}}}+\frac{\widehat{\theta}_{l,h,{\lfloor nt \rfloor}}(1-\widehat{\theta}_{l,h,{\lfloor nt \rfloor}})}{N_{l,h,{\lfloor nt \rfloor}}}} }.$$
The next theorem shows that the Wald-type test $({U}_{h,n,t})_t$ converges in distribution, as $n\to \infty$,
to a standard Wiener process under $H_0$ and to a drifted Wiener process under $H_1$.
\begin{thm}\label{thmSeqMon}
In the Skorohod's space, under the null hypothesis the process $(\sqrt{t} \, U_{h, n, t})_t$  converges in distribution to a standard Wiener process; whereas, under the alternative hypothesis, the process $(\sqrt{t}\, U_{h,n,t} -t \sqrt{n} \mu_{h})_t $ converges in distribution to a standard Wiener process, where $\mu_h={(\theta_{j,h}-\theta_{l,h})}/{\sqrt{v_{j,h}+v_{l,h}}}$ and
\begin{equation*}
\begin{split}
\lim_{n \to \infty} \frac{{\lfloor nt \rfloor}\widehat{\theta}_{j,h,{\lfloor nt \rfloor}}(1-\widehat{\theta}_{j,h,{\lfloor nt \rfloor}})}{{N_{j,h,{\lfloor nt \rfloor}}} }
\stackrel{a.s.}{=} \frac{\theta_{j,h}(1-\theta_{j,h})}{p_h\pi_{j,h}}=v_{j,h}.
\end{split}
\end{equation*}
\end{thm}
\begin{proof}
See Section \ref{app:seqMonitor}
 of the supplementary material.
\end{proof}
Therefore, from Theorem \ref{thmSeqMon}, for any set of chosen information times $0<t_1\leq t_2 \leq \ldots \leq t_K\leq1$, asymptotically
 for $n\to \infty$
\begin{itemize}
\item[i)] $(U_{h, n,t_1} ,\ldots , U_{h,n,t_K})$ is jointly normally distributed,
\item[ii)] $E[U_{h,n,t_i}]\approx\sqrt{nt_i} \mu_h$,
\item[iii)] $Cov(U_{h,n,t_i};U_{h,n,t_j})\approx\sqrt{{\lfloor nt_i \rfloor}/{\lfloor nt_j \rfloor}}$ for $0 < t_i \leq t_j\leq 1$.
\end{itemize}
Thus, asymptotically, the sequence of test statistics $\{U_{h,n,t_1} ,\ldots , U_{h,n,t_k}\}$ satisfy the canonical joint distribution defined in \cite{Jen00} and therefore standard group sequential methods could be applied adopting the IUD.

\section{Finite sample properties}\label{secsimulation}
In order to explore the properties of the IUD for finite samples and to compare its behavior with a stratified permuted block design (PBD) and a CARA procedure, we present the following simulation study. The IUD has been implemented by setting $f(x)=(1-x)^{-1}$  and for the update mechanism in \eqref{updatefunc1} with asymptotically vanishing borrowing, denoted by IUD1, we take into account $\psi$ in (\ref{psisim}) with $\psi_{\max}=10$ (sensitivity analyses for these choices are
reported in Sections \ref{subs:G6} and \ref{subs:G7} of the Supplementary Material). The procedure with update based on treatment similarity in \eqref{updatefunc2} (IUD2) is implemented with $c_i=1/(\ln(i))$ for $i>1$, while the model-based approach in \eqref{updatefunc3} will be denoted by IUD3. The PBD is implemented with block size 4. The CARA procedure is the one proposed in \cite{Zhang07} adopting, to ensure homogeneous comparisons, the same limiting allocation proportion considered here. In practice, such procedure corresponds to an interacting urn design without allowing for the borrowing of information (i.e., setting $\rho_{j,h,n}=1$).  
In the simulation study, we consider $J=2$ treatments and we assume that the incoming patients are uniformly distributed across $H=5$ strata. 

Additional simulation results are provided in Section \ref{sec:add_sim} of the supplementary material, including unbalanced patient distributions across strata, power and type I error, finite-sample behavior of the stratum-specific estimators, behavior of randomization probabilities and allocation
proportions, amount of borrowing, and sensitivity analyses.

Here, we take into account two measures of performance for information precision and one inspired by ethical demands. With regard to inference, under the IUD we consider the following criterion
$$\text{INF}_{n}= \sqrt{\sum_{h=1}^H \left[\left(P_{1,h,n}-P_{2,h,n}\right)-\left(\theta_{1,h}-\theta_{2,h}\right)\right]^2}$$
that measures the Euclidean distance between the estimated treatment difference in each stratum with respect to the true one, where $P_{j,h,n}$ is replaced by $\widehat{\theta}_{j,h,n}$ for PBD.  We denote by INF$_{h,n}$ the corresponding stratum-specific metric. While INF$_n$ is more appropriate when the contrasts are of interest, another measure of estimation efficiency is given by $$\text{RMSE}_n=\sqrt{\sum_{h=1}^H [(P_{1,h,n}-\theta_{1,h,n})^2+ (P_{2,h,n}-\theta_{2,h,n})^2]}.$$ Notice that despite providing a global summary value, both INF$_n$ and RMSE$_n$ are calculated by summing up strata-specific metrics, thus they do not neglect the presence of different sub-groups.

As an ethical criterion, at each stratum $h$ we take into account the proportion of assignments to the worst treatment, namely
\begin{equation*}
\text{PW}_{h,n}=\left\{
\begin{aligned}
&\frac{N_{1,h,n}}{N_{\centerdot,h,n}}, &\mbox{ if } \theta_{1,h}<\theta_{2,h} \\
&\frac{N_{2,h,n}}{N_{\centerdot,h,n}}, &\mbox{ if } \theta_{1,h}>\theta_{2,h}
\end{aligned}
\right.
\end{equation*}

The reported results are obtained with $M=10^4$ simulations, for sample sizes $n=75$, $100$ and $200$ and for different scenarios. The first five scenarios are simulated by fixing the parameters, while in $S_4$-$S_5$, $\boldsymbol{\theta}_1$ and $\boldsymbol{\theta}_2$ are assumed to be realizations of random variables. In the deterministic scenarios, $\boldsymbol{\theta}_1$ and $\boldsymbol{\theta}_2$ are non-random quantities chosen in order to cover the cases of small variations of the treatment efficacy among sub-groups ($S_3$) and clusters of strata with the same efficacy ($S_1$-$S_2$). As suitable benchmarks, we also add two baseline scenarios dealing with the two extreme situations of constant treatment effects across strata $S_B$ (where the need for borrowing is maximal) and $S_{\bar{B}}$ where no relationships between the stratum-specific effectiveness are present  (under which the informational sharing is worthless).\\

\begin{table}[ht!]
\caption{Experimental Scenarios with deterministic parameters}\label{tab_scen}
\centering
\begin{tabular}{c|cc}
\hline
& & \\
Scenario & $\boldsymbol{\theta}_1^\top$ & $\boldsymbol{\theta}_2^\top$  \\
\vspace{-0.2 cm}
& & \\
\hline
$S_{\bar{B}}$ & $(0.9,0.4,0.6,0.8,0.2)$ & $(0.45,0.85,0.75,0.6,0.95)$\\
$S_B$ & $(0.5,0.5,0.5,0.5,0.5)$ & $(0.1,0.1,0.1,0.1,0.1)$\\
\hline
$S_1$ & $(0.5, 0.5, 0.5, 0.3, 0.3)$ & $(0.3, 0.3, 0.3, 0.1, 0.1)$ \\
$S_2$ & $(0.3, 0.3, 0.3, 0.3, 0.3)$ & $(0.1,0.1,0.1,0.5,0.5)$ \\
$S_3$ & $(0.56, 0.5, 0.55, 0.44, 0.45)$ & $(0.45,0.55, 0.50, 0.42, 0.58)$ \\
\hline
\end{tabular}
\end{table}

Moreover, notice that scenarios $S_B$ and $S_1$ show a common situation of stochastic dominance between the two treatments, where the first drug performs better than the other in each stratum; whereas, under $S_2$ treatment $2$ has a higher success probability for strata $4$ and $5$.
For $S_3$, the mean efficacy over the strata of both treatments is equal to $0.5$, but treatment $1$ is preferable in strata $1$-$2$ and $4$, while treatment $2$ performs better in the sub-groups $3$ and $5$.

In the remaining two scenarios $S_4$ and $S_5$ (see Table \ref{tab_scen_2}), each $\boldsymbol{\theta}_j$ is assumed to be a vector of independent realizations of a random variable $\Theta_{j}$ following a Beta distribution $\mathcal{B}(\alpha_j,\beta_j)$; at each iteration $m$ of the Monte Carlo simulation, the $5$-dim vector $\left(\theta_{j,1}^{(m)},\ldots,\theta_{j,5}^{(m)}\right)$ is generated from the corresponding Beta distribution (thus, for each $j=1,2$, the parameters $\theta_{j,h}$ for $h=1, \ldots, 5$ are not fixed but vary across each simulation, as well as the preferability of a given treatment in each stratum).
\begin{table}[ht]
\caption{Experimental Scenarios with random parameters}\label{tab_scen_2}
\centering
\begin{tabular}{c|rr|rr}
\hline
Scenario & $(\alpha_1,\beta_1)^\top$ & $E[\Theta_{1}]$  & $(\alpha_2,\beta_2)^\top$ & $E[\Theta_{2}]$ \\
\hline
$S_4$ & $(49.5;49.5)$ & $0.5$ & $ (3.5;31.5)$ & $0.1$ \\
$S_5$ & $(49.5;49.5)$ & $0.5$ & $(49.5;49.5)$ & $0.5$ \\
\hline
\end{tabular}
\end{table}

Here, in order to provide homogeneous comparisons, parameters $(\alpha_j,\beta_j)$ (for $j=1,2$) have been chosen by fixing the standard deviation of both Beta distributions equal to $0.05$ (i.e., by letting $sd[\Theta_{1}]=sd[\Theta_{2}]=0.05$ for every $h=1,\ldots,5$) and to provide in $S_4$ each $E[\Theta_{j}]$ equals to the (deterministic) efficacy of scenario $S_B$ (namely $0.5$ and $0.1$ for treatment 1 and 2, respectively), while scenario $S_5$ represents a randomized variant of $S_3$, where $E[\Theta_{j}]=0.5$ for both treatments. Clearly, under $S_4$-$S_5$ no relationships of stochastic dominance are present between the two drugs.

\begin{figure}[!ht]\caption{INF$_n$ for experimental scenarios $S_4$ and $S_5$ in Table \ref{tab_scen_2}.}\label{inf2}
\includegraphics[scale=.85]{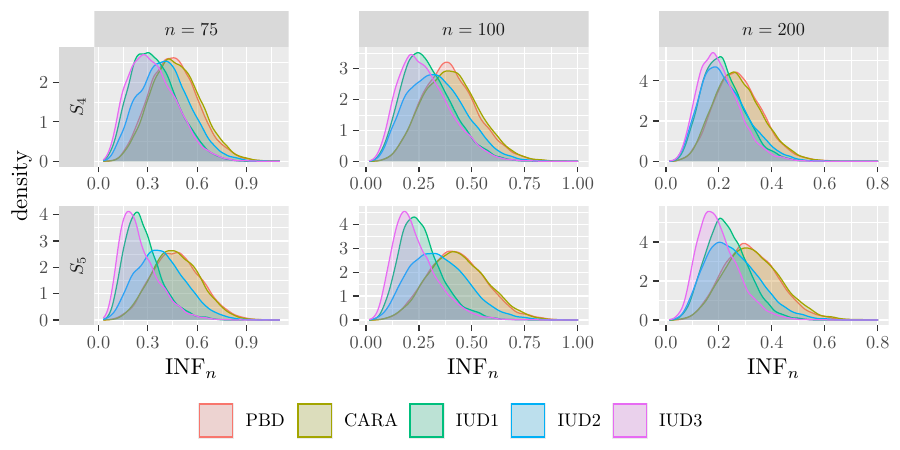}
\end{figure}

\begin{figure}[!ht]\caption{RMSE$_n$ for experimental scenarios $S_4$ and $S_5$ in Table \ref{tab_scen_2}.}\label{inf6}
\includegraphics[scale=.85]{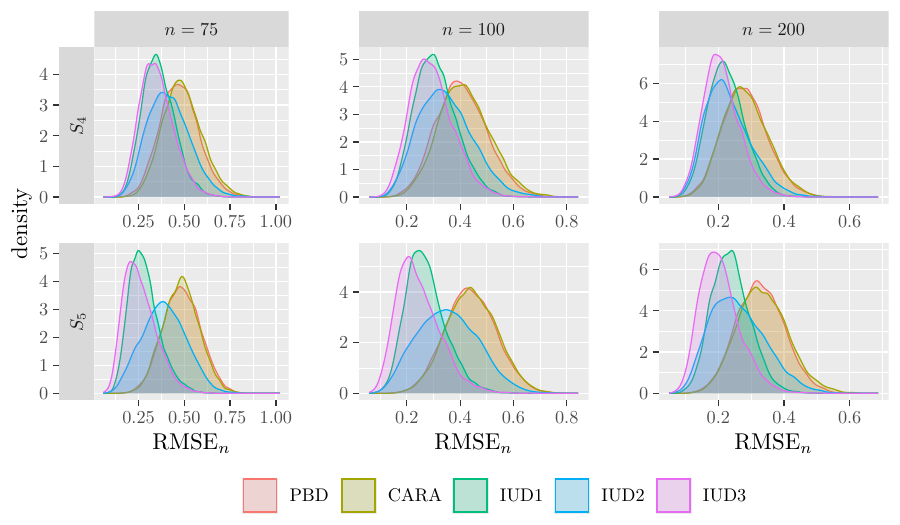}
\end{figure}

\begin{figure}[!ht]\caption{PW$_{h,n}$ for experimental scenarios $S_4$ and $S_5$ in Table \ref{tab_scen_2}.}\label{pwt2}
\includegraphics[scale=.9]{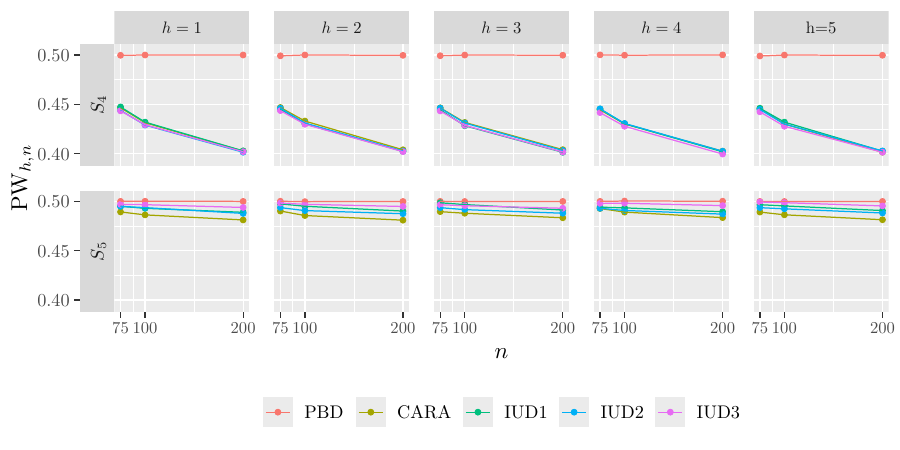}
\end{figure}

The following Figures \ref{inf1}, \ref{inf5} and  \ref{pwt1} show the behavior of $\text{INF}_{n}$, RMSE$_n$ and $\text{PW}_{n}$ in the scenarios of Table \ref{tab_scen} for $n=75$, $100$ and $200$, while the same metrics calculated for $S_4$ and $S_5$ are reported in Figures \ref{inf2}, \ref{inf6} and \ref{pwt2}.

 Regardless of the chosen update mechanism, the IUD always performs  better than PBD in terms of both inferential precision and ethical demands, except when the borrowing of information is inadequate ($S_{\bar{B}}$).  While PBD presents a preferable behavior in terms of the inferential metric, the IUD induces an ethical gain that increases with $n$. Indeed, due to its CARA feature, the IUD is able to detect the stratum-specific worst treatment and strongly penalize the assignments to it, until halving it with respect to the PBD (e.g., PW$_{5, 200}$ is around $50\%$ for CR versus $25\%$ for IUD2). The CARA procedure presents quite similar performance to the PBD in terms of the inferential metrics, while ensuring a much smaller number of patients assigned to the worst treatment. The IUDs represent a good trade-off between the inferential metrics and the ethical gain.
By comparing $S_B$ with $S_4$ and $S_3$ with $S_5$ we notice similar behavior for all the procedures except for the inferential metric calculated under IUD3, which performs better in $S_4$ and $S_5$ than in the corresponding deterministic scenarios.
\begin{figure}[!ht]\caption{INF$_n$ for experimental scenarios $S_{\bar{B}}$, $S_B$, $S_1$-$S_3$ in Table \ref{tab_scen}.}\label{inf1}
\includegraphics[scale=.85]{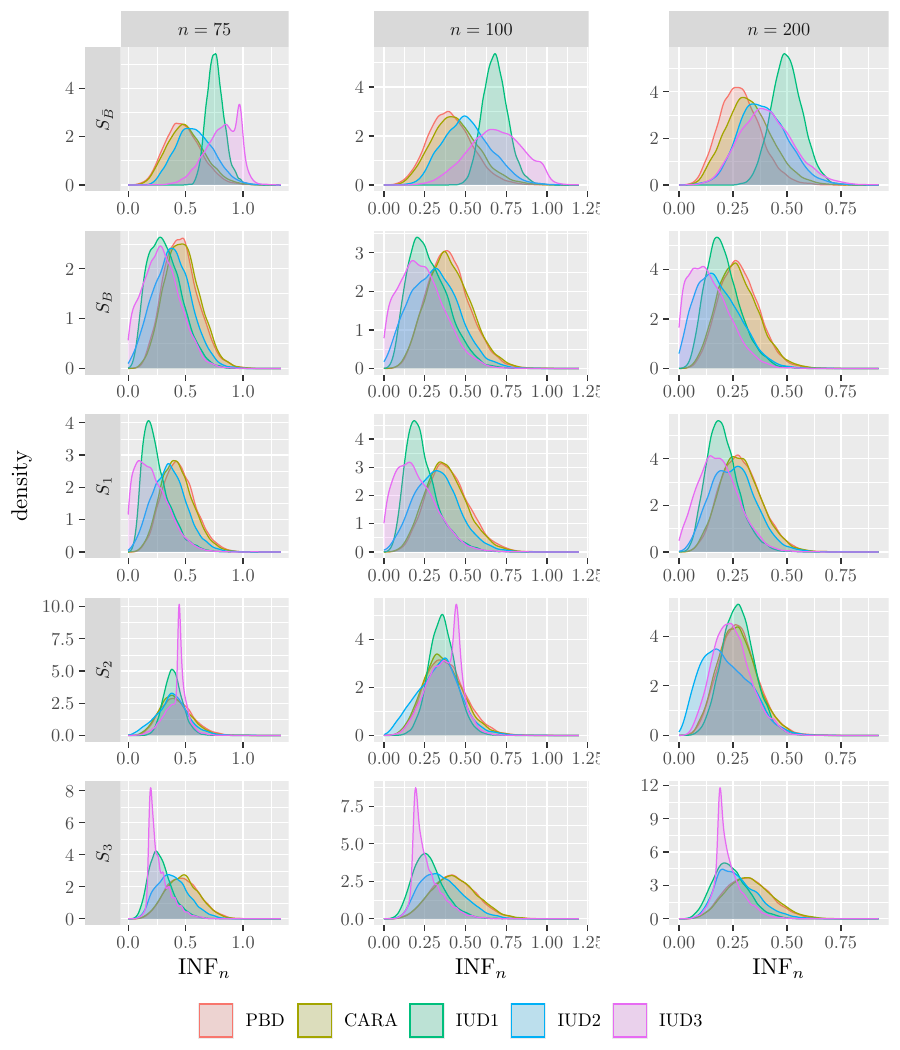}
\end{figure}

\begin{figure}[!ht]\caption{RMSE$_n$ for experimental scenarios $S_{\bar{B}}$, $S_B$, $S_1$-$S_3$ in Table \ref{tab_scen}.}\label{inf5}
\includegraphics[scale=.85]{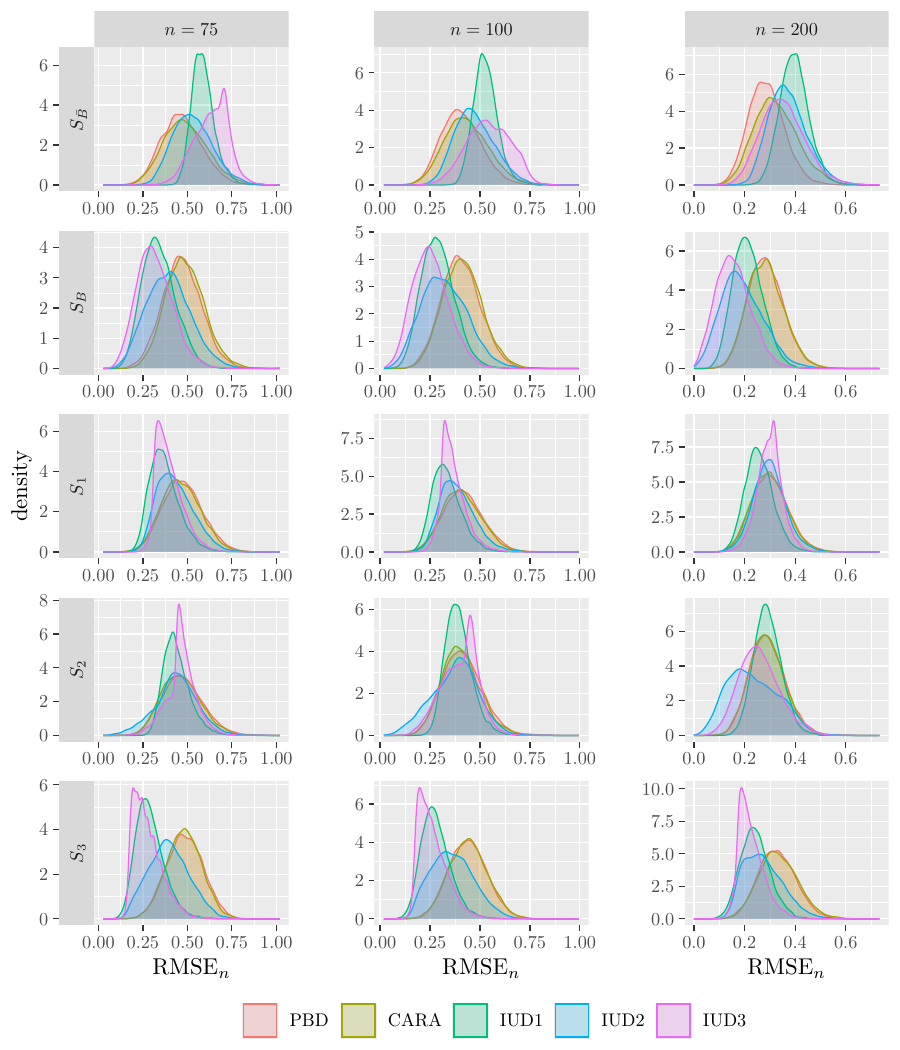}
\end{figure}

\begin{figure}[!ht]\caption{PW$_{h, n}$ for experimental scenarios $S_{\bar{B}}$, $S_B$, $S_1$-$S_3$ in Table \ref{tab_scen}.}\label{pwt1}
\includegraphics[scale=.9]{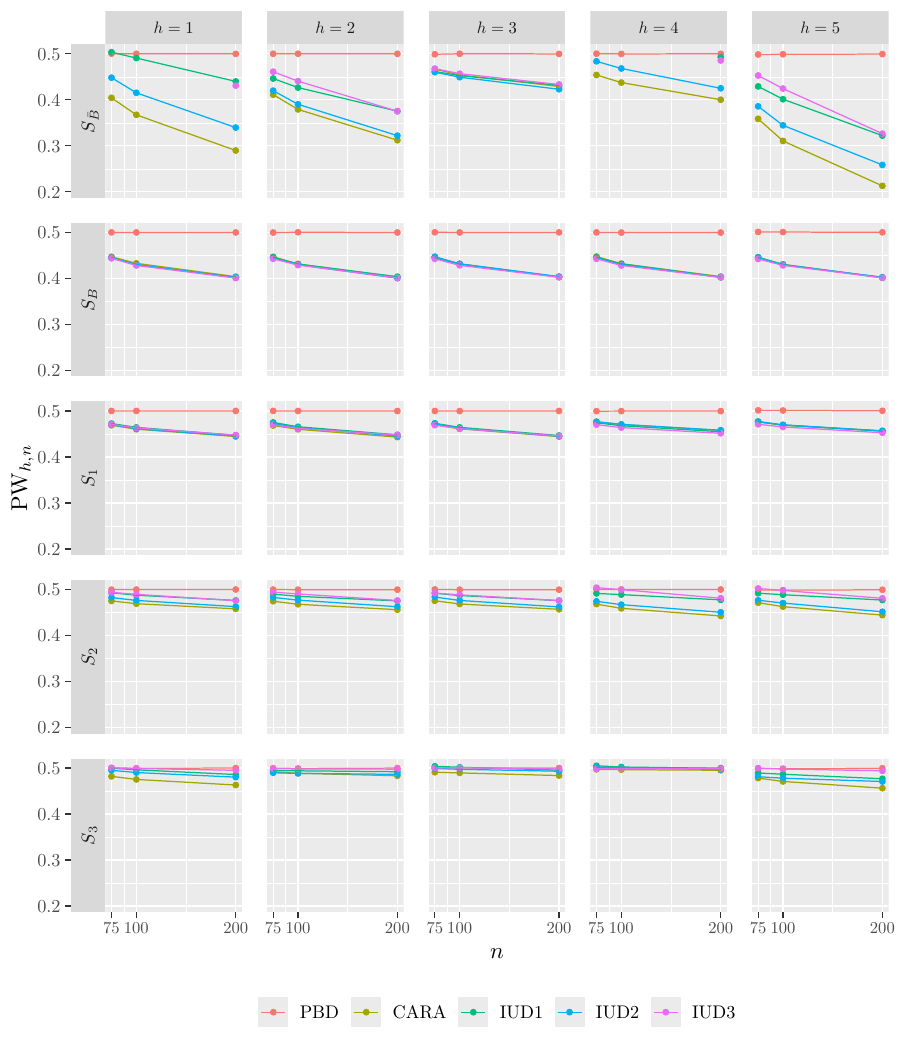}
\end{figure}

Concerning the amount of borrowing, IUD1 is associated with a higher degree of information sharing (which, however, can be tuned by $\psi_{\max}$): as far as the treatment efficacy is similar across strata, such a generalized borrowing works well. By contrasts, due to its properties, IUD2 identifies the clusters of strata to be aggregated, thus it is more appropriate from an ethical viewpoint then the other two mechanisms. Finally, since IUD3 governs the amount of borrowing with weights induced by the underlying model, it is characterized by the most flexible behavior across the considered scenarios, from both an inferential and an ethical viewpoint.

\section{Discussion}\label{sec:discussion}
In the context of stratified clinical trials for treatment comparisons, several approaches have been recently proposed to provide efficient studies for testing multiple treatments, possibly against a standard of care, in different sub-populations. When combined with response-adaptive randomization, their efficiency can be further improved from an ethical viewpoint. Within this framework, we suggest a novel design strategy aimed at sharing the information across different strata in order to i) provide more precise and robust estimates, especially in the early phase of the study or for strongly under-represented sub-groups and ii) skew the allocations towards the stratum-specific most promising treatment. The proposed interacting urns design (IUD) is a new CARA procedure based on a system of randomly updated interacting urns, that allows to control the informational exchange across strata and to adaptively skew the treatment allocation probability away from balance towards the better-performing treatment within each stratum.\\

We introduce three mechanisms for information sharing. The IUD1 is recommended when a more generalized borrowing of information across strata is appropriate and the treatment efficacies are expected to be homogeneous across strata (like, for instance, when testing a placebo or a treatment whose efficacy is known to be independent from the stratification variable). In IUD1, the degree of information sharing gradually decreases over time, allowing the procedure to become increasingly stratum-specific as more data are accumulated. Extending this framework, the IUD3  is most appropriate when treatment efficacies can be thought of as generated from a common Beta distribution. Under this mechanism, the borrowing of information is automatically adjusted, without requiring subjective choices. Finally, IUD2 is recommended when treatment efficacies are expected to be clustered across strata. Like for the other IUDs, the amount of borrowing vanishes asymptotically, unless there exist strata in which the treatment has the same efficacy; in this latter case the IUD2 aggregates the information, yielding substantial inferential benefits. 
For each of the proposed IUDs, we provide the theoretical properties and the corresponding asymptotic inference. Notice that the results are established 
under non-standard CARA designs and rely on estimators that account for information borrowing (so different from the classical MLEs), even if their asymptotic form resembles that of standard ones. Overall, these considerations suggest that the proposed IUD  is suitable for settings in which there is a clinically plausible rationale for borrowing information across strata. When no similarity across strata is expected, as in scenario $S_{\bar B}$, a poorer inferential performance of the IUD procedures is therefore expected. In this regard, the more adaptive mechanisms of IUD2 and IUD3 can partly mitigate inappropriate borrowing, whereas IUD1 is mainly recommended when stronger prior confidence in similarity across strata is available.

In addition, we demonstrate that the sequential test statistics asymptotically satisfy the canonical joint distribution of Jennison and Turnbull \cite{Jen00}. This allows the investigator to effectively use the accrued information and apply standard group sequential methodologies at suitably specified interim analyses; for instance, by adopting the $\alpha$-spending function approach to control type I errors, or by defining stopping boundaries to drop ineffective/efficacious treatments in specific strata and decide to stop the trial in such subgroups or continue with the remaining arms. This provides a theoretical foundation for sequential monitoring of CARA designs that solves the current dichotomy between response-adaptive randomization and group sequential methods in the presence of covariates (see, e.g., \cite{Wason14}). \\

Suitable extensions of our proposal to non-binary responses are obviously of great interest. The more natural implementation is to replace $Y$ with a monotonically increasing score function of patients' responses $s(Y)\in[0;1]$. This approach has been often applied to urn models for clinical trials with negative or heavy-tailed response distributions (e.g. see~\cite{AlGhRo, BaiHu, May09}). By letting $E[s(Y_i)\mid \delta_{j,i}=1,Z_i=h]=\theta_{j,h}$, the mathematical structure of the urn model would be exactly the same apart from the meaning of $S_{j,h,n}$ and $F_{j,h,n}$ (which may not be integers and so considered as successes or failures anymore). The results proved for IUD1-IUD2 remain valid, while IUD3 requires some technical changes to address specific model assumptions in the non-binary case. \\

From a theoretical perspective, this paper presents a general framework for the design and inference of stratified experiments with informational borrowing.  The proposed methodology is sufficiently flexible to accommodate specific characteristics and objectives of various fields of application (such as homogeneity tests between strata, non-inferiority tests within strata, dropping ineffective treatments or stopping the experimentation for efficacy in specific strata). These features make the proposal particularly relevant for complex multi-arm trial structures such as basket, umbrella, and platform trials. Clearly, further insights into the behavior of the proposed IUD in specific application domains should be explored in future research through a case-specific analysis to better understand the implications of these designs in diverse clinical contexts.

Finally, although in the present paper we assume that responses are available immediately after treatment, in real experimental contexts responses may be delayed. This issue can be accommodated within the proposed procedure by incorporating the delay mechanism into the urn reinforcement (updating the urns only when the corresponding responses become available). As shown by several authors (see, among others \cite{BaiHuRosenberger2002, CrimaldiLeisen2008, HZCC08, Zha07b}), the crucial issue concerns the amount of information actually collected during the recruitment period relative to the delay mechanism. Generally speaking, the asymptotic properties of response-adaptive designs are preserved provided that the delay mechanism is not dominant with respect to the subject entry process.

\section*{Acknowledgements}
G. Aletti is a member of the Italian Group “Gruppo Nazionale per il Calcolo Scientifico”
of the Italian Institute “Istituto Nazionale di Alta Matematica”. I. Crimaldi is a member of the
Italian Group “Gruppo Nazionale per l’Analisi Matematica, la Probabilità e le loro Applicazioni” of
the Italian Institute “Istituto Nazionale di Alta Matematica”. G. Aletti, A. Baldi Antognini, I. Crimaldi and R. Frieri
are partially supported by the European Union - NextGenerationEU through the Italian Ministry of University
and Research under the National Recovery and Resilience Plan (PNRR) - Mission 4 Education and
research - Component 2 From research to business - Investment 1.1 Notice Prin 2022 - DD N. 104, 2/2/2022, \textit{Optimal and adaptive designs for modern medical experimentation}, proposal code [2022TRB44L] - CUP [J53D23003270006]. A. Baldi Antognini is also supported
by EU funding within the NextGenerationEU-MUR PNRR Extended Partnership initiative on Emerging Infectious Diseases (Project
no. PE00000007, INF-ACT).

\newpage

{\centerline{\LARGE  Supplementary material}}


\appendix

\section{Proof of Theorem \ref{teoCons}}\label{proofTeoCons}
\noindent After some technical lemmas, we prove Theorem \ref{teoCons} for the update mechanisms \eqref{updatefunc1} and \eqref{updatefunc2} in Appendix \ref{iud45}, while the proof for mechanism \eqref{updatefunc3} is reported in Appendix \ref{iud3}. The first lemma regards the strong consistency of $\widehat{\theta}_{j,h,n}$.
\begin{lem}\label{lemmaTheta}
For every stratum $h=1,\ldots,H$, if $\lim_{n\rightarrow\infty}N_{j,h,n}\stackrel{a.s.}=\infty$ for a given treatment $j\in\{1, \dots, J\}$,
then $\lim_{n\rightarrow\infty}\widehat{\theta}_{j,h,n}\stackrel{a.s.}= \theta_{j,h}$.
\end{lem}
\begin{proof}
Let $M_{j,h, n}=\sum_{i=1}^n \mathbb{I}_{\{Z_i=h\}}\delta_{j,i}(Y_i-\theta_{j,h})=S_{j,h,n}-\theta_{j,h}N_{j,h,n}$,
then $M_{j,h, n}-M_{j,h, n-1}= \mathbb{I}_{\{Z_n=h\}}\delta_{j,n}(Y_n-\theta_{j,h})$, so that $M_{j,h}=(M_{j,h,n})_n$ is a martingale with respect to the filtration $(\mathcal{F}_n\vee\sigma(Z_{n+1},\boldsymbol{\delta}_{n+1}) )_n$ and
\begin{equation*}
\begin{split}
\langle M_{j,h}\rangle_n-\langle M_{j,h}\rangle_{n-1}=&E\left[(M_{j,h, n}-M_{j,h, n-1})^2 \vert \mathcal{F}_{n-1},Z_n,\boldsymbol{\delta}_n\right]\\
=&\mathbb{I}_{\{Z_n=h\}}\delta_{j,n}\theta_{j,h}(1-\theta_{j,h}).
\end{split}
\end{equation*}
Thus, $\langle M_{j,h}\rangle_n = \sum_{i=1}^n \mathbb{I}_{\{Z_i=h\}}\delta_{j,i} \theta_{j,h}(1-\theta_{j,h})=
\theta_{j,h}(1-\theta_{j,h})N_{j,h,n}$ and thus $\lim_{n\rightarrow\infty}\langle M_{j,h}\rangle_n \stackrel{a.s.}=\infty$, since $\lim_{n\rightarrow\infty}N_{j,h,n}\stackrel{a.s.}= \infty$; therefore, according to result \cite[12.14 (a)]{will},
$$\lim_{n\rightarrow\infty}\frac{M_{j,h, n}}{\langle M_{j,h}\rangle_n}=
\lim_{n\rightarrow\infty}\frac{S_{j,h,n}-\theta_{j,h}N_{j,h,n}}{\theta_{j,h}(1-\theta_{j,h})N_{j,h,n}} \stackrel{a.s.} = 0,$$
namely $\lim_{n\rightarrow\infty}\widehat{\theta}_{j,h,n}\stackrel{a.s.}=\theta_{j,h}$.
\end{proof}
The following lemma states that, if the assignment probabilities in a certain subset of possible treatments are eventually
far from zero, then the total number of patients allocated to that subset will grow linearly to $\infty$.
 \begin{lem}\label{lemma-infty}
At each step $n$, let
\begin{equation*}
Pr\left(\delta_{j,(n+1)}=1 | \mathcal{F}_n, Z_{n+1}=h\right)=\pi_{j,h,n}
\end{equation*} be a generic allocation function.
Then, for any given stratum $h$ and any non-empty subset $S \subseteq \{1,\dots,J\}$,
up to a negligible set,
\begin{equation*}
\begin{split}
\left\{\liminf_{n\to \infty} \textstyle\sum_{j\in S}\pi_{j,h,n}>0\right\} &\subseteq
\left\{\liminf_{n\to \infty} \tfrac{\sum_{j\in S} N_{j,h,n}}{n}> 0\right \}\\
&\subseteq \cup_{j\in S}\left\{\lim_{n\to\infty}N_{j,h,n}=\infty\right\}\,.
\end{split}
\end{equation*}
\end{lem}
\begin{proof} It is enough to prove the first containment, being the second one obvious.
For a non-empty set $S$, let
$$ M_{S,h,n}=\sum_{i=1}^n \mathbb{I}_{\{Z_i=h\}}\tfrac{\sum_{j\in S}(\delta_{j,i}-\pi_{j,h,i-1})}{i};$$
then $(M_{S,h,n})_n$ is a $L^2$-bounded martingale with respect to the filtration
$(\mathcal{F}_n\vee\sigma(Z_{n+1}))_n$, since
$$ E\Big[\sup_n M_{S,h,n}^2\Big]=O\Big(\sum_{i=1}^{\infty} i^{-2}\Big)<\infty.$$
Therefore, it is convergent almost surely and, by Kronecker's Lemma,
\begin{equation*}
\begin{split}
\frac{\sum_{j\in S} N_{j,h,n}}{n} - \frac{1}{n} \sum_{i=1}^n \mathbb{I}_{\{Z_i=h\}}\left(\sum_{j\in S}\pi_{j,h,i-1}\right)&=\\
\frac{1}{n} \sum_{i=1}^n i\left\{ \frac{\mathbb{I}_{\{Z_i=h\}}\sum_{j\in S}(\delta_{j,i}-\pi_{j,h,i-1})}{i}\right\}&
\stackrel{a.s.}\longrightarrow 0
\end{split}
\end{equation*}
and hence
$$ \liminf_n \frac{\sum_{j\in S}N_{j,h,n}}{n} \stackrel{a.s.}{=}\liminf_n \frac{1}{n} \sum_{i=1}^n \mathbb{I}_{\{Z_i=h\}} \left(\sum_{j\in S}\pi_{j,h,i-1}\right).$$
Now, define $\pi_{S,h,\infty}=\liminf_{n\to \infty} \sum_{j\in S} \pi_{j,h,n}$ and, for $\epsilon>0$,
consider the event $$B_\epsilon=\bigcup_n \bigcap_{i\geq n}\Big\{\sum_{j\in S}\pi_{j,h,i-1} > \pi_{S,h,\infty}-\epsilon/p_h\Big\}.$$
By definition of $\liminf$, $Pr(B_\epsilon)=1$ for each $\epsilon>0$. On the other hand,
by the above equality and since $\sum_{i=1}^n \mathbb{I}_{\{Z_i=h\}}/n\stackrel{a.s.}\to p_h>0$, then
$B_\epsilon$ is contained (up to a negligible set) in the event
$\{ \liminf_n \sum_{j\in S} N_{j,h,n}/n> \pi_{S,h,\infty}p_h-\epsilon \}$, for each $\epsilon>0$
$$\Pr\Big(\liminf_n \sum_{j\in S} \frac{N_{j,h,n}}{n}> \pi_{S,h,\infty}p_h-\epsilon\Big)=1,$$
namely $\Pr\big(\liminf_n \sum_{j\in S} N_{j,h,n}/n\geq \pi_{S,h,\infty}p_h\big)=1$. Thus, since $p_h>0$, we get
$$\{\pi_{S,h,\infty}>0\}\subseteq\Big\{\liminf_n \sum_{j\in S} N_{j,h,n}/n>0\Big\}$$ up to a negligible set.
\end{proof}

\subsection{IUD with update mechanisms \eqref{updatefunc1} and \eqref{updatefunc2}}\label{iud45}
The following lemma shows that, by adopting the IUD with update mechanisms \eqref{updatefunc1} and \eqref{updatefunc2}, $P_{j,h,n}$ is a strongly consistent estimator of $\theta_{j,h}$ provided that $\lim_{n\rightarrow\infty}N_{j,h,n}\stackrel{a.s.}= \infty$.
\begin{lem}\label{lemmaP}
Under the IUD with update mechanisms \eqref{updatefunc1} and \eqref{updatefunc2}, for every fixed stratum $h$, if $\lim_{n\rightarrow\infty}N_{j,h,n}\stackrel{a.s.}=\infty$ for a given treatment $j\in\{1, \dots, J\}$, then $\lim_{n\rightarrow\infty}P_{j,h,n}\stackrel{a.s.}= \theta_{j,h}$.
\end{lem}
\begin{proof}
Under the updating mechanism in \eqref{updatefunc1}, $\lim_{n\rightarrow\infty}\rho_{j,h,n}\stackrel{a.s.}= 1$ due to the boundedness of $\psi$; thus, $P_{j,h,n}$ is asymptotically equivalent to $\widehat{\theta}_{j,h,n}$, since $|P_{j,h,n}-\widehat{\theta}_{j,h,n}|=O(1-\rho_{j,h,n})\stackrel{a.s.}\to 0$. Taking now into account the updating mechanism \eqref{updatefunc2}, by Lemma~\ref{lemmaTheta} $\widehat{\theta}_{j,h,n}$ is a strongly consistent estimator of $\theta_{j,h}$; while,
for a given stratum $k\neq h$, $\lim_{n\rightarrow\infty}\widehat{\theta}_{j,k,n}\stackrel{a.s.}=\theta_{j,k}$ if $\lim_{n\rightarrow\infty}N_{j,k,n}\stackrel{a.s.}=\infty$, or there exists a (random) time $\tau$ such that $\lim_{n\rightarrow\infty}\widehat{\theta}_{j,k,n}\stackrel{a.s.}= \widehat{\theta}_{j,k,\tau}$ (potentially different from ${\theta}_{j, k}$). Since $c_n\to 0$, asymptotically the stratum $k$ will
belong to $A_{j,h,n}$ only if
$\theta_{j,k}=\theta_{j,h}$ or $\widehat{\theta}_{j,k,\tau}=\theta_{j,h}$. In both cases,  $P_{j,h,n}$ tends to be a mixture of two strongly consistent estimators of the same $\theta_{j,h}$ and so $\lim_{n\rightarrow\infty}P_{j,h,n}\stackrel{a.s.}=\theta_{j,h}$.
\end{proof}

\begin{proof}[Proof of Theorem \ref{teoCons} with updates \eqref{updatefunc1} and \eqref{updatefunc2}]
At stratum $h$ let
\begin{equation}\label{def-X}
X(\omega)=\{j: N_{j,h,n}(\omega)\rightarrow \infty\},
\end{equation}
then $X$ is a rv taking values in the power set $\wp(\{1,\dots,J\})$.
We will prove that $\Pr(X=\{1,\dots,J\})=1$, namely
$\Pr(\bigcap_{j=1}^J \{ N_{j,h,n}\to \infty\})=1$. For this purpose, firstly note that
$\Pr(X=\emptyset)=0$ because, by assumption, $N_{\centerdot,h,n}=\sum_{j=1}^JN_{j,h,n}\stackrel{a.s.}\sim n p_h \to \infty$.
Then it remains to prove that $\Pr(X=S)=0$ for each set $S\neq \emptyset$ with $S^c\neq\emptyset$.
Assume that $\Pr(X=S)>0$ for such a set $S$. Then, adopting updates \eqref{updatefunc1} and \eqref{updatefunc2}, by Lemmas \ref{lemmaTheta} and \ref{lemmaP}, up to a negligible set,
\begin{equation}\label{eq-fund}
\{X=S\}\subseteq \left(\cap_{j\in S}\{\lim_n P_{j,h,n}=\theta_{j,h}\}\right) \cap \left(\cap_{j\in S^c} \left\{\lim_n P_{j,h,n}=\tilde{\theta}_{j,h}\right\}\right),
\end{equation}
where $\tilde{\theta}_{j,h}$ denotes a random variable with values in $[0,1]$. Therefore, by adopting the allocation function $\pi_{j,h,n}$ in (\ref{randfunction}), the event $\{X=S\}$ is included, up to a negligible set, into
$$
\bigg\{\liminf_{n\to \infty}\sum_{j\in S^c} \pi_{j,h,n}=\lim_{n\to \infty} \sum_{j\in S^c}\pi_{j,h,n}=
\tfrac{ \sum\limits_{j\in S^c} f( \tilde{\theta}_{j,h} ) }{\sum\limits_{j\in S}f( \theta_{j,h} ) + \sum\limits_{j\in S^c} f( \tilde{\theta}_{j,h}) }>0\bigg\},
$$
since $f$ is continuous and strictly positive. However, from Lemma~\ref{lemma-infty}, this is possible only if $\Pr(X=S)=0$. Indeed, in terms of the rv $X$, this lemma (applied to $S^c\neq\emptyset$) states that $$\Big\{\liminf_{n\to \infty}\sum_{j\in S^c} \pi_{j,h,n}>0\Big\}$$ is contained (up to a negligible set) in
$\{X\cap S^c\neq \emptyset\}$ and so in $\{X\neq S\}$. Summing up, by contradiction, we have proven that $\Pr(X=S)=0$ for each $S\neq \emptyset$ with $S^c\neq \emptyset$. Hence, $\Pr(X=\{1,\dots,J\})=1$, namely $\lim_{n\rightarrow\infty}N_{j,h,n}\stackrel{a.s.}=\infty$ for every $j=1,\ldots,J$ and $h\in \{1,\ldots,H\}$; thus, by the previous lemmas, $\widehat{\theta}_{j,h,n}$ and $P_{j,h,n}$ are strongly consistent estimators of $\theta_{j,h}$ for every $j$ and $h$
and $\pi_{j,h,n}\stackrel{a.s.}\to \pi_{j,h}\in (0,1)$. Arguing as we did in Lemma \ref{lemmaTheta}, we can define
$\widetilde{M}_{j,h,n}=\sum_{i=1}^n \mathbb{I}_{\{Z_i=h\}}(\delta_{j,i}-\pi_{j,h,i-1})=N_{j,h,n}-\sum_{i=1}^n\mathbb{I}_{\{Z_i=h\}}\pi_{j,h,n}$,
so that $\widetilde{M}_{j,h}=(\widetilde{M}_{j,h,n})_n$ is a martingale with respect to the filtration $(\mathcal{F}_n\vee\sigma(Z_{n+1}))_n$
and therefore, as a consequence of result \cite[12.14 (a)]{will}, ${N_{j,h,n}}/{N_{\centerdot,h,n}}\stackrel{a.s.}\to \pi_{j,h}$.
\end{proof}

\subsection{IUD with update mechanism \eqref{updatefunc3}}\label{iud3}
By considering now the IUD with \eqref{updatefunc3}, the next lemma shows that the model-based updates are asymptotically bounded almost surely.
This result can be used in connection with Appendix~\ref{app-model-based-likelihood} and Lemma~\ref{teoBound} to show the strong consistency of $P_{j,h,n}$ for every treatment $j$ and stratum $h$.
\begin{lem}\label{teoBound_model_based}
Under the IUD with model-based update mechanism \eqref{updatefunc3},
\begin{equation}\label{eq-1bis}
\lim_{n\to \infty} N_{j,h,n}\stackrel{a.s.}=\infty,\qquad \text{for every } h\in\{1,\ldots,H\} \;\text{and } j\in\{1,\ldots,J\}
\end{equation}
and
\begin{equation}\label{eq-2}
\limsup_{n\to \infty} (\widehat{\alpha}_{j,n}+\widehat{\beta}_{j,n})<\infty, \qquad \text{for every } j\in\{1,\ldots,J\}.
\end{equation}
\end{lem}
\begin{proof}
At a given stratum $h$, let $X$ be defined as in \eqref{def-X}. We have to prove that $\Pr(X=S)=0$ for each $S\neq \emptyset$ with $S^c\neq\emptyset$. For update \eqref{updatefunc3}, \eqref{eq-fund} is not straightforward (due to the presence of the terms $\widehat{\alpha}_n$ and $\widehat{\beta}_n$
in the definition of $P_{j,h,n}$) and hence we adopt a different proof.\\
\indent Take a subset $S$ of $\{1,\dots,J\}$ with $S\neq \emptyset$ and set $\theta_{\max} = \max_{j,h} \theta_{j,h}$ (which is strictly smaller than $1$ by assumption).
By Lemma~\ref{lemmaTheta},
\begin{equation*}
\Pr\left(\{X=S\} \cap \left( \cup_{j\in S}\ \left\{ \lim_n \widehat{\theta}_{j,h,n} > \theta_{\max} \right\} \right) \right)=0.
\end{equation*}
Now, let $\theta_*=(\theta_{\max}+1)/2\in(\theta_{\max},1)$ and setting
\[\tau_{j,k} =\inf\left\{ n > \tau_{j,k-1}:
\widehat{\alpha}_{j,n}+\widehat{\beta}_{j,n} > k,
\tfrac{\widehat{\alpha}_{j,n}}{\widehat{\alpha}_{j,n}+\widehat{\beta}_{j,n}} \geq  \theta_*
\right\},\]
(with $\tau_{j,0}=0$ and $\inf\emptyset= \infty$), then
\[\{X=S\} \cap
\left( \cup_{j\in S} \left\{ \limsup_n P_{j,h,n} \geq \theta_* \right\}\right)
\subseteq
\{X=S\} \cap \left( \cup_{j\in S} \cap_{k} \left\{ \tau_{j,k} < \infty \right\} \right).
\]
On the other hand for $j\in S$, by Lemma~\ref{lemmaTheta}, Lemma \ref{lemma-likelihood} and \eqref{teoBound_ii} of Lemma~\ref{teoBound}
applied to $[\vN_{j,n}, \vS_{j,n}]$,
$\Pr\left(\{X=S\} \cap (\cap_{k} \{ \tau_{j,k} < \infty \}) \right) = 0$ and hence
\begin{equation}\label{eq:limsup_P_bounded}
\Pr\left( \{X=S\} \cap \left( \cup_{j\in S} \left\{ \limsup_n P_{j,h,n} \geq  \theta_*  \right\} \right) \right)= 0.
\end{equation}
Now we will show that $\Pr(X=S)=0$ when also $S^c\neq \emptyset$ by proving that
\begin{multline*}
\Pr\left(
\{X=S\} \cap  \left( \cap_{j\in S} \left\{ \limsup_n P_{j,h,n} < \theta_* \right\} \right) \right)
=\\
\Pr\left(\{X=S\} \cap \left( \cup_{j\in S} \left\{ \limsup_n P_{j,h,n} \geq \theta_* \right\} \right)^c \right)
= 0.
\end{multline*}
To this end, consider the allocation function $\pi_{j,h,n}$ in (\ref{randfunction}) and set $f_* = f(\theta_*)$,
from the monotonicity of $f$ it follows that
\begin{align*}
\{X=S\}&\cap \left(
\cap_{j\in S} \left\{ \limsup_n P_{j,h,n} < \theta_* \right\} \right)
\subseteq \\
\{X=S\}&\cap
\Big\{
\limsup_n \sum_{j\in S} \pi_{j,h,n} \leq \tfrac{|S|f_*}{|S|f_*+|S^c|f(0)}
\Big\}=\\
\{X=S\}&\cap
\Big\{
\liminf_n \sum_{j\in S^c} \pi_{j,h,n} \geq \tfrac{|S^c|f(0)}{|S|f_*+|S^c|f(0)}
\Big\},
\end{align*}
where $|S|$ denote the cardinality of $S$. By Lemma~\ref{lemma-infty} and since $f(0)>0$, the probability of the last event is zero when $S^c\neq \emptyset$.
Indeed, in terms of the rv $X$, this lemma (applied to $S^c$) states that the event
$\{\liminf_n \sum_{j\in S^c} \pi_{j,h,n} >0 \}$ is contained (up to a negligible set) in $\{X \neq S\}$. Summing up, we have proven that $\Pr(X = \{ 1,\ldots,J\})=1$. As a consequence, from \eqref{eq:limsup_P_bounded} with $ S = \{ 1,\ldots,J\}$, we get
$\Pr\left( \limsup_n P_{j,h,n} < \theta_*  \right) = 1$ for every $j$,
which implies that
\[\Pr \left( \liminf_n \pi_{j,h,n} \geq \tfrac{f(0)}{(J-1)f_*+f(0)} \right)=1,\quad \text{ for every } j\in\{1,\ldots,J\}.\]
By Lemma~\ref{lemma-infty} applied to $S=\{j\}$, at each $h\in\{1,\ldots,H\}$,
$\liminf_{n} N_{j,h,n}/n>0$ a.s. for every $j$,
which proves \eqref{eq-1bis} and,
combined with Lemma~\ref{lemmaTheta} and equation \eqref{teoBound_i} in Lemma~\ref{teoBound}, also proves \eqref{eq-2}.
\end{proof}

\begin{proof}[Proof of Theorem \ref{teoCons} with update \eqref{updatefunc3}]
From Equation \eqref{eq-1bis} and Lemma \ref{lemmaTheta},  we get the strong consistency of $\widehat{\theta}_{j,h,n}$ for every $h$ and $j$. Moreover, from \eqref{eq-1bis} and \eqref{eq-2}, we get $1-\rho_{j,h,n}=O(1/N_{j,h,n})\stackrel{a.s.}\to 0$ for every $h$ and $j$, so that $\lim_{n\rightarrow\infty}P_{j,h,n}\stackrel{a.s.}=\theta_{j,h}$ for every $h$ and $j$.
Finally, we can conclude exactly as we did in the proof of Theorem~\ref{teoCons} with updates \eqref{updatefunc1} and \eqref{updatefunc2}.
\end{proof}

\section{Proof of Theorem 4.2}\label{proofTeoAsiNorm}
We are going to apply \cite[Theorem C.1]{Cri23} (with $k_n=n$ and $t_n=1$ in its notation)
at each fixed stratum $h\in\{1,\ldots,H\}$. Recall that, from Theorem \ref{teoCons},
$\lim_{n\to \infty} {n^{-1} ( p_h\pi_{j,h})^{-1}}{N_{j,h,n}}\stackrel{a.s.}=1$ for every $j$.
Let us now define  a triangular array ${({\vT}_{n,i})}_{n\geq 1, 1\leq i \leq n}$ of $J$-dim vectors as
${\vT}_{n,i}=\left(T_{1,n,i}, \dots, T_{J,n,i}\right)^\top$, where
$T_{j,n,i}={\mathbb{I}_{[Z_i=h]}\delta_{j,i}(Y_i-\theta_{j,h})}/{\sqrt{n p_h\pi_{j,h}}}$. Note that $T_{j,n,i}=(M_{j,h,i}-M_{j,h,i-1})/\sqrt{n p_h\pi_{j,h}}$, where $(M_{j,h,n})_n$ is
the martingale defined in Lemma~\ref{lemmaTheta}. Therefore, for any fixed $n$,
${({\vT}_{n,i})}_{1\leq i \leq n}$ is a martingale difference array with respect to
the filtration $(\mathcal{G}_{n,i})_{i\geq 0}$, where here ${\mathcal G}_{n,i}=\mathcal{F}_i$
(note that they are trivially nested, i.e.~$\mathcal{G}_{n,i}\subseteq{\mathcal G}_{n+1,i}$). The above quoted theorem states, in particular, that the $J$-dim martingale
$\sum_{i=1}^n {\vT}_{n,i}$ converges in distribution to ${\mathcal N}(0, \boldsymbol{\Sigma}_{(h)})$,
provided the following conditions are satisfied:
\begin{itemize}
\item[(a)] $\sum_{i=1}^n{\vT}_{n,i}{\vT}_{n,i}^\top\stackrel{P}\to \boldsymbol{\Sigma}_{(h)}$,
\item[(b)] $E\left[\, \sup_{1\leq i\leq n} \sum_{j=1}^J |T_{j,n,i}|\, \right] \to 0$.
\end{itemize}
Since $|T_{j,n,i}|\leq 1/(\sqrt{n}\sqrt{p_h \min_j \{\pi_{j,h} \}})$ for every $j$, then condition (b) is obviously satisfied. Moreover,
\begin{equation}\label{matr}
\sum_{i=1}^n{\vT}_{n,i}{\vT}_{n,i}^\top=\text{diag}\left( \sum_{i=1}^n T_{j,n,i}^2 \right)_{j=1, \dots, J}.
\end{equation}
Since
\begin{equation}\label{eq-asy-sc}
\begin{aligned}
E\left[\mathbb{I}_{[Z_i=h]}\delta_{j,i}(Y_i-\theta_{j,h})^2 \vert \mathcal{F}_{i-1}\right]
&  =E\left[E[\mathbb{I}_{[Z_i=h]}\delta_{j,i}(Y_i-\theta_{j,h})^2 \vert \mathcal{F}_{i-1}\vee\sigma(Z_{i},\boldsymbol{\delta}_{i})] \vert \mathcal{F}_{i-1}\right]\\
&=E\left[\mathbb{I}_{[Z_i=h]}\delta_{j,i}E[(Y_i-\theta_{j,h})^2 \vert  \mathcal{F}_{i-1}\vee\sigma(Z_{i},\boldsymbol{\delta}_{i})] \vert \mathcal{F}_{i-1}\right]\\
&=E\left[\mathbb{I}_{[Z_i=h]}\delta_{j,i} \theta_{j,h}(1-\theta_{j,h}) \vert \mathcal{F}_{i-1}\right]\\
&= \theta_{j,h}(1-\theta_{j,h})E[\mathbb{I}_{[Z_i=h]}E[\delta_{j,i}  \vert \mathcal{F}_{i-1} \vee \sigma(Z_i)] \vert\mathcal{F}_{i-1}]\\
&= \theta_{j,h}(1-\theta_{j,h})\tfrac{ f( P_{j,h,i-1} ) }
{ \sum_{j=1}^J f( P_{j,h,i-1} )} E[\mathbb{I}_{[Z_i=h]}  \vert \mathcal{F}_{i-1}]\\
&\mathop{\longrightarrow}\limits_{i\to\infty}^{a.s.} \theta_{j,h}(1-\theta_{j,h}) p_h\pi_{j,h},
\end{aligned}
\end{equation}
we can apply \cite[Lemma B.1]{Cri23} to the sequence $\left(\mathbb{I}_{[Z_i=h]}\delta_{j,i}(Y_i-\theta_{j,h})^2 \right)_{i \geq 1}$ (with $\alpha_i=1$ and $\beta_n=n p_h\pi_{j,h}$ in its notation), so that, for the component $(j,j)$ of \eqref{matr},
\begin{equation*}
\sum_{i=1}^n T_{j,n,i}^2 = \frac{1}{n p_h\pi_{j,h}}\sum_{i=1}^n \mathbb{I}_{[Z_i=h]}\delta_{j,i}(Y_i-\theta_{j,h})^2
\stackrel{a.s.}\longrightarrow
\theta_{j,h}(1-\theta_{j,h}).
\end{equation*}
Hence, the matrix \eqref{matr} converges a.s. to $\boldsymbol{\Sigma}_{(h)}$ and condition (a) is satisfied. Then, as $n\rightarrow \infty$, $\sum_{i=1}^n {\vT}_{n,i}\stackrel{d}\hookrightarrow{\mathcal N}(0, \boldsymbol{\Sigma}_{(h)})$. Since
\begin{equation*}
\sum_{i=1}^n {T}_{j, n,i}=\sum_{i=1}^n \tfrac{\mathbb{I}_{\{Z_i=h\}}\delta_{j,i}(Y_i-\theta_{j,h})}{\sqrt{n  p_h\pi_{j,h}}},
\end{equation*}
then $\sqrt{n  p_h\pi_{j,h}}\sum_{i=1}^n {T}_{j, n,i}=S_{j,h,n}-\theta_{j,h}N_{j,h,n}$ and therefore
\begin{equation}\label{eq-trasl}
\sqrt{\tfrac{n p_h\pi_{j,h}}{N_{j,h,n}}}\sum_{i=1}^n {T}_{j, n,i}=\sqrt{N_{j,h,n}} (\widehat{\theta}_{j,h,n}-\theta_{j,h}),
\end{equation}
where $\lim_{n\to \infty}\sqrt{n p_h\pi_{j,h}/N_{j,h,n}}\stackrel{a.s.}= 1$. As a consequence, we obtain \eqref{eq:CLT-theta}.
\\
\indent In order to prove \eqref{eq:CLT-P}, observe that $1-\rho_{j,h,n}\stackrel{a.s.}=O(1/N_{j,h,n})$
under mechanisms \eqref{updatefunc1}, \eqref{updatefunc3} (where, by Lemma \ref{teoBound_model_based},
$\limsup_n(\widehat{\alpha}_{j,n}+\widehat{\beta}_{j,n})<\infty$ for each treatment $j$) and \eqref{updatefunc2} in the absence of similar strata; thus,
$\sqrt{N_{j,h,n}}|\widehat{\theta}_{j,h,n}-P_{j,h,n}| \leq \sqrt{N_{j,h,n}}O(1-\rho_{j,h,n}) \stackrel{a.s.}\longrightarrow 0$.
Under mechanism \eqref{updatefunc2} in the presence of similar strata, we recall that, for every stratum $h$ and treatment $j$, $A^*_{j,h}=\{k\neq h:\, \theta_{j,k}=\theta_{j,h}\}$. Consider the $h$th stratum as fixed and set $A^*_j=A^*_{j,h}\cup\{h\}$ and $\tilde{\theta}_j=\theta_{j,k}$ for all $k \in A^*_{j}$. Moreover, by letting ${\rho^*_{j,k,n}}={N_{j, k, n}}/{\sum_{l \in A^*_j}N_{j, l, n}}$, from Theorem \ref{teoCons}, $\lim_{n\to \infty}\rho^*_{j,k,n}\overset{a.s}= p_k\pi_{j,k}/\sum_{l \in A^*_j} p_l\pi_{j,l}$. Now, observe that
\begin{equation*}
\begin{split}
P^*_{j,h,n}=\tfrac{\sum_{k \in A^*_j} S_{j,k,n}}{\sum_{k \in A^*_j} N_{j,k,n}}=
\tfrac{\sum_{k \in A^*_j} N_{j,k,n}\widehat{\theta}_{j,k,n}}{\sum_{k \in A^*_j} N_{j,k,n}}=\tfrac{\sum_{k \in A^*_j} N_{j,k,n}(\widehat{\theta}_{j,k,n}-\tilde{\theta}_j)}{\sum_{k \in A^*_j} N_{j,k,n}}+\tilde{\theta}_j
\end{split}
\end{equation*}
and thus
\begin{equation*}
\sqrt{\textstyle\sum_{k \in A^*_j} N_{j,k,n}}(P_{j,h,n}^*-\tilde{\theta}_j) =
\textstyle\sum_{k \in A^*_j} \sqrt{\rho^*_{j,k,n}}\,\sqrt{N_{j,k,n}}(\widehat{\theta}_{j,k,n}-\tilde{\theta}_j),
\end{equation*}
where $$\sqrt{\rho_{j,k,n}^*}\sqrt{N_{j,k,n}}(\widehat{\theta}_{j,k,n}-\tilde{\theta}_j)
\stackrel{d}{\hookrightarrow}\mathcal N\left(0, \tfrac{\tilde{\theta}_j(1-\tilde{\theta}_j)}{{ p_h\pi_{j,h}}/{\sum_{k \in A^*_j} p_k\pi_{j,k}}}\right);$$
hence, from Slutsky's theorem,
$$\sqrt{\textstyle\sum_{k \in A^*_j} N_{j,k,n}}(P_{j,h,n}^*-\tilde{\theta}_j) \stackrel{d}{\hookrightarrow} \mathcal N\left(0,  \tilde{\theta}_j(1-\tilde{\theta}_j)\right). $$
Now, from Lemma \ref{lemmaTheta}, the martingale process $(M_{j,h,n})_n$ is such that $E[(M_{j,h,i}-M_{j,h, i-1})^2 |\mathcal{F}_{i-1}]=\theta_{j,h}(1-\theta_{j,h}) p_h\pi_{j,h}$. By applying the Law of Iterated Logarithm for martingales in \cite{Stout70}, following its notation let $s_n^2=\sum_{i=1}^n E[(M_{j,h, i}-M_{j,h, i-1})^2 |\mathcal{F}_{i-1}]=n \cdot \theta_{j,h}(1-\theta_{j,h}) p_h\pi_{j,h}$ and $u_i=\sqrt{2\ln\ln s_i^2}$; then, $K_i=u_i/s_i\to 0$ and $|M_{j,h, i}-M_{j,h, i-1}|\leq 1=K_i \, \cdot \, {s_i}/{u_i}.$ Therefore,
$$\limsup_{n\to \infty} M_{j,h, n}/(s_nu_n)\stackrel{a.s.}=1 \quad \text{and} \quad \liminf_{n\to\infty} M_{j,h, n}/(s_nu_n)\stackrel{a.s.}=-1.$$
Recalling that $N_{j,h,n}/n\stackrel{a.s.}\to  p_h\pi_{j,h}$, then
$$\limsup_{n\to \infty} {\tfrac{\sqrt{n}}{u_n}}(\widehat{\theta}_{j,h,n}-\theta_{j,h})\stackrel{a.s.}=c_{j,h}
\quad\mbox{and}\quad\liminf_{n\to \infty} {\tfrac{\sqrt{n}}{u_n}}(\widehat{\theta}_{j,h,n}-\theta_{j,h})\stackrel{a.s.}=-c_{j,h}\,,$$
where $c_{j,h}$ is the resulting constant.
Thus, for any fixed $\epsilon>0$ there exists
a (random) time $\tau_{j,h,\epsilon}$ such that
$|\widehat{\theta}_{j,h,n}-\theta_{j,h}|\leq (1+\epsilon)c_{j,h}u_n/\sqrt{n}$ for each $n\geq \tau_{j,h,\epsilon}$. This implies that, for each pair $(h,k)$ with $k\in A^*_h$, there exists a (random) time $\tau_{j,h,k,\epsilon}=\max\{\tau_{j,h,\epsilon},\tau_{j,k,\epsilon}\}$ such that
$|\widehat{\theta}_{j,h,n}-\widehat{\theta}_{j,k,n}|\leq 2(1+\epsilon)c_{j,h}u_n/\sqrt{n}$ for each $n\geq \tau_{j,h,k,\epsilon}$.
Then, choosing $c_n$ so that $c_n\to 0$ and $u_n/\sqrt{n}c_n \to 0$ (that is $c_n\sqrt{n}/u_n\to \infty$),
we can find a time $n_{j,h,\epsilon}$ such that $c_n\geq 2(1+\epsilon)c_{j,h}u_n/\sqrt{n}$ for every $n\geq n_{j,h,\epsilon}$. Therefore, we can conclude that
there exists a  (random) time $\tau^*_{j,h,k,\epsilon}=\max\{\tau_{j,h,k,\epsilon}, n_{j,h,\epsilon}\}$
such that $|\widehat{\theta}_{j,h,n}-\widehat{\theta}_{j,k,n}|\leq c_n$ for each $n\geq \tau^*_{j,h,k,\epsilon}$.
In other words, for every fixed stratum $h$, $A_{j,h,n}\cup\{h\}=A^*_j$ eventually and so  
$N_{j,h,n}+\sum_{k\in A_{j,h,n}} N_{j,k,n}=\sum_{k\in A^*_{j}} N_{j,k,n}$ and $P_{j,h,n}=P^*_{j,h,n}$ eventually, thus
$$\sqrt{N_{j,h,n}+\textstyle\sum_{k\in A_{j,h,n}} N_{j,k,n}}( P_{j,h,n} - \tilde{\theta}_j) \stackrel{d}\hookrightarrow \mathcal{N}(0,  \tilde{\theta}_j(1-\tilde{\theta}_j)).$$

\section{Proof of Theorem \ref{thmSeqMon} (sequential monitoring)}\label{app:seqMonitor}
Here we use the same notation as in Appendix~\ref{proofTeoAsiNorm}. For any fixed stratum $h$, let $\va=(a_1,\ldots,a_J)^\top$ be a vector of coefficients and consider the real triangular array $(\va^\top {\vT}_{n,i})_{n\geq 1, i\geq 1}$. By \cite[Theorem B.2 and Remark B.3]{Cri19Res} applied to this triangular array with $k_n(t)={\lfloor nt \rfloor}$,
$$\Big(\va^\top \sum_{i=1}^{{\lfloor nt \rfloor}} {\vT}_{n,i}\Big)_{t\geq 0}\stackrel{d}\hookrightarrow\widetilde{W}=(W_{V_t})_{t\geq 0},$$
in the Skorohod's space, where $W=(W_t)_{t\geq 0}$ denotes a standard Wiener process, provided the following conditions are satisfied:
\begin{itemize}
\item[(a)] $\sum_{i=1}^{{\lfloor nt \rfloor}} E\left[(\va^\top{\vT}_{n,i})^2 \mid \mathcal{G}_{n,i-1}\right]\stackrel{P}\to V_t$;
\item[(b)] $ \sum_{i=1}^{n} E\left[ \, |\va^\top{\vT}_{n,i}|^u\right]\to 0$ for some $u>2$.
\end{itemize}
Condition $(b)$ is obviously satisfied, since $|T_{j,n,i}|\leq 1/(\sqrt{np_h\min_j\{ \pi_{j,h}\}})$. Condition $(a)$ follows from the proof of Theorem \ref{teoAsiNorm}.
Indeed, by \eqref{eq-asy-sc} and recalling that ${\lfloor nt \rfloor}/n\to t$, then
\begin{equation*}
\begin{split}
\sum_{i=1}^{{\lfloor nt \rfloor}} E\left[(\va^\top{\vT}_{n,i})^2| \mathcal{G}_{n,i-1}\right] &=
\va^\top \left(\sum_{i=1}^{{\lfloor nt \rfloor}} E[{\vT}_{n,i}\vT^\top_{n,i}| \mathcal{F}_{i-1}]\right) \va\\
&\stackrel{a.s.}\longrightarrow V_t = t\, \va^\top \boldsymbol{\Sigma}_{(h)} \va =
t\,\sum_{j=1}^J a_j^2 \theta_{j,h}(1-\theta_{j,h})\,.
\end{split}
\end{equation*}
Note that the process $\widetilde{W}$  has the same distribution of $\left(\va^\top \widetilde{\vW}\right)_{t \geq 0}$, where
$$\widetilde{\vW}_t= \Big(
W_{\theta_{1,h}(1-\theta_{1,h})\, t}^{(1)}, \\
W_{\theta_{2,h}(1-\theta_{2,h})\, t}^{(2)}, \\
\dots \\
W_{\theta_{J,h}(1-\theta_{J,h})\, t}^{(J)}
 \Big)^\top,$$
with independent standard Wiener processes $\left(W_{t}^{(j)}\right)_{t\geq 0}$.  \\
\noindent For a fixed pair $j\neq l$ of treatments, define for each $n$
the stochastic process $\widetilde{U}_{h,n}=(\widetilde{U}_{h,n,t})_{t\geq 0}$ as
\begin{equation*}
 \widetilde{U}_{h,n,t} =
 \sqrt{{\lfloor nt \rfloor}}
 \tfrac{ \left(\widehat{\theta}_{j,h,{\lfloor nt \rfloor}}-\widehat{\theta}_{l,h,{\lfloor nt \rfloor}}\right) - (\theta_{j,h}-\theta_{l,h}) }
 { \sqrt{v_{j,h}+v_{l,h}}}\,,
\end{equation*}
where $v_{j,h}=\theta_{j,h}(1-\theta_{j,h})/p_h\pi_{j,h}$.
Setting $\widetilde{\va}=(v_{j,h}+v_{l,h})^{-1/2}(\ve_j-\ve_l)$, $\widetilde{U}_{h,n,t}$ can be rewritten as
\begin{equation*}
\widetilde{U}_{h,n,t}= \sqrt{{\lfloor nt \rfloor}}\,\widetilde{\va}^\top\left(\widehat{\vtheta}_{(h){\lfloor nt \rfloor}}-\vtheta_{(h)}\right)\,.
\end{equation*}
From \eqref{eq-trasl},
\begin{equation*}
\sqrt{t}\widetilde{U}_{h,n,t}=
\sqrt{\tfrac{nt}{{\lfloor nt \rfloor}}}\,
{\lfloor nt \rfloor}
\left(\diag(p_h\boldsymbol{\pi}_{(h)})\diag(\vN_{(h){\lfloor nt \rfloor}})^{-1}\va\right)^\top \sum_{i=1}^{{\lfloor nt \rfloor}}\vT_{n,i}
\end{equation*}
where
\begin{equation*}
\begin{split}
\va &= \tfrac{1}{\sqrt{p_h(v_{j,h}+v_{l,h})}} \big(\tfrac{\ve_j}{\sqrt{\pi_{j,h}}} - \tfrac{\ve_l}{\sqrt{\pi_{l,h}}}\big)\\
&=\sqrt{\lambda_h}
\tfrac{\ve_j}{\sqrt{\theta_{j,h}(1-\theta_{j,h})}} -
\sqrt{(1-\lambda_h)}\tfrac{\ve_l}{\sqrt{\theta_{l,h}(1-\theta_{l,h})}},
\quad\mbox{with } \lambda_h=v_{j,h}/(v_{j,h}+v_{l,h}).
\end{split}
\end{equation*}
Since
$\sqrt{nt/{\lfloor nt \rfloor}}\to 1$ and
 $${\lfloor nt \rfloor}\diag(p_h\boldsymbol{\pi}_{(h)})\diag(\vN_{(h){\lfloor nt \rfloor}})^{-1}\stackrel{a.s.}\to \vI_J,$$
the process $(\sqrt{t}\widetilde{U}_{h,n,t})_{t\geq 0}$ has the same asymptotic behavior
as $(\va^\top \sum_{i=1}^{{\lfloor nt \rfloor}} \vT_{n,i})_{t\geq 0}$, namely, for $n\to \infty$,
it converges in distribution in the Skorohod's space to $(\va^\top \widetilde{\vW}_t)_{t\geq 0}$,
that is to a standard Wiener process. Finally, if we define a strongly consistent estimator of $v_{j,h}$ as
$\widehat{v}_{j,h,n}= n N_{j,h, n}^{-1}\widehat{\theta}_{j,h,n}(1-\widehat{\theta}_{j,h,n})$,
we can write
\begin{equation*}
\sqrt{t} \, U_{h,n,t}=
\sqrt{t}\widetilde{U}_{h,n,t}
\sqrt{\tfrac{ v_{j,h}+v_{l,h} }{ (\widehat{v}_{j,h,{\lfloor nt \rfloor} } + \widehat{v}_{l,h,{\lfloor nt \rfloor} } ) } }
+
\sqrt{t} \sqrt{ {\lfloor nt \rfloor} } \tfrac{\theta_{j,h}-\theta_{l,h}}
{\sqrt{(\widehat{v}_{j,h, {\lfloor nt \rfloor}} + \widehat{v}_{l,h,{\lfloor nt \rfloor}} )}}
\end{equation*}
where the first term converges in distribution in the Skorohod's space to a standard Wiener process and
$$
\sqrt{t} \sqrt{ \tfrac{{\lfloor nt \rfloor}}{n} } \tfrac{\theta_{j,h}-\theta_{l,h}}
{\sqrt{(\widehat{v}_{j,h, {\lfloor nt \rfloor}} + \widehat{v}_{l,h,{\lfloor nt \rfloor}} )}}
\stackrel{a.s.}{\longrightarrow} t \, \frac{\theta_{j,h}-\theta_{l,h}}{\sqrt{v_{j,h}+v_{l,h}}}=t \mu_h.
$$
As a consequence, under the null hypothesis,
$(\sqrt{t} \, U_{h,n,t})_{t\geq 0}$ converges in distribution to a standard Wiener process and, under the alternative,
$(\sqrt{t}\, U_{h,n,t} -t \sqrt{n} \mu_h)_{t\geq 0}$ converges in distribution to a standard Wiener process.

\section{Likelihood for the model-based approach}\label{app-model-based-likelihood}
In this appendix, we are going to prove the functional form of the likelihood when the treatment effects are generated as in the model-based approach.
In this section, $\Gamma(x)$ and $B(\alpha,\beta)=\Gamma(\alpha)\Gamma(\beta)/\Gamma(\alpha+\beta)$ are the gamma  and the beta function, respectively. The parameters $\alpha$ and $\beta$ and their vectors $\valpha $ and $\vbeta$ are used in this appendix as function variables.
\begin{lem}\label{lemma-likelihood}
Under the IUD with update mechanism \eqref{updatefunc3}, let
$\valpha=(\alpha_1,\dots,\alpha_J)^\top$, $\vbeta=(\beta_1,\dots,\beta_J)^\top$,
$\vn_{j,n}=(n_{j,1,n},\dots, n_{j,H,n})^\top$,  $\vs_{j,n}=(s_{j,1,n},\dots, s_{j,H,n})^\top$,
$\vn_{n}=(\vn_{j,n})_{1\leq j\leq J}$ and $\vs_{n}=(\vs_{j,n})_{1\leq j\leq J}$;
the likelihood of $(\valpha,\vbeta)$
is
\begin{equation}\label{joint-likelihood}
\mathcal{L}_n(\valpha,\vbeta)
\propto \prod_{j=1}^J\prod_{h=1}^H BetaBin(n_{j,h,n},\alpha_{j},\beta_j)(s_{j,h,n})\,.
\end{equation}
where $BetaBin(n,\alpha,\beta)(s)$ denotes the Beta-Binomial distribution with parameters
$(n,\alpha,\beta)$ computed in $s$ and
$\propto$ means proportional with a proportionality factor that does not depend on $(\valpha,\vbeta)$.
\\
\indent Moreover, for each fixed treatment $j$, the joint distribution of the random vectors
$\vN_{j,n}$ and $\vS_{j,n}$ is of the form
\begin{equation}\label{marginal-distr}
\Pr\left(\vN_{j,n}=\vn_{j,n},\vS_{j,n}=\vs_{j,n};\alpha_j,\beta_j\right)\propto
\prod_{h=1}^H \frac{B(\alpha_j+s_{j,h,n},\beta_j+n_{j,h,n}-s_{j,h,n})}{B(\alpha_j,\beta_j)} \,,
\end{equation}
where $\propto$ means proportional with a factor that does not depend on $(\alpha_j,\beta_j)$.
\end{lem}
The MLEs $\widehat{\valpha}_n$ and $\widehat{\vbeta}_n$ can be derived by maximizing
\eqref{joint-likelihood}, which is equivalent to the maximization of the $J$ distinct ``single-products"
\begin{equation}\label{single-product}
\prod_{h=1}^H BetaBin(n_{j,h,n},\alpha_{j},\beta_j)(s_{j,h,n}),
\end{equation}
or, equivalently, of
\begin{equation}\label{marginal-distr-bis}
\prod_{h=1}^H \frac{B(\alpha_j+s_{j,h,n},\beta_j+n_{j,h,n}-s_{j,h,n})}{B(\alpha_j,\beta_j)}.
\end{equation}

\begin{proof} Setting $\vz=(z_1,\ldots,z_n)^\top$ and $\vtheta=(\theta_{j,h})_{j=1,\dots, J,\,h=1,\dots,H}$, then
\begin{equation*}
\Pr(\vDelta_n,\vY_n|\vtheta)=\sum_{\vz} \Pr(\vZ_n=\vz) \times \Pr(\vDelta_n,\vY_n|\vZ_n=\vz;\vtheta).
\end{equation*}
Let $j_i$ denote the treatment assigned to patient $i$,
namely such that $\delta_{j_i,i}=1$, then the $i$th assignment could be expressed through the canonical base via
$\vdelta_{i}={\ve}_{j_i}$ and $\Pr(\vDelta_n,\vY_n|\vZ_n=\vz;\vtheta)$ can be rewritten as follows
\begin{equation*}
\begin{split}
\prod_{i=1}^n& \Pr(Y_i=y_i|\vdelta_{i}={\ve}_{j_i};Z_i= z_i;\vtheta)\times \Pr(\vdelta_{i}={\ve}_{j_i}|\vY_{i-1}, \vDelta_{i-1},Z_i= z_i, \vZ_{i-1};\vtheta) \\
=&\bigg\{\prod_{i=1}^n \theta_{j_i,z_i}^{y_i} (1-\theta_{j_i,z_i})^{1-y_i} \bigg\}\prod_{i=1}^n \pi_{j_i,z_i,i-1}\,,
\end{split}
\end{equation*}
where $\pi_{j_i,z_i,i-1}$ is the allocation function in \eqref{randfunction}, which
does not explicitly depend on $\vtheta$. Then, the likelihood of $\vtheta$ could be written as
\begin{equation*}
\begin{split}
\mathcal{L}_n(\vtheta)
&=\bigg\{ \prod_{j=1}^J\prod_{h=1}^H
\theta_{j,h}^{s_{j,h,n}}(1-\theta_{j,h})^{n_{j,h,n}-s_{j,h,n}} \bigg\}\times
\mathcal{T}_n(\vn_n,\vs_n)\\
&\propto \prod_{j=1}^J\prod_{h=1}^H Bin(n_{j,h,n},\theta_{j,h})(s_{j,h,n}),
\end{split}
\end{equation*}
namely it is proportional to a product of Binomial distributions with parameters $(n_{j,h,n},\theta_{j,h,n})$ computed in $s_{j,h,n}$ (since $\mathcal{T}_n$ does not depend on $\vtheta$). Thus, the likelihood \eqref{joint-likelihood} is obtained by integrating with respect to the $\theta_{j,h}$. 
Finally, from \eqref{joint-likelihood} and recalling that 
$$BetaBin(n_{j,h,n},\alpha_j,\beta_j)(s_{j,h,n})=\binom{n_{j,h,n}}{s_{j,h,n}}\frac{B( \alpha_j+s_{j,h,n}, \beta_j+n_{j,h,n}-s_{j,h,n}) }{ B(\alpha_j,\beta_j) },$$
for each fixed treatment $j$ the distribution of $\vN_{j,n}$ and $\vS_{j,n}$ is of the form
\begin{equation*}
\begin{split}
&\Pr\left(\vN_{j,n}=\vn_{j,n},\vS_{j,n}=\vs_{j,n};\alpha_j,\beta_j\right)=\\
&q_n(\vn_{j,n},\vs_{j,n};(\alpha_l)_{l\neq j},(\beta_l)_{l\neq j})
\prod_{h=1}^H  \frac{B(\alpha_j+s_{j,h,n},\beta_j+n_{j,h,n}-s_{j,h,n})}{B(\alpha_j,\beta_j)},
\end{split}
\end{equation*}
where $q_n$ does not depend on $(\alpha_j,\beta_j)$ and this concludes the proof.
\end{proof}
In the following proposition we will show that, even when $\widehat{\alpha}_{j,n}+\widehat{\beta}_{j,n}$ is equal to infinite at a given time-step $n$, the estimator $P_{j,h,n}$ given in~\eqref{prop-urne-con-ipotesi} is naturally well defined as the overall average treatment effect.
\begin{prop}\label{prop:alpha_beta_infinite}
Under the IUD with update mechanism \eqref{updatefunc3}, if we have $\widehat{\alpha}_{j,n}+\widehat{\beta}_{j,n}=\infty$, then $\rho_{j,h,n}=0$ and
$$P_{j,h,n}=\frac{S_{j,\centerdot,n}}{N_{j,\centerdot,n}}=\frac{\sum_{h=1}^H S_{j,h,n}}{\sum_{h=1}^H N_{j,h,n}}.$$
\end{prop}
\begin{proof}
First note that $\rho_{j,h,n}$ is a continuous decreasing function of $\widehat{\alpha}_{j,n}+\widehat{\beta}_{j,n}$ and $\widehat{\alpha}_{j,n}+\widehat{\beta}_{j,n}=\infty$ implies $\rho_{j,h,n}=0$. For the second part of the proof, note that $\widehat{\alpha}_{j,n}+\widehat{\beta}_{j,n}=\infty$ means that there does not exist a
global maximum. Then, by Lemma~\ref{lem:log_likelihood_infinite}, the supremum is approached only when $\widehat{\alpha}_{j,n}/(\widehat{\alpha}_{j,n}+\widehat{\beta}_{j,n})= S_{j,\centerdot,n}/N_{j,\centerdot,n}$. The thesis follows.
\end{proof}
The next proposition, whose proof follows from the technical Lemma~\ref{lem:log_likelihood_finite}, states that the case $\widehat{\alpha}_{j,n}+\widehat{\beta}_{j,n}=\infty$ does not occur when the variability between the estimated treatment effects is not too small compared to the number of observations.
\begin{prop}
Under the IUD with update mechanism \eqref{updatefunc3}, if
\begin{equation}\label{eq:suff_variance}
\sum_{h=1}^H N_{j,h,n}^2 \left( \widehat{\theta}_{j,h,n} -\frac{S_{j,\centerdot,n}}{N_{j,\centerdot,n}} \right)^2>
N_{j,\centerdot,n}\tfrac{S_{j,\centerdot,n}}{N_{j,\centerdot,n}}\left(1-\tfrac{S_{j,\centerdot,n}}{N_{j,\centerdot,n}}\right),
\end{equation}
then $\widehat{\alpha}_{j,n}+\widehat{\beta}_{j,n}<\infty$.
\end{prop}
Note that if one defines the weights $w_{j,h,n}^{(1)} $ and $w_{j,h,n}^{(2)} $ proportional to $N_{j,h,n}$ and $N_{j,h,n}^2$, respectively, the left-hand term of \eqref{eq:suff_variance} may be rewritten as
\[
N_{j,\centerdot,n}\left[\sum_{h=1}^H w_{j,h,n}^{(1)} N_{j,h,n} \right]
\left[\sum_{h=1}^H w_{j,h,n}^{(2)} \left(\widehat{\theta}_{j,h,n} -\frac{S_{j,\centerdot,n}}{N_{j,\centerdot,n}} \right)^2\right]
\]
and hence \eqref{eq:suff_variance} is satisfied when the product of the weighted number of observations and the weighted variability between the estimated treatment effects is at least the variance of the combined treatment:
\[
\left[\sum_{h=1}^H w_{j,h,n}^{(1)} N_{j,h,n} \right]
\left[\sum_{h=1}^H w_{j,h,n}^{(2)} \left( \widehat{\theta}_{j,h,n} -\frac{S_{j,\centerdot,n}}{N_{j,\centerdot,n}} \right)^2\right]
>
\frac{S_{j,\centerdot,n}}{N_{j,\centerdot,n}} \left(1-\frac{S_{j,\centerdot,n}}{N_{j,\centerdot,n}}\right).
\]

\section{Some technical results related to beta functions}\label{proof_beta}
In this appendix we collect some technical results useful for the IUD with update mechanism \eqref{updatefunc3}.
To this aim, we recall the Kullback–Leibler divergence between two Bernoulli random variables of parameter $p$ and $q$:
\begin{equation}\label{eq:D_KL}
D_{KL}(p,q) = p \ln \left( \tfrac{p}{q} \right) + (1-p) \ln \left( \tfrac{1-p}{1-q} \right),
\end{equation}
which is always non-negative.

By Lemma \ref{lemma-likelihood}, for each fixed treatment $j$, the MLEs
$\widehat{\alpha}_{j,n}$ and $\widehat{\beta}_{j,n}$ are obtained by the maximization of the product
\eqref{single-product}, that is
by the maximization of \eqref{marginal-distr-bis}.
The first lemma studies the asymptotic behavior of the logarithm of this function, denoted by
$l(\alpha,\beta)$ in the whole appendix.

In all the results of this appendix,
the subscript $j$ is always omitted for the sake of simplicity, and the dependence on $n$
is necessary only for Lemma~\ref{teoBound},
where a sequence of maximization problems is studied for $l_n(\alpha,\beta)$.
We recall that $\alpha$ and $\beta$ are here always used as function variables.

Finally, we point out that in the following proof
the Stirling approximation is used both for gamma and beta functions and thus
the digamma function $\psi$ appears when we derive the logarithm of the beta function.

\begin{lem}\label{lem:log_likelihood_infinite}
Set
\begin{equation}\label{eq:LogL}
l(\alpha,\beta)=\sum_{h=1}^H \big[
\ln(B(\alpha+s_h,\beta+n_h-s_h))-\ln(B(\alpha,\beta)) \big],
\end{equation}
with $s_\centerdot = \sum_h s_h \geq 1$ and
$n_\centerdot- s_\centerdot = \sum_h (n_h-s_h) \geq 1$, then
$$
\lim_{\alpha+\beta\to \infty} l(\alpha,\beta)+ {n_\centerdot}\left(\mathcal{H}\left(\tfrac{s_\centerdot}{n_\centerdot}\right)+D_{KL}\left(
\tfrac{s_\centerdot}{n_\centerdot}, \tfrac{\alpha}{\alpha+\beta}\right)\right)=0,
$$
where $\mathcal{H}(z)$ is the Entropy of $z$, i.e. $\mathcal{H}(z)=z\ln(z)+(1-z)\ln(1-z)$
and \emph{$D_{KL}$} stands for the Kullback–Leibler divergence defined in \eqref{eq:D_KL}.
In particular,
\begin{enumerate}
\item if there exists $\alpha,\beta>0$ such that $l(\alpha,\beta)\geq -{n_\centerdot}\mathcal{H}\left(\tfrac{s_\centerdot}{n_\centerdot}\right)$, then a global maximum of $l(\alpha,\beta)$
is attained for $\alpha+\beta< \infty$;
\item if, for any $\alpha,\beta>0$,
$l(\alpha,\beta)<-{n_\centerdot}\mathcal{H}\left(\tfrac{s_\centerdot}{n_\centerdot}\right)$,
then
$$
\sup_{\alpha,\beta>0}l(\alpha,\beta) =
-{n_\centerdot}\mathcal{H}\left(\tfrac{s_\centerdot}{n_\centerdot}\right),
$$
and it is approached only for $\alpha+\beta\to \infty$ and $\tfrac{\alpha}{\alpha+\beta}\to\tfrac{s_\centerdot}{n_\centerdot}$.
\end{enumerate}
\end{lem}
\begin{proof}
First note that $l (\alpha,\beta) $ is continuous and such that
$$\lim_{\alpha\to 0} l (\alpha,\beta)  = \lim_{\beta\to 0} l (\alpha,\beta) = - \infty.$$
Then either $l (\alpha,\beta)$ has a global maximum or it attains its supremum for $\alpha+\beta \to \infty$.
By the Stirling approximation of the beta function,
\begin{align*}
l (\alpha,\beta)
& =
\sum_{h=1}^H \Big[
\alpha \ln \left( \tfrac{\alpha + s_h}{\alpha}\tfrac{\alpha+\beta}{\alpha + \beta + s_h + f_h} \right)
+
\beta \ln \left( \tfrac{\beta + f_h}{\beta}\tfrac{\alpha+\beta}{\alpha + \beta + s_h + f_h} \right)
\\
& \qquad +
s_h \ln \left( \tfrac{\alpha + s_h}{\alpha + \beta + s_h + f_h} \right)
+
f_h \ln \left( \tfrac{\beta + f_h}{\alpha + \beta + s_h + f_h} \right)
\\
& \qquad -
\tfrac{1}{2} \ln \left( \tfrac{\alpha + s_h}{\alpha}\tfrac{\beta + f_h}{\beta}\tfrac{\alpha+\beta}{\alpha + \beta + s_h + f_h} \right)
\Big] + o(1)
\\
& =
\sum_{h=1}^H \Big[
s_h - \tfrac{\alpha}{\alpha+\beta} (s_h + f_h)
+
f_h - \tfrac{\beta}{\alpha+\beta} (s_h + f_h)
\\
& \qquad +
s_h \ln \left( \tfrac{\alpha}{\alpha + \beta}  \right)
+
f_h \ln \left( \tfrac{\beta}{\alpha + \beta} \right)
\Big] + o(1)
\\
& =
\sum_{h=1}^H \Big[
s_h \ln \left( \tfrac{\alpha}{\alpha + \beta}  \right)
+
f_h \ln \left( \tfrac{\beta}{\alpha + \beta} \right)
\Big] + o(1)
\\
& =
-
 {n_\centerdot}\left(\mathcal{H}\left(\tfrac{s_\centerdot}{n_\centerdot}\right)+D_{KL}\left(
\tfrac{s_\centerdot}{n_\centerdot}, \tfrac{\alpha}{\alpha+\beta}\right)\right) + o(1).
\end{align*}
Then,
\[
\limsup_{\alpha+\beta\to\infty} l(\alpha,\beta) \leq
-
 {n_\centerdot}\mathcal{H}\left(\tfrac{s_\centerdot}{n_\centerdot}\right).
\]
In addition, \emph{$D_{KL}(p,q)$} increases when $q$ steps out from $p$, then
\begin{multline*}
\liminf_{\alpha+\beta\to\infty} l(\alpha,\beta)
\leq  - {n_\centerdot}
\Bigg[\mathcal{H}\left(\tfrac{s_\centerdot}{n_\centerdot}\right)
\\
+
\max
\left\{
 D_{KL}\left(
\tfrac{s_\centerdot}{n_\centerdot}, \liminf_n \tfrac{\alpha_n}{\alpha_n+\beta_n}\right)
,
D_{KL}\left(
\tfrac{s_\centerdot}{n_\centerdot}, \limsup_n \tfrac{\alpha_n}{\alpha_n+\beta_n}\right)
\right\}
\Bigg].
\end{multline*}
The third relation, that implies the thesis, is the following
\[\alpha+\beta \to \infty, \tfrac{\alpha}{\alpha+\beta}\to\tfrac{s_\centerdot}{n_\centerdot}
\qquad
\Longrightarrow
\qquad
l(\alpha,\beta)
\longrightarrow
 {n_\centerdot}\mathcal{H}\left(\tfrac{s_\centerdot}{n_\centerdot}\right). \qedhere
\]
\end{proof}
The next result provides a sufficient condition for the existence of a global maximum of $l(\alpha,\beta)$.
\begin{lem}\label{lem:log_likelihood_finite}
If
\begin{equation}\label{eq:suff_variance2}
\sum_{h=1}^H
n_h^2 \left(
\tfrac{s_h}{n_h}
-
\tfrac{s_\centerdot}{n_\centerdot}
\right)^2
>
n_\centerdot
\tfrac{s_\centerdot}{n_\centerdot}
\left( 1 -\tfrac{s_\centerdot}{n_\centerdot}
\right),
\end{equation}
then a global maximum of $l(\alpha,\beta)$ in (\ref{eq:LogL}) is attained for $\alpha+\beta< \infty$.
\end{lem}
\begin{proof}
By Lemma~\ref{lem:log_likelihood_infinite}, it is sufficient to prove that
there exists $(\alpha,\beta)$ such that \( l(\alpha,\beta)\geq -n_\centerdot \mathcal{H}(\tfrac{s_\centerdot}{n_\centerdot})\).
We set $\alpha = \tfrac{s_\centerdot}{n_\centerdot} x$ and $\beta = (1 - \tfrac{s_\centerdot}{n_\centerdot}) x$ and we show the
assertion above for $x$ sufficiently large. Define $g(x) = l\left (\tfrac{s_\centerdot}{n_\centerdot} x,\left(1-\tfrac{s_\centerdot}{n_\centerdot}\right) x \right) $.
By Lemma~\ref{lem:log_likelihood_infinite},
\(
\lim_{x\to\infty} g(x) = -n_\centerdot \mathcal{H}(\tfrac{s_\centerdot}{n_\centerdot})
\),
and therefore it is sufficient to prove that $g'(x)<0$ for $x$ sufficiently large.
Recall that
\[
\frac{\partial \ln ( B(\alpha,\beta) ) }{\partial \alpha} = \psi(\alpha)  - \psi(\alpha+\beta)  ,
\qquad
\frac{\partial \ln ( B(\alpha,\beta) ) }{\partial \alpha} = \psi(\beta)  - \psi(\alpha+\beta)  ,
\]
where $\psi$ is the digamma function, for which
$\psi(x + m) - \psi(x) = \sum_{j=0}^{m-1} \frac{1}{x+j}$
for any $x>0$ and $m\in\mathbb{N}$. Denote $f_h= n_h-s_h$ and $f_\centerdot = \sum_lf_l$, then
\begin{align*}
g'(x)
& =
\sum_{h=1}^H \tfrac{s_\centerdot}{n_\centerdot} \left( \psi \left(\tfrac{s_\centerdot}{n_\centerdot}x + s_h \right) - \psi(x+n_h) - \psi \left(\tfrac{s_\centerdot}{n_\centerdot}x\right) + \psi (x) \right)
\\
& \qquad +
\left(1-\tfrac{s_\centerdot}{n_\centerdot}\right) \left( \psi \left( \left(1-\tfrac{s_\centerdot}{n_\centerdot} \right)x + f_h\right) - \psi(x+n_h) - \psi \left(\left(1-\tfrac{s_\centerdot}{n_\centerdot}\right)x\right) + \psi (x) \right)
\\
& =
\sum_{h=1}^H
\left[
\tfrac{s_\centerdot}{n_\centerdot} \sum_{j=0}^{s_h-1}
\frac{1}{\tfrac{s_\centerdot}{n_\centerdot}x + j}
+
\left(1-\tfrac{s_\centerdot}{n_\centerdot}\right)
\sum_{j=0}^{f_h-1}
\frac{1}{\left(1-\tfrac{s_\centerdot}{n_\centerdot}\right)x + j}
-
\sum_{j=0}^{n_h-1} \frac{1}{x + j}
\right]
\\
& =
\sum_{h=1}^H
\frac{1}{x}
\left[
\sum_{j=0}^{s_h-1}
\frac{1}{1 + \frac{j}{\tfrac{s_\centerdot}{n_\centerdot}x}}
+
\sum_{j=0}^{f_h-1}
\frac{1}{1 + \frac{j}{\left(1-\tfrac{s_\centerdot}{n_\centerdot}\right)x}}
-
\sum_{j=0}^{n_h-1} \frac{1}{1 + \frac{j}{x}}
\right]
\\
& =
\sum_{h=1}^H
\frac{1}{2x^2}
\left[
-
\frac{{n_\centerdot}}{{s_\centerdot}} (s_h-1)s_h
-
\frac{{n_\centerdot}}{{f_\centerdot}} (f_h-1)f_h
+
(n_h-1)n_h
\right] + o \Big(\frac{1}{x^2}\Big)
\\
& =
\frac{1}{2x^2}
\frac{{n_\centerdot}}{{s_\centerdot}}
\frac{{n_\centerdot}}{{f_\centerdot}}
\Big[
-
\frac{{f_\centerdot}}{{n_\centerdot}}
\left({\textstyle\sum_h s_h^2} - s_\centerdot \right)
-
\frac{{s_\centerdot}}{{n_\centerdot}}
({\textstyle\sum_h f_h^2} - f_\centerdot)
\\
& \qquad
+
\frac{{f_\centerdot}}{{n_\centerdot}}
\frac{{s_\centerdot}}{{n_\centerdot}}
\left(
{\textstyle\sum_h \left(s_h+f_h\right)^2} - {n_{\centerdot}}
\right)
\Big] + o \left(\frac{1}{x^2}\right)
\\
& =
-\frac{1}{2x^2}
\frac{{n_\centerdot}}{{s_\centerdot}}
\frac{{n_\centerdot}}{{f_\centerdot}}
\Big[
\left(\tfrac{{f_\centerdot}}{{n_\centerdot}} \right)^2
{\textstyle\sum_h s_h^2}
+
\left(\tfrac{{s_\centerdot}}{{n_\centerdot}} \right)^2
{\textstyle\sum_h f_h^2}
- 2 \tfrac{{s_\centerdot}}{{n_\centerdot}} \tfrac{{f_\centerdot}}{{n_\centerdot}}
{\textstyle\sum_h s_hf_h}
-
\frac{{f_\centerdot}}{{n_\centerdot}}
\frac{{s_\centerdot}}{{n_\centerdot}} {n_{\centerdot}}
\Big] \\
&+ o \left(\frac{1}{x^2}\right)
\\
& =
-\frac{1}{2x^2}
\bigg\{
\frac{{n_\centerdot}}{{s_\centerdot}}
\frac{{n_\centerdot}}{{f_\centerdot}}
\left[
\sum_{h=1}^H
n_h^2 \left(
\tfrac{s_h}{n_h}
-
\tfrac{{s_\centerdot}}{{n_\centerdot}}
\right)^2
-
\frac{{f_\centerdot}}{{n_\centerdot}}
\frac{{s_\centerdot}}{{n_\centerdot}} {n_{\centerdot}}
\right] + o (1)
\bigg\}.
\end{align*}
By \eqref{eq:suff_variance2}, the term inside the square brackets is strictly positive, which concludes the proof.
\end{proof}

By \eqref{marginal-distr},
the next lemma can be applied, for each fixed treatment $j$, to
$[\vN_{j,n}, \vS_{j,n}]$ in order to obtain the corresponding result for
the MLEs $\widehat{\alpha}_{j,n}$ and $\widehat{\beta}_{j,n}$.
In this case, the general result studies exactly the asymptotic finiteness of MLEs.
The dependence on the behavior of the two leading random vectors
$(S_{(1),n},\, N_{(1),n})^\top$ and $(S_{(2),n},N_{(2),n})^\top$ is given.
For completeness, in their definition, we declare that $\arg\max_h $ is the minimum $\arg\max$ if more than one $\max$ is present, even if this particular choice does not modify the result.

\begin{lem}
\label{teoBound}
For any $n\geq 1$, let $\vN_n = (N_{1,n},\dots, N_{H,n})$ and $\vS_n = (S_{1,n},\dots, S_{H,n})$ be integer-valued random vectors
with $N_{h,n} \geq 1$, $S_{h,n} \leq N_{h,n}$ and joint distribution proportional to
\[\prod_{h=1}^H \frac{B(\alpha+s_{h},\beta+n_{h}-s_{h})}{B(\alpha,\beta)},\]
with a proportionality factor that does not depend on the parameters $(\alpha,\beta)$.
Define, for any $n$, the random vector
$$
\begin{aligned}
(S_{(1),n},&\, N_{(1),n}, S_{(2),n},N_{(2),n})^\top =\\
&\left\{(S_{h_1,n},N_{h_1,n},S_{h_2,n},N_{h_2,n})^\top:\
h_1=\arg\max_h N_{h,n},\
h_2=\arg\max_{h\neq h_1} N_{h,n}\right\}
\end{aligned}$$
and let $\widehat{\alpha}_n$ and $\widehat{\beta}_n$ be the MLEs of $\alpha$ and $\beta$ based on $\vS_n$ and $\vN_n$. Thus,
\begin{equation}
\label{teoBound_i}
\begin{scriptsize}
\Pr \left(
\limsup_{n\rightarrow + \infty}(\widehat{\alpha}_{n}+\widehat{\beta}_{n}) = \infty
,
\ln(N_{(1),n}) = o(N_{(2),n})
,
\left(
\tfrac{S_{(1),n}}{N_{(1),n}} - \tfrac{S_{(2),n}}{N_{(2),n}}
\right)
\nrightarrow 0
\right) = 0
\end{scriptsize}
\end{equation}
and, for any random increasing sequence $(n_k)_k$,
\begin{equation}
\begin{scriptsize}
\label{teoBound_ii}
\Pr \left(
\min
\left(
N_{(1),n_k},\widehat{\alpha}_{n_k}+\widehat{\beta}_{n_k}
\right)
\to \infty
,
\left(
\tfrac{\widehat{\alpha}_{n_k}}{\widehat{\alpha}_{n_k}+\widehat{\beta}_{n_k}}-
\tfrac{S_{(1),n_k}}{N_{(1),n_k}}
\right)
\nrightarrow 0
\right) = 0 .
\end{scriptsize}
\end{equation}
\end{lem}
\begin{proof}
In the sequel we denote by $F_{h,n}=N_{h,n}-S_{h,n}$ and by $l_n(\alpha,\beta)$ the log-likelihood function.
For any positive sequences $(\alpha_n)_{n}$ and $(\beta_n)_{n}$, we can consider the difference
$$
l_{n}(\alpha_n,\beta_n)-l_{n}(1,1)=
\sum_{h=1}^H\ln\left(\frac{B(\alpha_n+S_{h,n},\beta_n+F_{h,n})}{B(\alpha_n,\beta_n)B(1+S_{h,n},1+F_{h,n})}\right).
$$
Naturally, if we take $\alpha_n=\widehat{\alpha}_{n}$ and $\beta_n=\widehat{\beta}_{n}$ we would have by definition that
\begin{equation}\label{eq:result_liminf}
\liminf_{n\rightarrow\infty} \left(l_{n}(\alpha_n,\beta_n)-l_{n}(1,1)\right)\geq 0.
\end{equation}
Regarding \eqref{teoBound_i}, we will prove that,
\[
\Pr\left(
\limsup_{n\rightarrow\infty}(\widehat{\alpha}_{n}+\widehat{\beta}_{n}) = \infty
,
\ln(N_{(1),n}) = o(N_{(2),n})
,
\tfrac{S_{(1),n}}{N_{(1),n}} - \tfrac{S_{(2),n}}{N_{(2),n}}
\nrightarrow 0
,
\eqref{eq:result_liminf}
\right) = 0.
\]
To this end, since
$\ln(B(1+S_{h,n},1+F_{h,n}))=\ln(B(S_{h,n},F_{h,n}))+O(1)$, from the Stirling approximation of the gamma function,
\begin{equation*}
\begin{split}
l_{n}(\alpha_n,\beta_n)-l_{n}(1,1)&=
\sum_{h=1}^H\ln\left(\frac{B(\alpha_n+S_{h,n},\beta_n+F_{h,n})}{B(\alpha_n,\beta_n)B(S_{h,n},F_{h,n})}\right) + O(1)
\\
&= \sum_{h=1}^H A_{h,n} - \frac{1}{2}\sum_{h=1}^H B_{h,n} +O(1)\,,
\end{split}
\end{equation*}
with
\begin{align*}
A_{h,n}&=(\alpha_n+S_{h,n})\ln(\alpha_n+S_{h,n})+(\beta_n+F_{h,n})\ln(\beta_n+F_{h,n})\\
& \qquad -(\alpha_n+S_{h,n}+\beta_n+F_{h,n})\ln(\alpha_n+S_{h,n}+\beta_n+F_{h,n})\\
& \qquad -\left[\alpha_n\ln(\alpha_n)+\beta_n\ln(\beta_n)-(\alpha_n+\beta_n)\ln(\alpha_n+\beta_n)\right]\\
& \qquad -\left[S_{h,n}\ln(S_{h,n})+F_{h,n}\ln(F_{h,n})-(N_{h,n})\ln(N_{h,n})\right]\\
&=\alpha_n\ln\Bigg(\frac{\frac{\alpha_n+S_{h,n}}{\alpha_n+\beta_n+N_{h,n}}}{\frac{\alpha_n}{\alpha_n+\beta_n}}\Bigg)+
\beta_n\ln\Bigg(\frac{\frac{\beta_n+F_{h,n}}{\alpha_n+\beta_n+N_{h,n}}}{\frac{\beta_n}{\alpha_n+\beta_n}}\Bigg)\\
& \qquad +S_{h,n}\ln\Bigg(\frac{\frac{\alpha_n+S_{h,n}}{\alpha_n+\beta_n+N_{h,n}}}{\frac{S_{h,n}}{N_{h,n}}}\Bigg)+
F_{h,n}\ln\Bigg(\frac{\frac{\beta_n+F_{h,n}}{\alpha_n+\beta_n+N_{h,n}}}{\frac{F_{h,n}}{N_{h,n}}}\Bigg)\\
&=-(\alpha_n+\beta_n)D_{KL}\Big(
\tfrac{\alpha_n}{\alpha_n+\beta_n},
\tfrac{\alpha_n+S_{h,n}}{\alpha_n+\beta_n+N_{h,n}}
\Big)
-N_{h,n} D_{KL}\Big(
\tfrac{S_{h,n}}{N_{h,n}},
\tfrac{\alpha_n+S_{h,n}}{\alpha_n+\beta_n+N_{h,n}}
\Big)
\end{align*}
and
\begin{align*}
B_{h,n}&=\ln(\alpha_n+S_{h,n})+\ln(\beta_n+F_{h,n})-\ln(\alpha_n+S_{h,n}+\beta_n+F_{h,n})\\
& \qquad -\left[\ln(\alpha_n)+\ln(\beta_n)-\ln(\alpha_n+\beta_n)\right]\\
& \qquad -\left[\ln(S_{h,n})+\ln(F_{h,n})-\ln(N_{h,n})\right]\\
&=\ln\big(\tfrac{1}{\alpha_n}+\tfrac{1}{S_{h,n}}\big)+\ln\big(\tfrac{1}{\beta_n}+\tfrac{1}{F_{h,n}}\big)-
\ln\big(\tfrac{1}{\alpha_n+\beta_n}+\tfrac{1}{N_{h,n}}\big)\\
&=-\ln\big(\min\{\alpha_n,S_{h,n}\}\big)
-\ln\big(\min\{\beta_n,F_{h,n}\}\big)
+\ln\big(\min\{\alpha_n+\beta_n,N_{h,n}\}\big)\\
&\qquad +\ln\big(1+\tfrac{\min\{\alpha_n,S_{h,n}\}}{\max\{\alpha_n,S_{h,n}\}}\big)+\ln\big(1+\tfrac{\min\{\beta_n,F_{h,n}\}}{\max\{\beta_n,F_{h,n}\}}\big)\\
&\qquad \qquad-
\ln\big(1+\tfrac{\min\{\alpha_n+\beta_n,N_{h,n}\}}{\max\{\alpha_n+\beta_n,N_{h,n}\}}\big)\\
&=-\ln\big(\min\{\alpha_n,S_{h,n}\}\big)
-\ln\big(\min\{\beta_n,F_{h,n}\}\big)
+\ln\big(\min\{\alpha_n+\beta_n,N_{h,n}\}\big)\\
&+O(1)\\
&=-\ln\left(\min (\min\{\alpha_n,S_{h,n}\},\min\{\beta_n,F_{h,n}\} )\right)\\
&\qquad -\ln\big(\tfrac{\max (\min\{\alpha_n,S_{h,n}\},\min\{\beta_n,F_{h,n}\} )}{\min\{\alpha_n+\beta_n,N_{h,n}\}}\big)+O(1)\\
&\geq-\ln\left(\min (\min\{\alpha_n,S_{h,n}\},\min\{\beta_n,F_{h,n}\} )\right) +O(1)\\
&\geq -\ln\left(\min\{\alpha_n+\beta_n,N_{h,n}\}\right) +O(1).
\end{align*}
Hence, setting $M_{h,n}=\min\{\alpha_n+\beta_n,N_{h,n}\}$,
we obtain that
\begin{multline*}
l_{n}(\alpha_n,\beta_n)-l_{n}(1,1) \leq
\sum_{h=1}^H O\left(\ln\left(M_{h,n}\right)\right)\\
-\sum_{h=1}^H M_{h,n}
\Big[ D_{KL}\left(
\tfrac{\alpha_n}{\alpha_n+\beta_n},
\tfrac{\alpha_n+S_{h,n}}{\alpha_n+\beta_n+N_{h,n}}
\right)
+ D_{KL}\left(
\tfrac{S_{h,n}}{N_{h,n}},
\tfrac{\alpha_n+S_{h,n}}{\alpha_n+\beta_n+N_{h,n}}
\right)
\Big].
\end{multline*}
Now, denoting by $M_{(1),n}$ and $M_{(2),n}$ the first and the second highest $M_{h,n}$ over $h=1,\dots, H$, respectively, we obtain
that the quantity $\frac{l_{n}(\alpha_n,\beta_n)-l_{n}(1,1)}{M_{(2),n}}$ is bounded above by
\begin{multline*}
\begin{aligned}
&
-\sum_{h=1}^H\frac{M_{h,n}}{M_{(2),n}}
\Big[ D_{KL}\left(
\tfrac{\alpha_n}{\alpha_n+\beta_n},
\tfrac{\alpha_n+S_{h,n}}{\alpha_n+\beta_n+N_{h,n}}
\right)
+ D_{KL}\left(
\tfrac{S_{h,n}}{N_{h,n}},
\tfrac{\alpha_n+S_{h,n}}{\alpha_n+\beta_n+N_{h,n}}
\right)
\Big]
\\
& \qquad +
\sum_{h=1}^H O\left(\tfrac{\ln\left(M_{h,n}\right)}{M_{(2),n}}\right)\\
&\leq
-\frac{M_{(1),n}}{M_{(2),n}}
\Big[ D_{KL}\left(
\tfrac{\alpha_n}{\alpha_n+\beta_n},
\tfrac{\alpha_n+S_{(1),n}}{\alpha_n+\beta_n+N_{(1),n}}
\right)
+ D_{KL}\left(
\tfrac{S_{(1),n}}{N_{(1),n}},
\tfrac{\alpha_n+S_{(1),n}}{\alpha_n+\beta_n+N_{(1),n}}
\right)
\Big]
\\
& \qquad -\Big[ D_{KL}\left(
\tfrac{\alpha_n}{\alpha_n+\beta_n},
\tfrac{\alpha_n+S_{(2),n}}{\alpha_n+\beta_n+N_{(2),n}}
\right)
+ D_{KL}\left(
\tfrac{S_{(2),n}}{N_{(2),n}},
\tfrac{\alpha_n+S_{(2),n}}{\alpha_n+\beta_n+N_{(2),n}}
\right)
\Big]
\\
& \qquad + o(1) + O\left(\tfrac{\ln\left(M_{(1),n}\right)}{M_{(2),n}}\right).
\end{aligned}
\end{multline*}
On
\[\left\{\limsup_{n\rightarrow\infty}(\widehat{\alpha}_{n}+\widehat{\beta}_{n}) = \infty \right\} \cap \left\{ \ln(N_{(1),n}) = o(N_{(2),n}) \right\}\]
we get $\lim_n M_{(2),n}=\infty$ and $O\left(\frac{\ln (M_{(1),n})}{M_{(2),n}}\right)= o(1)$, hence all the $\limsup D_{KL}$ must be zero to verify \eqref{eq:result_liminf}.
However, this is possible only if
$$
\tfrac{S_{(1),n}}{N_{(1),n}}
/
\tfrac{\alpha_{n}}{\alpha_{n}+\beta_{n}}
\longrightarrow 1
\qquad\mbox{and}\qquad
\tfrac{S_{(2),n}}{N_{(2),n}}
/
\tfrac{\alpha_{n}}{\alpha_{n}+\beta_{n}}
\longrightarrow 1,
$$
which  imply
$
\frac{S_{(1),n}}{N_{(1),n}}
/
\frac{S_{(2),n}}{N_{(2),n}}
\longrightarrow 1,
$
that concludes the proof of \eqref{teoBound_i}.\\

Regarding \eqref{teoBound_ii}, we will prove that,
\[
\Pr\left(
\min
\big\{
N_{(1),n_k},\widehat{\alpha}_{n_k}+\widehat{\beta}_{n_k}
\big\}
\to \infty
,
\left(
\tfrac{\widehat{\alpha}_{n_k}}{\widehat{\alpha}_{n_k}+\widehat{\beta}_{n_k}}-
\tfrac{S_{(1),n_k}}{N_{(1),n_k}}
\right)
\nrightarrow 0
,
\eqref{eq:result_liminf}
\right) = 0.
\]
To this end, we can follow the same arguments as before, dividing the difference
$l_{n}(\alpha_n,\beta_n)-l_{n}(1,1)$ by $M_{(1),n}$ instead of by $M_{(2),n}$.
Indeed, on the event
\[\big\{
\min
\big\{
N_{(1),n_k},\widehat{\alpha}_{n_k}+\widehat{\beta}_{n_k}
\big\}
\to \infty
\big\}
\]
we get $M_{(1),n}\rightarrow \infty$ and $\ln\left(M_{h,n}\right) / M_{(1),n} \rightarrow0$ $\forall h$, hence only the first two $\limsup D_{KL}$ must be zero to verify \eqref{eq:result_liminf}.
However, this is possible only if
$$
\tfrac{S_{(1),n}}{S_{(1),n}+F_{(1),n}}
/
\tfrac{\alpha_{n}}{\alpha_{n}+\beta_{n}}
\nrightarrow 1,
$$
that concludes the proof of \eqref{teoBound_ii}.
\end{proof}

\subsection{By-product result: MLEs for Beta-Binomial distributions with different number of trials}
We here apply the above Lemma~\ref{teoBound} to a general sequence of vectors of independent Beta-Binomial random variables with different number of trials. To the best of our knowledge, the analysis of the MLEs in this case is missing in the literature.\\
Let $\boldsymbol{\theta}=(\theta_{1},\dots,\theta_H$) be a vector collecting $H$ i.i.d. realizations of a Beta random variable with parameters $\alpha,\beta>0$.
For any  time-step $n$, let $S_{1,n},\dots, S_{H,n}$ be a sample of $H$ independent Binomial random variables with parameters $\theta_1,\ldots,\theta_H$ and $n_{1,n},\dots, n_{H,n}\geq 1$, respectively.
Therefore, for any $n$, $(S_{1,n},\dots, S_{H,n})$ represents a sample of independent Beta-Binomial random variables with different number of trials.
As in Lemma~\ref{teoBound} we define, for any $n$, the quantities
$$
\begin{aligned}
(S_{(1),n},&\, n_{(1),n}, S_{(2),n},n_{(2),n}) =\\
&\Big\{(S_{h_1,n},n_{h_1,n},S_{h_2,n},n_{h_2,n}):\
h_1=\arg\max_h n_{h,n},\
h_2=\arg\max_{h\neq h_1} n_{h,n}\Big\}.
\end{aligned}$$
 The setting above typically arises when, at each time-step, a different number of trials may be set for each
$h$ and we observe the number of successes of those trials.
The following result characterizes the boundedness property of the MLEs of $\alpha$ and $\beta$.
\begin{prop}
For any $n\geq 1$, let $S_{1,n},\dots, S_{H,n}$ be defined as above and
let $\widehat{\alpha}_{n}$ and $\widehat{\beta}_{n}$ be the MLEs of
$\alpha$ and $\beta$, respectively.
\begin{itemize}
\item[(i)] If $\ln(n_{(1),n}) = o(n_{(2),n})$,
then $$\limsup_{n\rightarrow\infty}(\widehat{\alpha}_{n}+\widehat{\beta}_{n})<\infty;$$

\item[(ii)] If $\Pr(\limsup_{n\rightarrow\infty}(\widehat{\alpha}_{n}+\widehat{\beta}_{n})=\infty)>0$ and $n_{(1),n}\to \infty$,
along any subsequence such that
$(\widehat{\alpha}_{n_k}+\widehat{\beta}_{n_k})\to \infty$ it follows that
 $$\frac{\widehat{\alpha}_{n_k}}{\widehat{\alpha}_{n_k}+\widehat{\beta}_{n_k}}-
\frac{S_{(1),n_k}}{n_{(1),n_k}}\rightarrow 0.$$
\end{itemize}
\end{prop}
\begin{proof} Let us denote by $\Theta_h$ the random variable corresponding to the $h$-th sample from the above mentioned Beta random variable with parameters $\alpha,\,\beta$. Therefore, $\Theta_1,\dots, \Theta_H$ are i.i.d. random variables.
Since the compound distribution of the binomial distribution, where the parameter of the success is drawn from a beta distribution, is a beta-binomial one, then for any $n$ the joint probability distribution of $\vN_n$ and $\vS_n$ is
\begin{equation*}
\begin{split}
\prod_{h=1}^H &{n_h\choose s_h}\frac{B(s_h+\alpha,n_h-s_h+\beta)}{B(\alpha,\beta)}
\mathbbm{1}_{[n_h=n_{h,n}]}=\\
&\left(\prod_{h=1}^H
{n_h\choose s_h}\mathbbm{1}_{[n_h=n_{h,n}]}\right)
\prod_{h=1}^H
\frac{B(s_h+\alpha,n_h-s_h+\beta)} {B(\alpha,\beta)}\,.
\end{split}
\end{equation*}
Then we may apply Lemma~\ref{teoBound} to prove (i) and (ii).
\\
\indent {\em  Proof of (i)}.  Note that $\ln(n_{(1),n}) = o(n_{(2),n})$ is a deterministic event.
When it is satisfied, then $n_{(2),n}\to \infty$ and $n_{(1),n}\to \infty$. Since for any $h$ such that $n_{h,n}\to \infty$
\[\Pr \left( \textstyle\limsup_n \left| \tfrac{S_{h,n}}{n_{h,n}} - \Theta_{h} \right| = 0 \right) = 1,\]
then, by the triangular inequality, for any $k$
$$\Pr \left( \liminf_n \left| \tfrac{ S_{(1),n} }{ n_{(1),n} } - \tfrac{ S_{(2),n} }{ n_{(2),n} }  \right| \geq \frac{1}{k} \right) \geq
\Pr \left( \inf_{h_1\neq h_2} | \Theta_{h_1} - \Theta_{h_2} | > \frac{1}{k} \right).$$
Since $\Pr( \Theta_{h_1} = \Theta_{h_2}) = 0$ for all $h_1\neq h_2$, then $\lim_k \Pr \left( \inf_{h_1\neq h_2} | \Theta_{h_1} - \Theta_{h_2} | > \frac{1}{k} \right) = 1$
and hence
\[\Pr \left( \liminf_n \left| \tfrac{S_{(1),n}}{n_{(1),n}} - \tfrac{S_{(2),n}}{n_{(2),n}}  \right| > 0 \right) = \lim_{k}
\Pr \left( \liminf_n \left| \tfrac{S_{(1),n}}{n_{(1),n}} - \tfrac{S_{(2),n}}{n_{(2),n}}  \right| \geq \frac{1}{k} \right) = 1.\]
Thus, equation \eqref{teoBound_ii} in Lemma~\ref{teoBound} reads $\Pr\big(\limsup_{n\rightarrow\infty}(\widehat{\alpha}_{n}+\widehat{\beta}_{n})=\infty\big) = 0$. Statement (ii) could be simply proved by applying equation \eqref{teoBound_ii} in Lemma~\ref{teoBound}.
\end{proof}

\section{Equal treatments with model-based approach}\label{sec:equal_treatments_model_based}

In this section, we present a generalization of the model-based approach in order to allow,
for any given treatment $j$,
some of the success-probabilities of different strata $\theta_{j,1},\dots,\theta_{j, H}$ to have the same value.
Indeed, in the formulation of the model-based mechanism presented in Section~\ref{modelbased}
the success-probabilities of two strata have probability zero to be equal as they are independent realizations from a Beta distribution.
This additional flexibility can be achieved by introducing in the model a preliminary step in which,
for any given treatment $j$, the original $H$ strata are clustered in $H_c\leq H$ subsets
that form a partition $\mathcal{P}_j$ of the set $\{1,\dots, H\}$. The strata that belong
to the same subset will share the same probability of success and we will assume that
these subset-specific success-probabilities  $\theta_{j,1},\dots,\theta_{j,H_c}$ will be again i.i.d. with
 Beta distribution ${\mathcal B}(\alpha_j,\beta_j)$. Therefore, in order to appropriately modify the urn proportion
$P_{j,h,n}$ in \eqref{prop-urne-con-ipotesi} to implement this new variant, we need to define how to cluster different strata in the same subset at any given time-step $n$, i.e. how to define the partition $\mathcal{P}_{j,n}$ at any time $n$. Ideally, we want to aggregate strata
when their current estimates $\widehat{\theta}_{j,h,n}$
are close enough to provide evidence that the corresponding success-probabilities $\theta_{j,h}$ are equal.
Then, analogously to the mechanism based on treatment-similarity presented in Section~\ref{treatment-similarity}, we decide to adopt a decreasing sequence of positive real numbers converging to zero, $(c_n)_{n\geq1}$ with $c_n\rightarrow 0$, which here represents the maximum distance between any pair of consecutive estimated success probabilities of a given treatment within the same subset. In other words, if we sort the strata according to their estimated success probabilities (from the lowest to the highest) $\widehat{\theta}_{j,(1),n},\dots,\widehat{\theta}_{j,(H),n}$, then the strata in position $(h-1)$ and $(h)$ will be in the same subset of the partition $\mathcal{P}_{j,n}$ if and only if $|\widehat{\theta}_{j,(h-1),n}-\widehat{\theta}_{j,(h),n}|\leq c_n$.
Then, denoting by $I_{j,1,n},\dots,I_{j,H_{c,n},n}$ the $H_{c,n}$ subsets of $\mathcal{P}_{j,n}$
and by $I_{j,n}(h)$ the only subset containing $h$ (i.e. $I_{j,n}(h)=I_{j,h',n}$ iff $h\in I_{j,h',n}$),
the urn proportion $P_{j,h,n}$ in \eqref{prop-urne-con-ipotesi} can be replaced by
$$
P_{j,h,n}=
\frac{\widehat{\alpha}_{j,n}
+\sum_{k\in I_{j,n}(h)} S_{j,k,n}}
{\widehat{\alpha}_{j,n} + \widehat{\beta}_{j,n} + \sum_{k\in I_{j,n}(h)} N_{j,k,n}},
$$
where $\widehat{\alpha}_{j,n}$ and $\widehat{\beta}_{j,n}$ indicate, respectively, the values of $\alpha_j$ and $\beta_j$ that maximize the likelihood proportional to
\begin{equation*}
\prod_{h'=1}^{H_{c,n}}
\frac{B\big(\alpha_j+\sum_{k\in I_{j,h',n}}s_{j,k, n},\beta_j+\sum_{k\in I_{j,h',n}}(n_{j,h, n}-s_{j,h, n})\big)}{B(\alpha_j,\beta_j)}
\,,
\end{equation*}
so that $\widehat{\alpha}_{j,n}$ and $\widehat{\beta}_{j,n}$ will depend only on the aggregated quantities.
By applying Lemma~\ref{teoBound} on this new version of the likelihood, we can get an analogous result of~\eqref{teoBound_ii}.
This result allows us to apply Lemma~\ref{teoBound_model_based}, which is used in Theorem~\ref{teoCons} to prove that all the $N_{j,h,n}$ diverge to infinity (linearly with $n$), and the almost sure consistency of both the estimators $\widehat{\theta}_{j,h,n}$ and $P_{j,h,n}$. Indeed, in analogy with the treatment-similarity Section~\ref{treatment-similarity},
since $c_n$ tends to zero this procedure asymptotically does not aggregate
the information on a treatment $j$ between the strata $k$ and $h$ if $\theta_{j,h}\neq\theta_{j,k}$ and
$N_{j,h,n}, N_{j,k,n}\rightarrow \infty$ (i.e.
$I_{j,n}(h)$ is definitely included in $\{h\}\cup A^*_{j,h}$, where we recall that $A^*_{j,h}=\{k\neq h:\, \theta_{j,k}=\theta_{j,h}\}$).
Finally, if we also assume $\lim_{n\rightarrow\infty} c_n\sqrt{ n/\ln(\ln(n)) }=\infty$, then we can apply Theorem~\ref{teoAsiNorm}, so obtaining the CLTs on both
$\widehat{\theta}_{j,h,n}$ and $P_{j,h,n}$.

\section{Additional simulation results}\label{sec:add_sim}

In this section, we report additional simulation results. In
particular, Section \ref{subs:G1} investigates the behavior of the proposed procedures
under an unbalanced distribution of patients across strata; Section \ref{subs:G2} reports
additional results on power and type I error rate. Section \ref{subs:G3} illustrates the
finite-sample behavior of the stratum-specific estimator $\widehat{\theta}_{j,h,n}$. Section \ref{subs:G4} shows the evolution of the
stratum-specific randomization probabilities and treatment allocation proportions; Section \ref{subs:G5} focuses on the amount of borrowing induced by the three IUDs, measured through $1-\rho_{j,h,n}$. Finally, Sections \ref{subs:G6} and \ref{subs:G7}
provide sensitivity analyses with respect to the allocation function $f$ and to
the parameter $\psi_{\max}$ in IUD1, respectively.

\subsection{Unbalanced distribution of patients across strata}\label{subs:G1}

\begin{figure}[tbhp]\caption{ INF$_{h,n}$ (left column) and PW$_{h,n}$ (right column) for $S_1$ with $n=200$ and $\mathbf{p}=(0.3, 0.3, 0.05, 0.05, 0.3)^\top$ }\label{unb1}
\includegraphics[scale=.85]{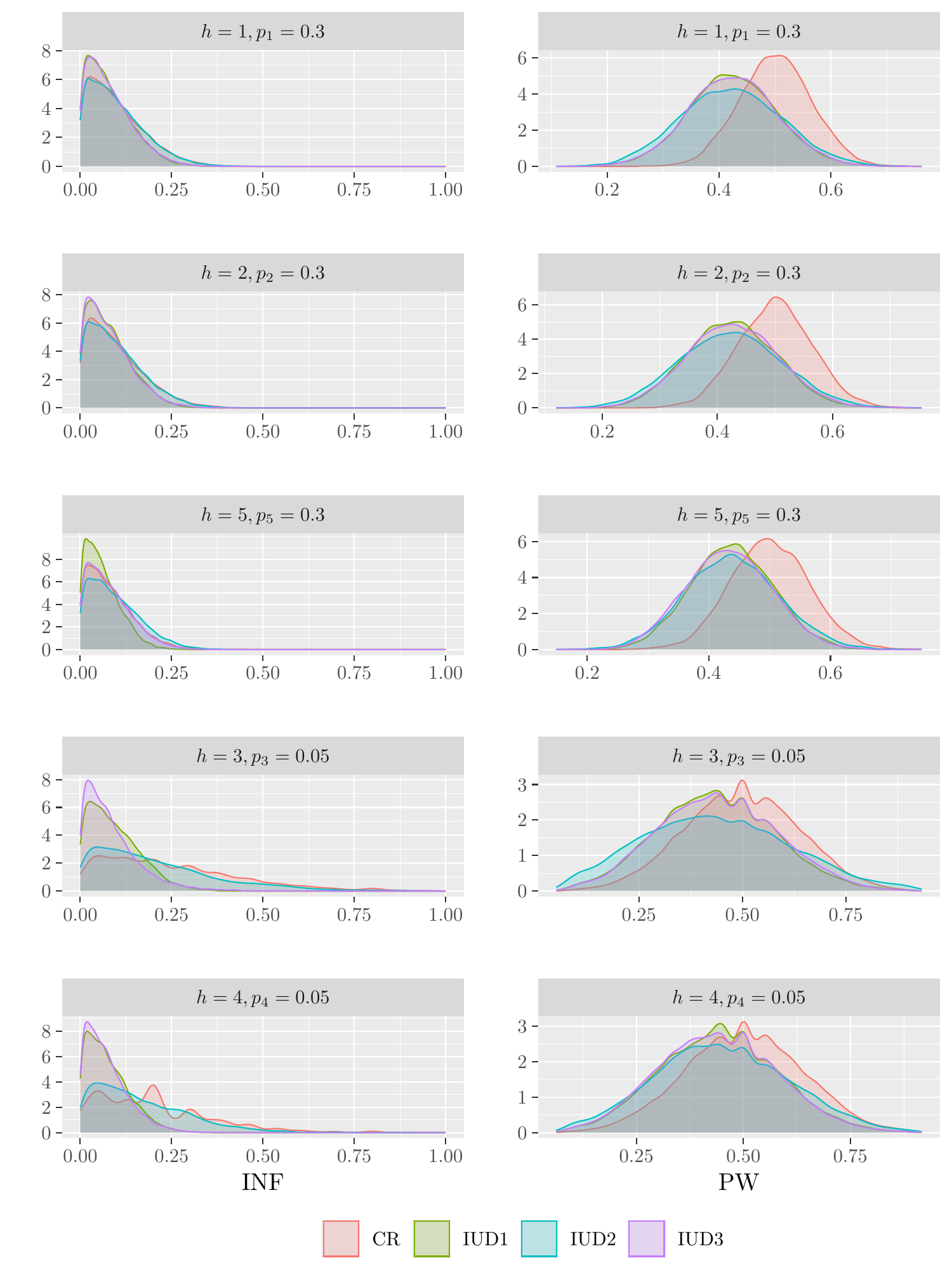}
\end{figure}

We now investigate the IUD's features, including comparisons with the PBD, under an unbalanced distribution of patients across strata.  The stratum-specific results for the inferential metric based on the contrasts  
and the percentage of assignments to the worst treatment with $n=200$ in scenario $S_1$ are reported in Figure \ref{unb1} for $\mathbf{p}=(0.3, 0.3, 0.05, 0.05, 0.3)^\top$. The corresponding means are included in Table \ref{tab_med}. In the case of under-represented strata (as for 3 and 4 here, see the last two rows of the figure), the sharing of information across the right sub-groups enhances the inferential performance when adopting $P_{j,h,n}$ to estimate the stratum specific success probabilities. 
Note that, in the right column of Figure \ref{unb1}, the expected percentage of assignments to the worst treatment of the PBD, i.e. 0.5, is represented by the red vertical line.

\begin{table}[h]\caption{Simulated average $\sqrt{\left[\left(P_{1,h,n}-P_{2,h,n}\right)-\left(\theta_{1,h}-\theta_{2,h}\right)\right]^2}$ for $n=200$ in scenario $S_1$ and unbalanced distribution of patients across strata.}\label{tab_med}
\centering
\begin{tabular}{cccccc}
\hline
 & $h=1$ & $h=2$ & $h=3$ & $h=4$ & $h=5$ \\
\hline
IUD1 & 0.083 & 0.083 & 0.083 & 0.068 & 0.068 \\
IUD2 & 0.103 & 0.103 & 0.102 & 0.087 & 0.085 \\
IUD3 & 0.075 & 0.076 & 0.074 & 0.068 & 0.067 \\
PBD  & 0.117 & 0.117 & 0.116 & 0.094 & 0.095 \\
\hline
\end{tabular}
\end{table}

\subsection{Power and type I error rate}\label{subs:G2}

\begin{figure}[tbhp]\caption{ Simulated average power POW$_{h,n}$ for IUD (with $c_i=1/(\ln(i))-0.15$ and $\psi_{\max}=6$) and PBD}\label{fig_power}
\includegraphics[scale=.8]{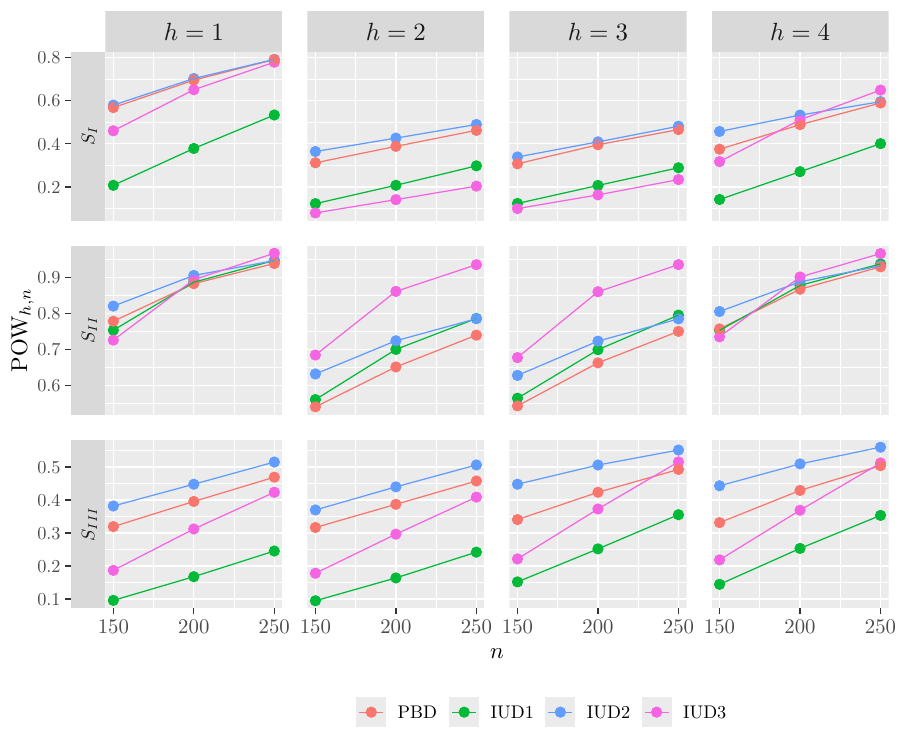}
\end{figure}

As discussed in Remark \ref{rem_test}, we provide the theoretical results that enable testing a variety of hypotheses that can be tailored to the specific experimental context at hand. As an example, we present the simulated average power POW$_{h,n}$ for the stratum-specific one-tailed test, adopting the test statistic in \eqref{eq_ts}, where for the IUD, the estimated $\theta$'s are replaced by the corresponding $P_{j,h,n}$ 
taking into account $H=4$ strata (assuming uniform distribution of patients across strata) and a significance level of $5\%$. For values of $\theta$ described in Table \ref{tab_scen_pow}, the results are displayed in Figure \ref{fig_power}. Within $S_I$, the values of $\boldsymbol{\theta}_1$ do not call  for sharing information across strata while $\boldsymbol{\theta}_2$ induces  borrowing across clusters of strata: indeed, in this case, IUD1 and IUD3 exhibit lower power, while IUD2 is associated with higher values of POW$_{h,n}$, similar to the PBD. In $S_{II}$ all the IUDs perform well, with a gain of up to approximately 25 percentage points on the PBD. The IUD1 tends to show similar power to the PBD for $n \geq 150$, due to its upper bound on the information sharing. Finally, in $S_{III}$, in which $\boldsymbol{\theta}_1$ suggests a  generalized borrowing and $\boldsymbol{\theta}_2$ suggests sharing information only across strata 1-2 and 3-4, we observe that the performance of IUD1 and IUD3 deteriorates while, thanks to the correct aggregation of information, IUD2 is associated with a power gain over all the other designs. With regard to type I error rate, the values in Table \ref{tab:typeIerr}, show that IUD1 and IUD3 are slightly conservative while IUD2 is closer to the nominal level, showing a slight inflation when the values of the treatment effects are closer, due to the increasing aggregation of strata.\\

\begin{table}[ht!]
\caption{Description of simulating parameters for Figure \ref{fig_power}. }\label{tab_scen_pow}
\centering
\begin{tabular}{c|cc}
\hline
& & \\
Scenario & $\boldsymbol{\theta}_1^\top$ & $\boldsymbol{\theta}_2^\top$  \\
\vspace{-0.2 cm}
& & \\
\hline
$S_{I}$ & $(0.2,0.7,0.5,0.9)$ & $(0.5,0.5,0.7,0.7)$\\
$S_{II}$ & $(0.6, 0.6, 0.6, 0.7)$ & $(0.2, 0.3, 0.3, 0.3)$ \\
$S_{III}$ & $(0.5, 0.5, 0.6, 0.6)$ & $(0.3,0.3,0.8,0.8)$ \\
\hline
\end{tabular}
\end{table}

\begin{table}[h]\caption{Type I error rate for $n=250$.} \label{tab:typeIerr}
\centering
\begin{tabular}{ccccc}
\hline
  & {IUD1} & {IUD2} & {IUD3} & {PBD} \\
\hline
$\boldsymbol{\theta}_1=\boldsymbol{\theta}_2=(0.5,0.5,0.7,0.7)^\top$\\
\hline
$h=1$ & 0.02 & 0.07 & 0.02 & 0.05 \\
$h=2$ & 0.02 & 0.06 & 0.01 & 0.05 \\
$h=3$ & 0.02 & 0.07 & 0.02 & 0.04 \\
$h=4$ & 0.03 & 0.07 & 0.02 & 0.04 \\
\hline
$\boldsymbol{\theta}_1=\boldsymbol{\theta}_2=(0.3,0.3,0.8,0.8)^\top$\\
\hline
$h=1$ & 0.02 & 0.05 & 0.03 & 0.05 \\
$h=2$ & 0.02 & 0.05 & 0.04 & 0.05 \\
$h=3$ & 0.03 & 0.06 & 0.05 & 0.04 \\
$h=4$ & 0.03 & 0.06 & 0.05 & 0.04 \\
\hline
\end{tabular}
\end{table}

\subsection{Behavior of stratum specific estimator $\hat{\theta}_{j,h,n}$}\label{subs:G3}

In this section we illustrate the finite-sample behavior of $\widehat{\theta}_{j,h,n}$   (its asymptotic properties were already established in Theorem \ref{teoCons}). 

\begin{figure}[h] \caption{Behavior of $\widehat{\theta}_{j,h,n}$ as $n$ increases for the scenarios in Table \ref{tab_scen} of the main paper.}
    \label{fig:thetaP}
    \centering
    \includegraphics[width=1\linewidth]{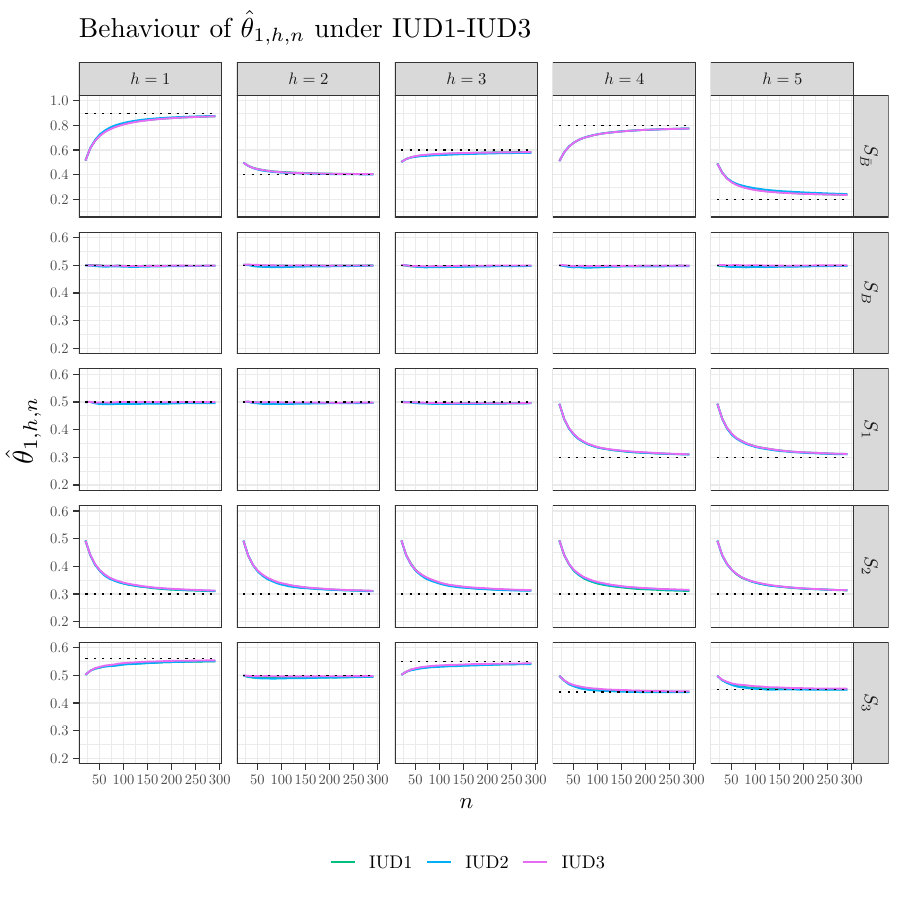}
\end{figure}

Figure \ref{fig:thetaP} focuses on treatment 1 and reports the finite-sample behavior of the stratum-specific empirical estimator $\widehat{\theta}_{1,h,n}$ in the scenarios of Table \ref{tab_scen} of the main paper, assuming uniform distribution of patients across strata. The dotted black lines denote the true values $\theta_{1,h}$. The curves show that $\widehat{\theta}_{1,h,n}$ moves toward the corresponding true stratum-specific value as the sample size increases. It is worth noting that, under the uniform entry mechanism and with $H=5$ strata, the largest sample size considered here, $n=300$, corresponds to approximately 60 patients per stratum.

\subsection{Stratum-specific randomization probabilities and treatment allocation proportions of IUDs}\label{subs:G4}
\begin{figure}[h]
    \centering
     \caption{Simulated average $\pi_{1,h,n}$ as $n$ increases, under scenarios of Table \ref{tab_scen} of the main paper.}
    \label{fig:pigreco}
    \includegraphics[width=1\linewidth]{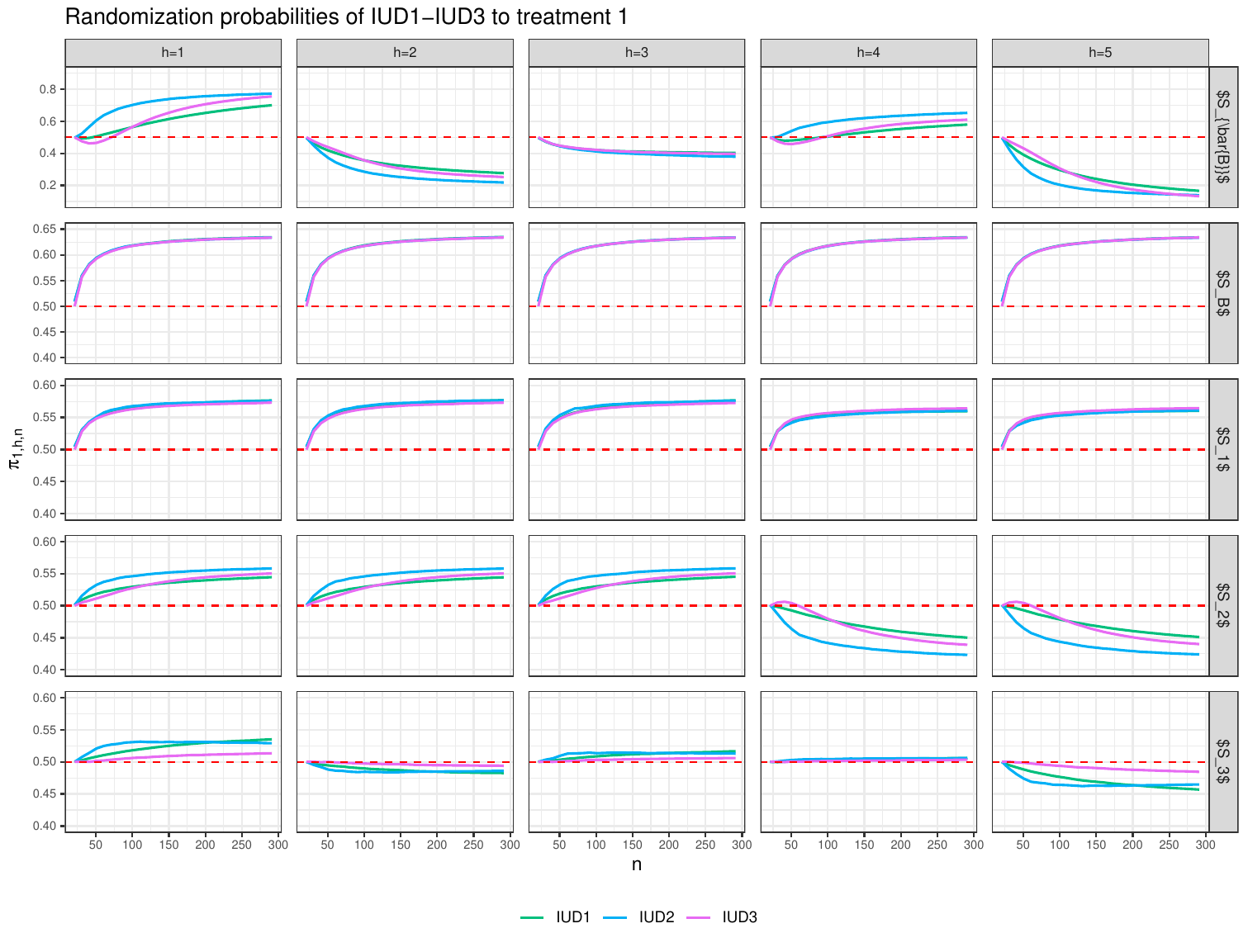}
\end{figure}

\begin{figure}[h] \caption{Simulated average $N_{1,h,n}/N_{\centerdot, h, n}$ as $n$ increases, under scenarios of Table \ref{tab_scen} of the main paper.}
    \label{fig:allocprob}
    \centering
    \includegraphics[width=1\linewidth]{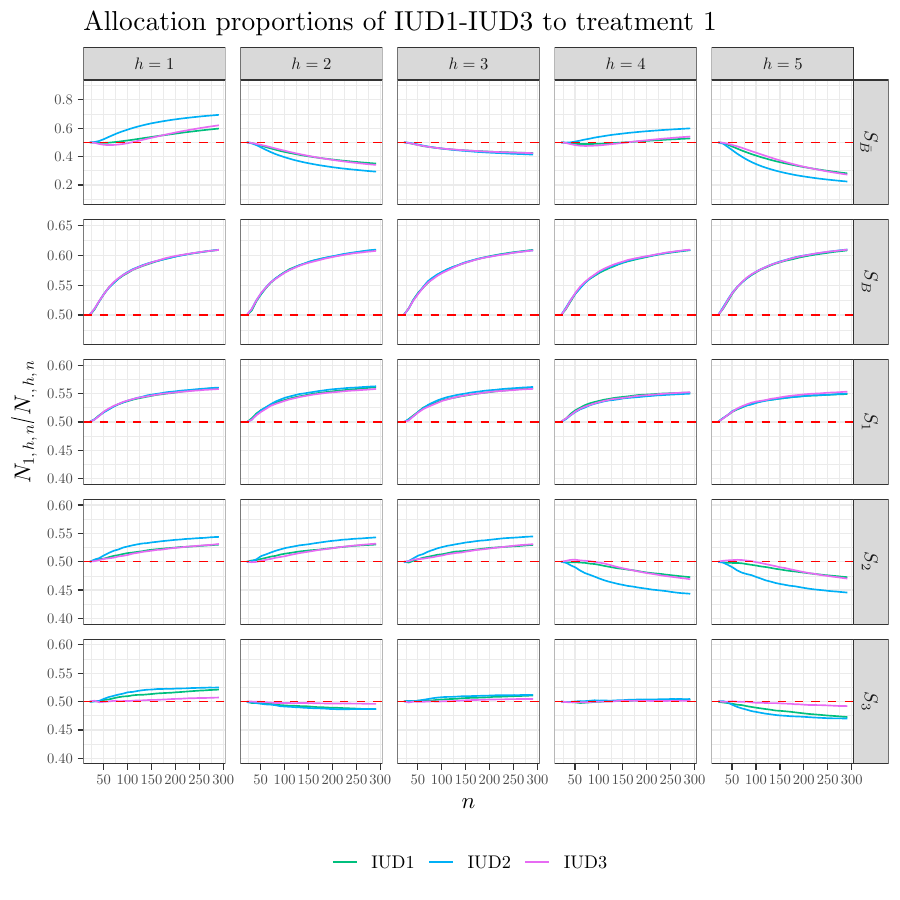}
\end{figure}

In this section, under the scenarios described in Table \ref{tab_scen} of the main paper and uniform distribution of patients across strata, we show how the randomization probabilities to treatment 1, $\pi_{1,h,n}$, and the stratum-specific allocation proportions,
$N_{1,h,n}/N_{\centerdot,h,n}$, evolve as $n$ increases under the three proposed IUDs (see Figures \ref{fig:pigreco} and \ref{fig:allocprob}, respectively).
Since $J=2$, the corresponding quantities for treatment 2 are obtained as their
complements.

The two quantities are expected to exhibit a similar limiting behavior, as Theorem \ref{teoCons} proves that
$N_{1,h,n}/N_{\centerdot,h,n}$ converges almost surely to $\pi_{1,h}$. Thus, both the randomization probabilities and the allocation proportions move above $0.5$ in the strata
where treatment 1 is preferable and below $0.5$ in the strata where treatment 2 is preferable, with differences among IUD1-IUD3 mainly affecting how quickly and how strongly
the allocations move away from balance.
 
\clearpage

\subsection{Amount of borrowing $1-\rho_{j,h,n}$}\label{subs:G5}

\begin{figure}[h] \caption{Behavior of 
$1-\rho_{1,h,n}$ in the benchmark scenarios $S_B$ and
$S_{\bar B}$ (see Table \ref{tab_scen} of the main paper). }
    \label{fig:amountborrowing}
    \centering   
    \includegraphics[width=1\linewidth]{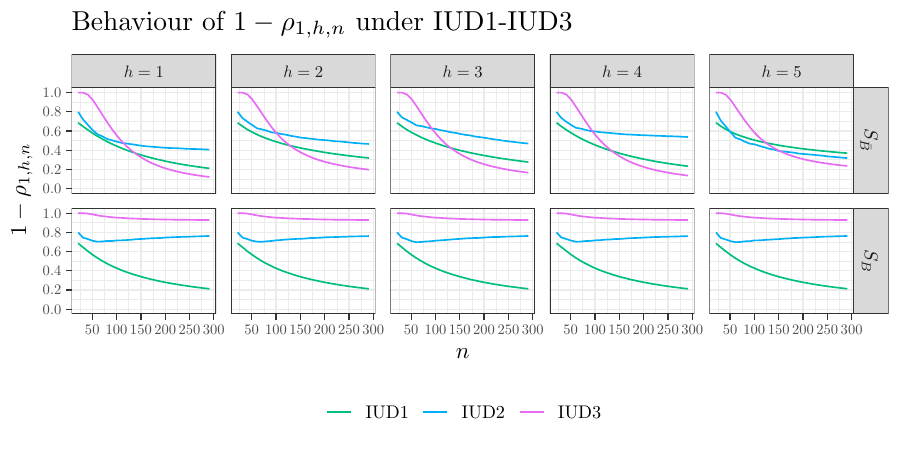}
\end{figure}

To illustrate the amount of borrowing induced by the three proposed IUDs, we report the behavior of $1-\rho_{1,h,n}$, which measures the weight given to the information sharing across strata (values close to
zero indicate limited borrowing, whereas larger values correspond to stronger borrowing). We focus on the two benchmark scenarios $S_B$ and $S_{\bar B}$, which represent opposite borrowing settings. The results in Figure \ref{fig:amountborrowing} show that the amount of borrowing depends on both the update mechanism and the underlying scenario. In $S_{\bar B}$, where borrowing is expected to provide no benefit, $1-\rho_{1,h,n}$ generally decreases as the sample size increases, reflecting the increasing role of the stratum-specific information. Conversely, in $S_B$, where borrowing is expected to be beneficial, the amount of borrowing remains relatively large, especially for IUD2 and IUD3.

\subsection{Sensitivity to different choices of $f$ in the randomization probabilities \eqref{randfunction}}\label{subs:G6}

\begin{figure}[h]
\caption{Behavior of the simulated averaged $\pi_{1,h,n}$ for the proposed design as $f$ in \eqref{randfunction} changes.}
    \label{fig:choicef}
    \centering
\includegraphics[width=1\linewidth]{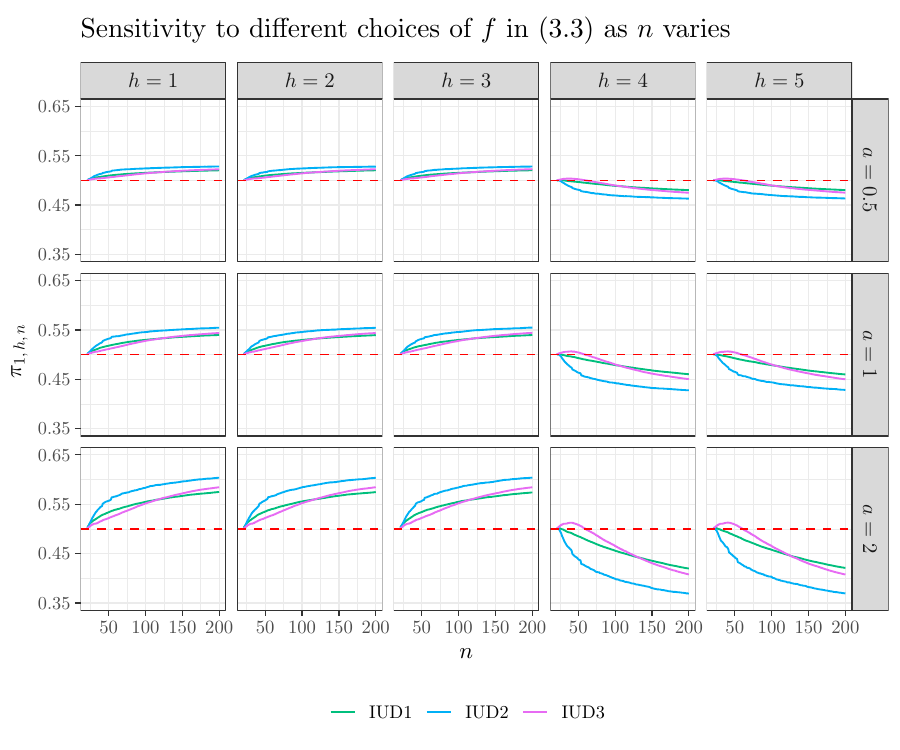}
\end{figure}

 In this section we perform a sensitivity analysis to assess the impact of
the allocation function $f(\cdot)$ on the behavior of the proposed designs. For illustrative purposes, we focus on scenario $S_2$ (see Table \ref{tab_scen} of the main paper) and we consider $f(x)=(1-x)^{-a}$, with different values of $a$. The choice $a=1$ corresponds to the allocation function used in the main simulation study, whereas smaller values of $a$ induce a more conservative behavior, with allocation probabilities closer to balance, and larger values of $a$ produce a more aggressive skewing toward the treatment currently estimated as more promising (recalling that in scenario $S_2$, treatment 1 is preferable in strata $1-3$, whereas treatment 2 is preferable in strata $4$ and $5$).
Figure \ref{fig:choicef} reports  $\pi_{1,h,n}$, under different choices of $a$, as $n$ varies.  As expected, increasing $a$ leads to a stronger departure from balanced allocation, while smaller
values keep the randomization probabilities closer to $0.5$.

Clearly, this family is only one possible choice for the function $f$. Other increasing functions satisfying $f(0)>0$ could be adopted, such as exponential
allocation functions $f(x)=\exp(a \cdot x)$, with $a>0$.

\subsection{Sensitivity of IUD1 to different choices of $\psi_{\max}$}\label{subs:G7}

\begin{figure}[h]
    \centering
      \caption{Sensitivity of IUD1 to different choices of $\psi_{\max}$}  \label{fig:PSIMAX}
\includegraphics[width=1\linewidth]{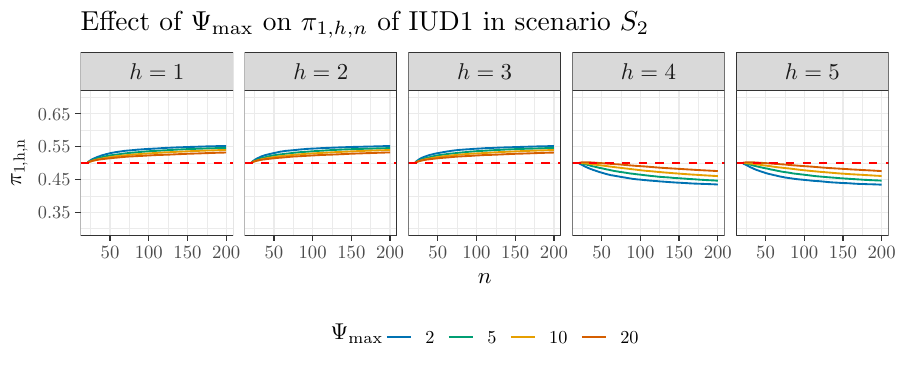}
\end{figure}

To further investigate the role of
$\psi_{\max}$ in IUD1, Figure \ref{fig:PSIMAX} reports the behavior of $\pi_{1,h,n}$, under different values of
$\psi_{\max}$. In Scenario $S_2$ treatment 1 is preferable in
strata $1-3$, whereas treatment 2 is preferable in strata $4$ and $5$. As
expected, the randomization probability to treatment 1 increases above $0.5$ in
the first three strata and decreases below $0.5$ in the last two strata. Smaller
values of $\psi_{\max}$ lead to a more stratum-specific behavior, while larger
values induce stronger borrowing from the other strata and keep the allocation
probabilities closer to $0.5$.

\begin{thebibliography}{10}

\bibitem{urne1}
G.~Aletti, I.~Crimaldi, and A.~Ghiglietti.
\newblock Synchronization of reinforced stochastic processes with a
  network-based interaction.
\newblock {\em The Annals of Applied Probability}, 27:3787--3844, 2017.

\bibitem{urne2}
G.~Aletti, I.~Crimaldi, and A.~Ghiglietti.
\newblock Interacting reinforced stochastic processes: statistical inference
  based on the weighted empirical means.
\newblock {\em Bernoulli}, 26:1098--1138, 2020.

\bibitem{urne3}
G.~Aletti, I.~Crimaldi, and A.~Ghiglietti.
\newblock Networks of reinforced stochastic processes: a complete description
  of the first-order asymptotics.
\newblock {\em Stochastic Processes and their Applications}, 176:104427, 2024.

\bibitem{AlGhRo}
G.~Aletti, A.~Ghiglietti, and W.~F. Rosenberger.
\newblock Nonparametric covariate-adjusted response-adaptive design based on a
  functional urn model.
\newblock {\em Ann. Statist.}, 46(6B):3838--3866, 2018.

\bibitem{BaiHu}
Z.-D. Bai and F.~Hu.
\newblock Asymptotics in randomized urn models.
\newblock {\em The Annals of Applied Probability}, 15(1B):914--940, 2005.

\bibitem{BaiHuRosenberger2002}
Z.~D. Bai, F.~Hu, and W.~F. Rosenberger.
\newblock Asymptotic properties of adaptive designs for clinical trials with
  delayed response.
\newblock {\em The Annals of Statistics}, 30(1):122--139, 2002.

\bibitem{baldi2015adaptive}
A.~Baldi~Antognini and A.~Giovagnoli.
\newblock {\em Adaptive Designs for Sequential Treatment Allocation}.
\newblock Chapman \& Hall/CRC Biostatistics, 2015.

\bibitem{Berry13}
S.~M. Berry, K.~R. Broglio, S.~Groshen, and D.~A. Berry.
\newblock Bayesian hierarchical modeling of patient subpopulations: Efficient
  designs of phase ii oncology clinical trials.
\newblock {\em Clinical Trials}, 10(5):720--734, 2013.

\bibitem{chang-pollard}
J.~T. Chang and D.~Pollard.
\newblock Conditioning as disintegration.
\newblock {\em Statistica Neerlandica}, 51(3):287--317, 1997.

\bibitem{Chu18}
Y.~Chu and Y.~Yuan.
\newblock A bayesian basket trial design using a calibrated bayesian
  hierarchical model.
\newblock {\em Clinical Trials}, 15(2):149--158, 2018.

\bibitem{BLAST}
Y.~Chu and Y.~Yuan.
\newblock Blast: Bayesian latent subgroup design for basket trials accounting
  for patient heterogeneity.
\newblock {\em Journal of the Royal Statistical Society Series C, Applied
  Statistics}, 67:723--740, 2018.

\bibitem{Cri19Res}
I.~Crimaldi, P.~{Dai Pra}, P.-Y. Louis, and I.~G. Minelli.
\newblock Synchronization and functional central limit theorems for interacting
  reinforced random walks.
\newblock {\em Stochastic Processes and their Applications}, 129(1):70--101,
  2019.

\bibitem{CrimaldiLeisen2008}
I.~Crimaldi and F.~Leisen.
\newblock Asymptotic results for a generalized {P}{\'o}lya urn with
  ``multi-updating'' and applications to clinical trials.
\newblock {\em Communications in Statistics -- Theory and Methods},
  37(17):2777--2794, 2008.

\bibitem{Cri23}
I.~Crimaldi, L.~Pierre-Yves, and I.~Minelli.
\newblock Statistical test for an urn model with random multidrawing and random
  addition.
\newblock {\em Stochastic Processes and their Applications}, 158:342--360,
  2023.

\bibitem{Cun17}
K.~Cunanan, A.~Iasonos, R.~Shen, C.~Begg, and M.~G\"{o}nen.
\newblock An efficient basket trial design.
\newblock {\em Statistics in Medicine}, 36(10):1568--1579, 2017.

\bibitem{fdaMP}
FDA.
\newblock Master protocols: Efficient clinical trial design strategies to
  expedite development of oncology drugs and biologics guidance for industry,
  2022.
\newblock Available on line at \url{https://www.fda.gov/media/120721/download}.

\bibitem{park19}
J.~H~Park, E.~Siden, M.~Zoratti, L.~Dron, O.~Harari, J.~Singer, R.~Lester,
  K.~Thorlund, and E.~Mills.
\newblock Systematic review of basket trials, umbrella trials, and platform
  trials: a landscape analysis of master protocols.
\newblock {\em Trials}, 20:1--10, 2019.

\bibitem{Hobbes18}
B.~P. Hobbes and R.~Landin.
\newblock Bayesian basket trial design with exchangeability monitoring.
\newblock {\em Statistics in Medicine}, 37:3557--3572, 2018.

\bibitem{Hu06}
F.~Hu and W.~F. Rosenberger.
\newblock {\em The Theory of Response-Adaptive Randomization in Clinical
  Trials}.
\newblock New York: John Wiley \& Sons, 2006.

\bibitem{HZCC08}
F.~Hu, L.-X. Zhang, S.~H. Cheung, and W.~S. Chan.
\newblock Doubly adaptive biased coin designs with delayed responses.
\newblock {\em The Canadian Journal of Statistics}, 36:541--559, 2008.

\bibitem{Jen00}
C.~Jennison and B.~W. Turnbull.
\newblock {\em Group Sequential Methods with Applications to Clinical Trials}.
\newblock Chapman \& Hall, New York, 2000.

\bibitem{Lu21}
C.~Lu, X.~Li, K.~Broglio, P.~Bycott, Q.~Jiang, X.~Li, A.~McGlothlin, H.~Tian,
  and J.~Ye.
\newblock Practical considerations and recommendations for master protocol
  framework: basket, umbrella and platform trials.
\newblock {\em Therapeutic Innovation \& Regulatory Science}, 55:1145--1154,
  2021.

\bibitem{May09}
C.~May and N.~Flournoy.
\newblock Asymptotics in response-adaptive designs generated by a two-color,
  randomly reinforced urn.
\newblock {\em The Annals of Statistics}, 37(2):1058--1078, 2009.

\bibitem{Neu16}
B.~Neuenschwander, S.~Wandel, S.~Roychoudhury, and S.~Bailey.
\newblock Robust exchangeability designs for early phase clinical trials with
  multiple strata.
\newblock {\em Pharmaceutical Statistics}, 15(2):123--134, 2016.

\bibitem{Saville16}
B.~R. Saville and S.~M. Berry.
\newblock Efficiencies of platform clinical trials: a vision of the future.
\newblock {\em Clinical Trials}, 13:358--366, 2016.

\bibitem{Stout70}
W.~F. Stout.
\newblock A martingale analogue of kolmogorov's law of the iterated logarithm.
\newblock {\em Wahrscheinlichkeitstheorie verw Gebiete}, 15:279--290, 1970.

\bibitem{Thall03}
P.~F. Thall, J.~Wathen, N.~Bekele, L.~Baker, and B.~RS.
\newblock Hierarchical bayesian approaches to phase ii trials in diseases with
  multiple subtypes.
\newblock {\em Statistics in Medicine}, 22:763--780, 2003.

\bibitem{Ventz17}
S.~Ventz, W.~T. Barry, G.~Parmigiani, and L.~Trippa.
\newblock Bayesian response-adaptive designs for basket trials.
\newblock {\em Biometrics}, 55:905--915, 2017.

\bibitem{Wason14}
J.~M.~S. Wason and L.~Trippa.
\newblock A comparison of bayesian adaptive randomization and multi-stage
  designs for multi-arm clinical trials.
\newblock {\em Statistics in Medicine}, 33:2206--2221, 2014.

\bibitem{WeiLac88}
L.~J. Wei and L.~J. M.
\newblock Properties of the urn randomization in clinical trials.
\newblock {\em Controlled Clinical Trials}, 9:345–--364, 1988.

\bibitem{will}
D.~Williams.
\newblock {\em Probability with Martingales}.
\newblock Cambridge University Press, 1991.

\bibitem{Wood17}
J.~Woodcock and L.~LaVange.
\newblock Master protocols to study multiple therapies, multiple diseases, or
  both.
\newblock {\em The New England Journal of Medicine}, 277:62--70, 2017.

\bibitem{Zha07b}
L.-X. Zhang, W.~S. Chan, S.~H. Cheung, and F.~Hu.
\newblock A generalized drop-the-loser urn for clinical trials with delayed
  responses.
\newblock {\em Statistica Sinica}, 17:387--409, 2007.

\bibitem{Zhang07}
L.~X. Zhang, F.~Hu, S.~H. Cheung, and W.~S. Chan.
\newblock Asymptotic properties of covariate-adjusted reponse-adaptive designs.
\newblock {\em The Annals of Statistics}, 35:1166--1182, 2007.

\bibitem{Zheng23}
H.~Zheng, M.~J. Grayling, P.~Mozgunov, T.~Jaki, and J.~M.~S. Wason.
\newblock Bayesian sample size determination in basket trials borrowing
  information between subsets.
\newblock {\em Biostatistics}, 24(4):1000--1016, 2023.

\bibitem{Zheng23biom}
H.~Zheng, T.~Jaki, and J.~M. Wason.
\newblock Bayesian sample size determination using commensurate priors to
  leverage preexperimental data.
\newblock {\em Biometrics}, 79(2):669--683, 2023.

\bibitem{Zheng22}
H.~Zheng and J.~M.~S. Wason.
\newblock Borrowing of information across patient subgroups in a basket trial
  based on distributional discrepancy.
\newblock {\em Biostatistics}, 23:120--135, 2022.

\bibitem{Zhou24}
T.~Zhou and Y.~Ji.
\newblock Bayesian methods for information borrowing in basket trials: an
  overview.
\newblock {\em Cancers}, 16:1--19, 2024.

\bibitem{Zhu10}
H.~Zhu and F.~Hu.
\newblock Sequential monitoring of response-adaptive randomized clinical
  trials.
\newblock {\em The Annals of Statistics}, 38(4):2218--2241, 2010.

\bibitem{Zhu19}
H.~Zhu and F.~Hu.
\newblock Sequential monitoring of covariate-adaptive randomized clinical
  trials.
\newblock {\em Statistica Sinica}, 29:265--282, 2019.

\end{thebibliography}

\end{document}